\RequirePackage{fix-cm} 
\documentclass[a4paper, twoside, reqno, 12pt]{amsart}
\usepackage{fixltx2e}   

\usepackage{etex}

\usepackage[T1]{fontenc}

\usepackage{esint}
\usepackage{dsfont}
\usepackage{xspace}
\usepackage{amsgen}
\usepackage{amsthm}
\usepackage{amssymb}
\usepackage{amsmath}
\usepackage{wasysym}
\usepackage{upgreek}
\usepackage{amsfonts}
\usepackage{stmaryrd}
\usepackage{nicefrac}
\usepackage{mathtools}

\usepackage[mathcal, mathscr]{euscript}

\usepackage{mathrsfs}
\DeclareMathAlphabet{\mathscrbf}{OMS}{mdugm}{b}{n}

\usepackage[lined, boxed, linesnumbered, english]{algorithm2e}

\usepackage{multicol}
\usepackage{multirow}

\usepackage{a4wide}

\usepackage[dvipsnames, table]{xcolor}
\definecolor{bckg}{RGB}{20.8, 20.8, 20.8}
\definecolor{oneblue}{rgb}{0.0, 0.0, 0.85}
\definecolor{Lightblue}{RGB}{214, 214, 214}
\definecolor{bluepigment}{rgb}{0.2, 0.2, 0.6}
\definecolor{charcoal}{rgb}{0.21, 0.27, 0.31}
\definecolor{denimblue}{rgb}{0.08, 0.38, 0.74}
\definecolor{Lightgray}{rgb}{0.89, 0.89, 0.89}
\definecolor{darkgrey}{rgb}{0.273, 0.281, 0.30}
\definecolor{darkelectricblue}{rgb}{0.33, 0.41, 0.47}

\usepackage[sort&compress, comma, square, numbers]{natbib}

\usepackage{psfrag}
\usepackage{graphicx}
\usepackage{subfigure}
\usepackage{morefloats}
\usepackage{indentfirst}

\usepackage{acronym}
\usepackage{microtype}
\usepackage[labelsep=period,%
            labelfont={bf,sf,color=bluepigment},%
            justification=raggedright]{caption}

\usepackage[perpage, symbol]{footmisc}

\usepackage[usenames, pdf]{pstricks}
\usepackage{epsfig}
\usepackage{pst-grad} 
\usepackage{pst-plot} 

\usepackage{url}
\usepackage[colorlinks,
           urlcolor=oneblue,
           linkcolor=denimblue,
           citecolor=NavyBlue,
           bookmarksopen=false,
           pdfpagemode=UseNone,
           pagebackref]{hyperref}

\usepackage[explicit]{titlesec}

\titleformat{\section}[block]
  {\color{NavyBlue}\Large\sffamily\bfseries}
  {}
  {0.0em}
  {\colorbox{bckg!5}{\strut\parbox{\dimexpr\linewidth-2\fboxsep\relax}{\thesection. #1}}}
  [\vspace*{0.33em}]

\titleformat{name=\section,numberless}[block]
  {\color{NavyBlue}\Large\sffamily\bfseries}
  {}
  {0.0em}
  {\colorbox{bckg!5}{\strut\parbox{\dimexpr\linewidth-2\fboxsep\relax}{#1}}}
  [\vspace*{0.33em}]

\titleformat{\subsection}
  {\color{NavyBlue}\large\sffamily\bfseries}
  {}
  {0.0em}
  {\colorbox{bckg!5}{\parbox{\dimexpr\linewidth-2\fboxsep\relax}{\thesubsection. #1}}}
  [\vspace*{0.33em}]

\titleformat{name=\subsection,numberless}
  {\color{NavyBlue}\Large\sffamily\bfseries}
  {}
  {0em}
  {\colorbox{bckg!5}{\parbox{\dimexpr\linewidth-2\fboxsep\relax}{#1}}}
  [\vspace*{0.33em}]

\titleformat{\subsubsection}
  {\color{bluepigment}\sffamily\normalsize\bfseries}
  {\thesubsubsection}
  {0.5em}
  {#1}
  [\vspace*{0.33em}]

\titleformat{\paragraph}[runin]
  {\color{bluepigment}\sffamily\small\bfseries}
  {}
  {0em}
  {#1}

\titlespacing{\section}{0.0em}{1.5em plus 2pt minus 2pt}%
{1.0em plus 2pt minus 2pt}[0em]
\titlespacing{\subsection}{0.5em}{1.5em plus 2pt minus 2pt}%
{1.0em}[0em]
\titlespacing{\subsubsection}{0.5em}{1.5em plus 2pt minus 2pt}%
{1.0em plus 2pt minus 2pt}[0em]

\usepackage{titletoc}

\contentsmargin{0.5em}
\newlength{\tocsep} 
\titlecontents{section}[\tocsep]
  {\addvspace{10pt}\bfseries\sffamily}
  {\contentslabel[\thecontentslabel]{\tocsep}}
  {}
  {\ \titlerule*[0.75pc]{.}\ \thecontentspage}
  []
\titlecontents{subsection}[\tocsep]
  {\addvspace{8pt}\sffamily}
  {\contentslabel[\thecontentslabel]{\tocsep}}
  {}
  {\ \titlerule*[0.5pc]{.}\ \thecontentspage}
  []
\titlecontents*{subsubsection}[\tocsep]
  {\addvspace{2pt}\footnotesize\sffamily}
  {}
  {}
  {\ \titlerule*[0.35pc]{.}\ \thecontentspage}
  [\\*]

\makeatletter
\def\@setauthors{%
  \begingroup
  \def\thanks{\protect\thanks@warning}%
  \trivlist
  \centering\footnotesize \@topsep30\p@\relax
  \advance\@topsep by -\baselineskip
  \item\relax
  \author@andify\authors
  \def\\{\protect\linebreak}%
  \textsc{\normalsize\textcolor{darkelectricblue}{\authors}}%
  \ifx\@empty\contribs
  \else
    ,\penalty-3 \space \@setcontribs
    \@closetoccontribs
  \fi
  \endtrivlist
  \endgroup
}
\def\@settitle{\begin{center}%
  \baselineskip14\p@\relax
    \bfseries
    \textsc{\Large\textcolor{charcoal}{\@title}}
  \end{center}%
}
\makeatother

\usepackage{enumitem}
\setlist[description]{%
  topsep=30pt,               
  itemsep=5pt,               
  font={\bfseries\sffamily\color{NavyBlue}}, 
}

\usepackage{fancyhdr}
\usepackage{lastpage}

\newcommand*\Title{\textcolor{bluepigment}{On the solution of coupled transport}}
\newcommand*\Authors{\textcolor{bluepigment}{J.~Berger, S.~Gasparin, D.~Dutykh \& N.~Mendes}}
\newcommand*{\plogo}{\textcolor{gray}{{\texttt{arXiv.org} / \textsc{hal}}}} 

\numberwithin{equation}{section}

\newcommand{\ie}{\emph{i.e.}\xspace}
\newcommand{\eg}{\emph{e.g.}\xspace}

\newcommand{\SG}{\textsc{Scharfetter}--\textsc{Gummel}}

\newcommand{\CN}{\textsc{Crank}--\textsc{Nicolson}}
\newcommand{\Eu}{\textsc{Euler}}
\newcommand{\COMSOL}{\textsc{Comsol\;\texttrademark}}

\newcommand*\unit[1]{ \, \mathsf{#1} \,}

\newcommand{\scal}{\boldsymbol{\cdot}}
\newcommand*\pd[2]{\dfrac{\partial #1}{\partial #2}}
\newcommand*\od[2]{\dfrac{\mathrm{d} #1}{\mathrm{d} #2}}
\renewcommand{\div}{\grad\scal}
\newcommand{\grad}{\boldsymbol{\nabla}}
\newcommand{\eqdef}{\mathop{\stackrel{\,\mathrm{def}}{:=}\,}}

\renewcommand{\O}{\mathcal{O}}
\newcommand*\egal{\ = \ }
\newcommand*\plus{\ + \ }
\newcommand*\moins{\ - \ }
\newcommand{\f}{\mathrm{f}}
\newcommand*\e[1]{\cdot 10^{\,#1}}
\newcommand{\dt}{\Delta t}
\newcommand{\dx}{\Delta x}

\newcommand{\expo}[1]{\mathrm{e}^{\,#1}}
\newcommand{\ii}{\mathrm{i}}

\newcommand{\R}{\mathds{R}}

\newcommand{\half}{{\textstyle{1\over2}}}

\newcommand{\ai}[1]{a_{\,#1}}

\newcommand{\di}[1]{d_{\,#1}}
\newcommand{\h}[1]{h_{#1}}
\newcommand{\ja}[1]{\boldsymbol{j}_{\,\mathrm{a}#1}}
\newcommand{\jc}{\boldsymbol{j}_{\,\mathrm{c}}}
\newcommand{\jd}[1]{\boldsymbol{j}_{\,\mathrm{d}#1}}
\newcommand{\jq}{\boldsymbol{j}_{\,\mathrm{q}}}
\newcommand{\Pc}{P_{\,c}}
\newcommand{\Ps}{P_{\,s}}
\newcommand{\Pv}{P_{\,v}}
\newcommand{\Rv}{R_{\,v}}
\newcommand{\uinf}{u^{\,\infty}}
\newcommand{\vi}{\mathsf{v}}
\newcommand{\vinf}{v^{\,\infty}}

\newcommand{\Thi}[1]{\Theta_{\,#1}}

\begin{document}

\title[\Title]{On the solution of coupled heat and moisture transport in porous material}

\author[J.~Berger]{Julien Berger$^*$}
\address{\textbf{J.~Berger:} LOCIE, UMR 5271 CNRS, Universit\'e Savoie Mont Blanc, Campus Scientifique, F-73376 Le Bourget-du-Lac Cedex, France}
\email{Berger.Julien@univ-smb.fr}
\urladdr{https://www.researchgate.net/profile/Julien\_Berger3/}
\thanks{$^*$ Corresponding author}

\author[S.~Gasparin]{Suelen Gasparin}
\address{\textbf{S.~Gasparin:} LAMA, UMR 5127 CNRS, Universit\'e Savoie Mont Blanc, Campus Scientifique, F-73376 Le Bourget-du-Lac Cedex, France and Thermal Systems Laboratory, Mechanical Engineering Graduate Program, Pontifical Catholic University of Paran\'a, Rua Imaculada Concei\c{c}\~{a}o, 1155, CEP: 80215-901, Curitiba -- Paran\'a, Brazil}
\email{suelengasparin@hotmail.com}
\urladdr{https://www.researchgate.net/profile/Suelen\_Gasparin/}

\author[D.~Dutykh]{Denys Dutykh}
\address{\textbf{D.~Dutykh:} Univ. Grenoble Alpes, Univ. Savoie Mont Blanc, CNRS, LAMA, 73000 Chamb\'ery, France and LAMA, UMR 5127 CNRS, Universit\'e Savoie Mont Blanc, Campus Scientifique, F-73376 Le Bourget-du-Lac Cedex, France}
\email{Denys.Dutykh@univ-smb.fr}
\urladdr{http://www.denys-dutykh.com/}

\author[N.~Mendes]{Nathan Mendes}
\address{\textbf{N.~Mendes:} Thermal Systems Laboratory, Mechanical Engineering Graduate Program, Pontifical Catholic University of Paran\'a, Rua Imaculada Concei\c{c}\~{a}o, 1155, CEP: 80215-901, Curitiba -- Paran\'a, Brazil}
\email{Nathan.Mendes@pucpr.edu.br}
\urladdr{https://www.researchgate.net/profile/Nathan\_Mendes/}



\begin{titlepage}
\thispagestyle{empty} 
\noindent
{\Large Julien \textsc{Berger}}\\
{\it\textcolor{gray}{LOCIE--CNRS, Universit\'e Savoie Mont Blanc, France}}
\\[0.02\textheight]
{\Large Suelen \textsc{Gasparin}}\\
{\it\textcolor{gray}{Pontifical Catholic University of Paran\'a, Brazil}}\\
{\it\textcolor{gray}{LAMA--CNRS, Universit\'e Savoie Mont Blanc, France}}
\\[0.02\textheight]
{\Large Denys \textsc{Dutykh}}\\
{\it\textcolor{gray}{LAMA--CNRS, Universit\'e Savoie Mont Blanc, France}}
\\[0.02\textheight]
{\Large Nathan \textsc{Mendes}}\\
{\it\textcolor{gray}{Pontifical Catholic University of Paran\'a, Brazil}}
\\[0.10\textheight]

\colorbox{Lightblue}{
  \parbox[t]{1.0\textwidth}{
    \centering\huge\sc
    \vspace*{0.7cm}
    
    \textcolor{bluepigment}{On the solution of coupled heat and moisture transport in porous material}

    \vspace*{0.7cm}
  }
}

\vfill 

\raggedleft     
{\large \plogo} 
\end{titlepage}


\newpage
\thispagestyle{empty} 
\par\vspace*{\fill}   
\begin{flushright} 
{\textcolor{denimblue}{\textsc{Last modified:}} \today}
\end{flushright}


\newpage
\maketitle
\thispagestyle{empty}


\begin{abstract}

Comparisons of experimental observation of heat and moisture transfer through porous building materials with numerical results have been presented in numerous studies reported in literature. However, some discrepancies have been observed, highlighting underestimation of sorption process and overestimation of desorption process. Some studies intend to explain the discrepancies by analysing the importance of hysteresis effects as well as carrying out sensitivity analyses on the input parameters as convective transfer coefficients. This article intends to investigate the accuracy and efficiency of the coupled solution by adding advective transfer of both heat and moisture in the physical model. In addition, the efficient \textsc{Scharfetter} and \textsc{Gummel} numerical scheme is proposed to solve the system of  advection--diffusion equations, which has the advantages of being well--balanced and asymptotically preserving. Moreover, the scheme is particularly efficient in terms of accuracy and reduction of computational time when using large spatial discretisation parameters. Several linear and non-linear cases are studied to validate the method and highlight its specific features. At the end, an experimental benchmark from the literature is considered. The numerical results are compared to the experimental data for a pure diffusive model and also for the proposed model. The latter presents better agreement with the experimental data. The influence of the hysteresis effects on the moisture capacity is also studied, by adding a third differential equation.


\bigskip\bigskip
\noindent \textbf{\keywordsname:} Heat and moisture in porous material; Advection--diffusion system equations; \SG ~numerical scheme;  benchmarking sorption--desorption experimental data; moisture sorption hysteresis \\


\end{abstract}


\newpage
\tableofcontents
\thispagestyle{empty}


\newpage
\section{Introduction}

Models to represent physical phenomena of heat and moisture transfer in porous media have been carried since the fifties, with the works of \textsc{Philip} and \textsc{De Vries} \cite{Philip1957} and \textsc{Luikov} \cite{Luikov1966}. In the area of building physics, a detailed review of numerical models has been reported in \cite{Woloszyn2008, Mendes2017}.

Since the robustness of a model relies on its accuracy to predict the physical phenomena, several studies report on the comparison of the results of the numerical model with experimental data. In \cite{James2010}, gypsum boards with an initial moisture content is submitted to an adsorption phase during $24 \ \mathsf{h}$ and then to a desorption phase during $24 \ \mathsf{h}$ also. Several building materials have been considered for similar investigations: spruce plywood and cellulose insulation in \cite{Talukdar2007, Talukdar2007a}, hemp concrete in \cite{Lelievre2014}, Calcium Silicate in \cite{VanBelleghem2011}, wood fiberboard in \cite{Perre2015}. Interested readers may consult \cite{Busser2018} for a complete review on such comparison within the context of building physics.

However, these studies highlight some discrepancies when confronting the numerical predictions with the experimental data. Particularly, results of numerical simulations underestimate the adsorption process and/or overestimate the desorption process. In other words, the experimental moisture front rushes faster than the simulation predicts. To reduce the discrepancies, some studies improved the physical model by incorporating the hysteresis of the moisture sorption material capacity as for instance in \cite{Kwiatkowski2009, Lelievre2014, Colinart2016}. In \cite{Rouchier2017} the authors estimate new material properties to reduce the discrepancies. Nevertheless, the estimated properties have no physical sense since the estimated vapor resistance was lower than one. In \cite{Olek2016} a non-\textsc{Fickian} moisture diffusion model was developed for wood-based materials. In these studies, it is important to note that  the physical model considers only the diffusion process in the moisture transfer. Thus, these models neglect the moisture transfer by advection, which corresponds to the transport of moisture due to an air velocity occurring through the porous matrix. Within the context of transfer phenomena in soils, many models include advective phenomenon, \eg \cite{Simunek2009, Sun2015, Assouline2003}.

In building materials, advection of moisture may also occur. Indeed, a difference of air pressure is observed between the inside and outside parts of a building facade and induces an air velocity through the porous materials. In the case of the above mentioned experimental studies, the air velocity is probably induced by a difference in the boundary vapor pressure. In \cite{Berger2017a}, the physical model was improved by considering moisture transfer by diffusion and advection. However, the coupling with heat transfer through porous material was neglected. This assumption certainly needs to be reconsidered, particularly in the context of building physics, where the temperature has daily and seasonally variations. Therefore, the first objective of this work is to improve the physical model proposed in \cite{Berger2017a}, by including the energy conservation equations, and analyze the effect of this improvement when comparing it to the experimental data from \cite{James2010}.

When dealing with non-linear advection-diffusion equations, it is of capital importance to obtain an accurate solution at low computational costs. An accurate and fast numerical method may be particularly advantageous when it is required to solve inverse problems or performing the sensitivity analysis, where numerous computations of the direct problem are needed. Indeed, when using the unconditionally stable implicit \Eu ~or \CN ~schemes, several sub-iterations are necessary at each time step to treat the non-linearities of the problem \cite{Gasparin2017}. To address this issue, the second objective of this work is to explore the use of the innovative \SG ~numerical scheme for a system of coupled parabolic differential equations. This scheme was studied in \cite{Berger2017a} and interesting results were shown with a very accurate solution obtained at a low computational cost. Since these results were obtained for a single non-linear equation, it is necessary to extend them for the case of a system. The analysis will be performed by comparing the results to analytical solutions and to the one obtained using a commercial software (\COMSOL).

Therefore, the paper, in Section~\ref{sec:HM_transfer}, presents a mathematical model of coupled heat and moisture transfer in porous material, considering both diffusion and advection mechanisms. Then, in Section~\ref{sec:NM_scalar_equation}, the \SG ~numerical scheme is briefly recalled for the scalar case and validated introducing analytical solution considering non-linear material properties. A comparison with with the result of a commercial software, widely used in Building Physics community, will be realized. The properties of the numerical scheme to solve a system of coupled advection--diffusion equations are provided in Section~\ref{sec:NM_coupled_equations}, while Section~\ref{sec:Exp_comp} shows the results of the numerical model confronted with the experimental data to discuss the importance of advection in porous building materials. Section~\ref{sec:conclusion} addresses the final remarks.


\section{Heat and moisture transfer in porous materials}
\label{sec:HM_transfer}

The physical framework involves heat and moisture transfer in porous material. The moisture includes vapor water, denoted by index $1\,$, and liquid water, denoted by index $2\,$. The porous matrix of the material is indexed by $0\,$. We assume that the temperature is much greater than the freezing point and therefore liquid solid phase change is not considered.


\subsection{Moisture transfer}

The transfer of moisture in the porous matrix is driven by a convective flow $\jc \ \mathsf{[\,kg/(s.m^{\,2})\,]}\,$, including both diffusive $\jd{}$ and advective $\ja{}$ fluxes:
\begin{align*}
\jc \egal \jd{} \plus \ja{} \,.
\end{align*}
The advection flux occurs due to the air motion through the pores. The differential equation describing liquid or vapor transport can be formulated as: 

\begin{align}\label{eq:moisture_transport}
  \pd{w_{\,i}}{t} \egal - \, \div \Bigl(\, \jd{\,,\,i} \plus \ja{\,,\,i} \,\Bigr) \plus I_{\,i} \,, && i \egal \bigl\{\, 1 \,,\, 2 \,\bigr\} \,,
\end{align}
where $I_{\,i}\ \mathsf{[\,kg/(s.m^{\,3})\,]}$ is the volumetric term of source ($I_{\,i} \ \geqslant \ 0$) or sink ($I_{\,i} \ < \ 0$) and $w_{\,i}\ \mathsf{[\,kg/m^{\,3}\,]}$ is the volumetric concentration of substance $i\,$. On one hand, the quantity $I_{\,2}$ defines the source term of liquid water occurring by condensation of vapor water into liquid. On the other hand, the quantity $I_{\,1}$ defines the source term of vapor water appearing by evaporation of liquid water. Since it is assumed that water is not present in its solid phase, by definition we have: 
\begin{align*}
I_{\,1} \plus I_{\,2} \egal 0 \,.
\end{align*}
Moreover, if we assume that the mass of vapor is negligible compared to the liquid one ($w_{\,1} \ \ll \ w_{\,2})\,$, then the time variation of $w_{\,1}$ can also be supposed to vanish:
\begin{align*}
  \pd{w_{\,1}}{t} \ \approx \ 0 \,.
\end{align*}
Thus, by applying $i \egal 1$ to Eq.~\eqref{eq:moisture_transport}, we obtain:
\begin{align*}
  0 \egal - \, \div \Bigl(\, \jd{\,,\,1} \plus \ja{\,,\,1} \,\Bigr) \plus I_{\,1} \,,
\end{align*}
which is equivalent to
\begin{align}\label{eq:vapour_transport}
  I_{\,1} \egal \div \Bigl(\, \jd{\,,\,1} \plus \ja{\,,\,1} \,\Bigr) \,.
\end{align}

The diffusive fluxes are given by: 
\begin{align*}
  \jd{\,,\,1} & \egal -\,k_{\,1} \, \grad \, \Pv \,, \\
  \jd{\,,\,2} & \egal -\,k_{\,2} \, \grad \, \Pc \,,
\end{align*}
where $k_{\,1} \ \mathsf{[\,s\,]}$ and $k_{\,2}\ \mathsf{[\,s\,]}$ are the vapor and liquid permeability of the material, respectively. Both depend on the saturation degree of the material. It should be noted that the unity of the vapor permeability is in $[\,\mathsf{s}\,]$ since it is expressed considering the vapor pressure $\Pv$ gradient (and not the vapor mass content $w_{\,1}$). In addition, the dispersion effects on the moisture transport, inducing a modification of the diffusion coefficient due to variation of the velocity in the pores \cite{DeMarsily1986}, is neglected. This hypothesis will be confirmed in Section~\ref{sec:Exp_comp}. The quantities $\Pv$ and $\Pc$ in $\ \mathsf{[\,Pa\,]}$ are the vapor and capillary pressure. In order to use the vapor pressure as the driving potential \cite{Funk2008}, we consider the physical relation, known as \textsc{Kelvin}'s equation, between $\Pv$ and $\Pc\,$:
\begin{align*}
  \Pc & \egal \rho_{\,2} \, \Rv \, T \, \ln\left(\dfrac{\Pv}{\Ps(T)}\right)\,, \\
  \pd{\Pc}{\Pv} & \egal \rho_{\,2} \, \dfrac{\Rv\, T}{\Pv} \,,
\end{align*}
where $\rho_{\,2}\ \mathsf{[\,kg/m^{\,3}\,]}$ is the liquid water density and $\Rv\ \mathsf{[\,J/(kg.K)\,]}$ is the water vapor constant. Thus, neglecting the variation of the capillary pressure with temperature, we have:
\begin{align*}
  \pd{\Pc}{x} \egal \pd{\Pc}{\Pv} \cdot \pd{\Pv}{x} \plus \pd{\Pc}{T} \cdot \pd{T}{x} \ \simeq \ \rho_{\,2} \, \dfrac{\Rv \, T}{\Pv} \cdot \pd{\Pv}{x} \,.
\end{align*}
The diffusive flux of liquid water can then be written as: 
\begin{align*}
  \jd{\,,\,2} & \egal -\,k_{\,2} \, \rho_{\,2} \; \dfrac{\Rv \, T}{\Pv} \; \grad \, \Pv \,.
\end{align*}

The total diffusive flux of moisture can be expressed as: 
\begin{align*}
  \jd{\,,\,m} & \egal \jd{\,,\,1} \plus \jd{\,,\,2} \egal -\,k_{\,m} \, \grad \, \Pv \,,
\end{align*}
where $k_{\,m}\ \eqdef\ k_{\,1} \plus k_{\,2} \, \rho_{\,2} \; \dfrac{\Rv \, T}{\Pv} \ \mathsf{[\,s\,]}\,$. The advective fluxes of moisture in the capillary material are expressed as: 
\begin{align*}
  & \ja{\,,\,1} \egal w_{\,1\,k} \, \boldsymbol{\vi} \,, 
  && \ja{\,,\,2} \egal w_{\,2\,k} \, \boldsymbol{\vi} \,,
\end{align*}
where $\vi\ \mathsf{[\,m/s\,]}$ is the molar average velocity \cite{Whitaker1986, Whitaker1986a}. It is assumed that the velocity is equal for both water and vapor phase. Parameters $w_{\,1\,k}$ and $w_{\,2\,k}\,$, in $\ \mathsf{[\,kg/m^{\,3}\,]}\,$, are the volumetric concentration of moving vapor and liquid mass, respectively. It is assumed that the air motion has no influence on the liquid substance, $\ja{\,,\,2} \ \equiv \ 0 \ \mathsf{[\,kg/(s.m^{\,2})\,]}\,$. Moreover, the quantity of vapor $w_{\,1\,k}$ can be expressed as:
\begin{align*}
  w_{\,1\,k} \egal \dfrac{w_{\,1}}{b_{\,1}}\,,
\end{align*}
where $b_{\,1} \ \mathsf{[\,-\,]}$ is the ratio of the volume of vapor $V_{\,1} \ \mathsf{[\,m^{\,3}\,]}$ to the total volume of capillaries $V\,$. Thus, we have:
\begin{align*}
  b_{\,1}\ \eqdef\ \dfrac{V_{\,1}}{V} \,.
\end{align*}
It is also assumed that there is no variation of the capillaries volume, meaning that the shrinkage and expansion effects, due to variation of the moisture content, are neglected. Using the perfect gas law, we obtain: 
\begin{align*}
  \ja{\,,\,1} \egal \dfrac{\Pv}{R_{\,v} \, T} \; \boldsymbol{\vi} \,.
\end{align*}
Defining the advection coefficient $\boldsymbol{a}_{\,m} \ \eqdef \ \dfrac{\boldsymbol{\vi}}{R_{\,v} \, T} \ \mathsf{[\,s/m\,]}\,$, the moisture advective flow is given by: 
\begin{align*}
  \ja{\,,\,m} & \egal \ja{\,,\,1} \plus \ja{\,,\,2} \ \simeq \ \ja{\,,\,1}  \egal \boldsymbol{a}_{\,m} \, \Pv \,.
\end{align*}

We define the total moisture content as $w_{\,m} \ \eqdef \ w_{\,1} \plus w_{\,2} \ \mathsf{[\,kg/m^{\,3}\,]}\,$. It can be related to the relative humidity using the material sorption curve $w_{\,m} \egal f (\,\phi\, )\,$. It is assumed that the material sorption curve is almost invariant with the temperature \cite{Rouchier2013}. Therefore, we can write:
\begin{align*}
  \pd{w_{\,m}}{t} \ \simeq \ \dfrac{\f^{\,\prime}(\phi)}{\Ps} \ \pd{\Pv}{t}\,,
\end{align*}
where $\Ps\ \mathsf{[\,Pa\,]}$ is the saturation pressure. We denote the moisture capacity $c_{\,m} \ \eqdef \ \dfrac{\f^{\,\prime}(\phi)}{\Ps}$ $\mathsf{[\,kg/(m^{\,3}.Pa)\,]}\,$. By summing Eq.~\eqref{eq:moisture_transport} for $i \egal \bigl\{\, 1 \,,\,2 \,\bigr\}\,$, knowing that $\sum_{i = 1}^2 \, I_{\,i} \egal 0\,$, we obtain the differential equation of moisture transfer in porous material: 
\begin{align*}
  c_{\,m}\;\pd{\Pv}{t} \egal \div \Bigl(\, k_{\,m} \, \grad \, \Pv \moins \boldsymbol{a}_{\,m} \, \Pv \,\Bigr)\,.
\end{align*}


\subsection{Heat transfer}
\label{sec:heat_transfer_sec}

The heat transfer equation is obtained from the first law of thermodynamics. The volumetric concentration of the total enthalpy $\h{}\ \mathsf{[\,J/kg\,]}$ equals the divergence of the enthalpy flux, heat conduction and heat advection, expressed as:
\begin{align}\label{eq:energy_conservation}
  \pd{}{t}\;\biggl(\, \h{\,0} \, \rho_{\,0} \plus \sum_{i=1}^2 \h{\,i} \, w_{\,i} \,\biggr) \egal - \, \div \Bigl(\, \jq \plus \sum_{i=1}^2 \h{\,i} \, \bigl(\, \ja{\,,\,i} \plus \jd{\,,\,i} \,\bigr) \,\Bigr) \,,
\end{align}
where $\rho_{\,0}\ \mathsf{[\,kg/m^{\,3}\,]}$ is the material dry-basis specific mass. The heat flux $\jq\ \mathsf{[\,W/m^{\,2}\,]}$ is driven by the conduction and advection phenomena: 
\begin{align*}
  \jq \egal - \, k_{\,q} \, \grad T \plus \boldsymbol{a}_{\,q} \, T \,,
\end{align*}
where $k_{\,q}\ \mathsf{[\,W/(m.K)\,]}$ is the thermal conductivity of the material depending on the moisture content and $\boldsymbol{a}_{\,q} \ \eqdef \ \rho_{\,a} \, c_{\,a} \, \boldsymbol{\vi}\ \mathsf{[\,J/(K.m^{\,2}.s)\,]}$ is the heat advection coefficient. Parameters $\rho_{\,a}\ \mathsf{[\,kg/m^{\,3}\,]}$ and $c_{\,a}\ \mathsf{[\,J/(m^{\,3}.K)\,]}$ are the density and specific heat capacity of the dry air, correspondingly. It should be noted that the heat capacity of the vapor phase is included in the term $h_{\,1} \, \ja{\,,\,1}\,$. Assuming a constant total volume, Eq.~\eqref{eq:energy_conservation} becomes: 
\begin{align}\label{eq:heat_transport}
  \biggl(\, c_{\,0} \, \rho_{\,0}  \plus \sum_{i\,=\,1}^{\,2} c_{\,i} \, w_{\,i} \,\biggr)\;\pd{T}{t} \plus \sum_{i\,=\,1}^{\,2} \h{\,i} \; \pd{w_{\,i}}{t} \egal & \ - \, \div \, \jq \moins \sum_{i\,=\,1}^{\,2} \h{\,i} \, \div \, \bigl(\, \ja{\,,\,i} \plus \jd{\,,\,i} \,\bigr) \nonumber \\
  & \moins \sum_{i\,=\,1}^{\,2} \bigl(\, \grad  h_{\,i} \,\bigr) \, \scal \, \bigl(\, \ja{\,,\,i} \plus \jd{\,,\,i} \,\bigr) \,.
\end{align}

Then, by summing Eq.~\eqref{eq:moisture_transport} ($i \ \leftarrow \ 1$), multiplied by $h_{\,1}$ and Eq.~\eqref{eq:moisture_transport} ($i \ \leftarrow \ 2$), multiplied by $h_{\,2}\,$, we get: 
\begin{align*}
  \sum_{i\,=\,1}^{\,2} \h{\,i} \;\pd{w_{\,i}}{t} \egal -\, \sum_{i=1}^2 \h{\,i} \, \div \, \bigl(\, \ja{\,,\,i} \plus \jd{\,,\,i} \,\bigr) \plus \sum_{i=1}^2 \h{\,i} \, I_{\,i} \,.
\end{align*}

We denote by $r_{\,1\,2}\ \eqdef\ \h{\,1} \moins \h{\,2}\ \mathsf{[\,J/kg\,]}$ the latent heat of evaporation. We also denote by $c_{\,q}\ \eqdef\ \rho_{\,0} \, c_{\,0} \, \plus \sum_{i\,=\,1}^{\,2} c_{\,i} \, w_{\,i}\ \mathsf{[\,J/(m^{\,3}.K)\,]}$ the total volumetric heat capacity, including the contributions of the material, the liquid water and the vapor phase. Consequently, using Eq.~\eqref{eq:vapour_transport}, Eq.~\eqref{eq:heat_transport} becomes:
\begin{align*}
  c_{\,q} \;\pd{T}{t} & \egal - \, \div \, \jq \moins r_{\,12} \, \div \Bigl(\, \ja{\,,\,1} \plus \jd{\,,\,1} \,\Bigr) \moins \sum_{i=1}^2 \grad \bigl(\, c_{\,i} \, T \,\bigr) \, \scal \, \bigl(\, \ja{\,,\,i} \plus \jd{\,,\,
  i} \,\bigr) \,.
\end{align*}
The last term is assumed negligible \cite{Luikov1966}. This assumption is verified in Section~\ref{sec:Exp_comp}. Considering the expression of the fluxes, we obtain the following differential equation of heat transfer in porous material:
\begin{align*}
  c_{\,q} \;\pd{T}{t} & \egal \, \div \, \Bigl(\, k_{\,q} \, \grad T \moins \boldsymbol{a}_{\,q} \, T  \,\Bigr) \plus r_{\,12} \, \div \Bigl(\, k_{\,1} \, \grad \Pv \moins \boldsymbol{a}_{\,m} \, \Pv \,\Bigr)\,.
\end{align*}

For the sake of clarity, we introduce the coefficients $\boldsymbol{a}_{\,q\,m} \ \eqdef \ r_{\,1\,2} \, \boldsymbol{a}_{\,m}\ \mathsf{[\,W.s^{\,2}/(kg.m)\,]}$ and $k_{\,q\,m} \ \eqdef\ r_{\,1\,2} \, k_{\,1}\ \mathsf{[\,W/(m.K)\,]}\,$. Then, the energy governing differential equation becomes:
\begin{align*}
  c_{\,q} \;\pd{T}{t} & \egal \, \div \, \Bigl(\, k_{\,q} \, \grad T \moins  \boldsymbol{a}_{\,q} \, T  \,\Bigr) \plus \div \Bigl(\, k_{\,qm} \, \grad \Pv \moins \boldsymbol{a}_{\,qm} \, \Pv \,\Bigr) \,.
\end{align*}


\subsection{Initial and boundary conditions}

At the interface of the material with the ambient air, the vapor flux at the interface is proportional to the vapor pressure difference  between the surface and the ambient vapor pressure $\Pv^{\,\infty}\,$:
\begin{align}\label{eq:BC_vapor}
  \jd{\,,\,1} \plus \ja{\,,\,1} \egal \alpha_{\,m} \, \Bigl(\, \Pv \moins \Pv^{\,\infty} \, \Bigr) \cdot \boldsymbol{n}\,,
\end{align}
where $\alpha_{\,m}\ \mathsf{[\,s/m\,]}$ is the surface moisture transfer coefficient.

For the liquid phase, within building physics applications, the flux at the interface is imposed by the ambient air conditions: 
\begin{align}\label{eq:BC_liq}
  \jd{\,,\,2} \egal \boldsymbol{g}_{\,\infty} \,.
\end{align}
If the bounding surface is in contact with the outside building air, then $g_{\,\infty}\ \mathsf{[\,kg/(s.m^{\,2})\,]}$ corresponds to the liquid flux from wind driven rain \cite{Mendes2017}. If the bounding surface is in contact with the inside building air, then $g_{\,\infty} \egal 0\,$.

By summing Eq.~\eqref{eq:BC_vapor} and Eq.~\eqref{eq:BC_liq}, we obtain the boundary condition for the moisture transfer equation:
\begin{align*}
  \ja{\,,\,m} \plus \jd{\,,\,m} & \egal \alpha_{\,m} \, \Bigl(\, \Pv \moins \Pv^{\,\infty} \, \Bigr) \cdot \boldsymbol{n} \plus  \boldsymbol{g}_{\,\infty}\,.
\end{align*}

For the heat transfer, the heat flux $\jq\ \mathsf{[\,W/m^{\,2}\,]}$ occurring by diffusion and advection is proportional to the temperature difference between the surface and the ambient air $T^{\,\infty}\,$:
\begin{align*}
  \jq  \egal \alpha_{\,q} \, \Bigl(\, T \moins T^{\,\infty} \, \Bigr) \cdot \boldsymbol{n} \,,
\end{align*}
where $\alpha_{\,q}$ is the surface heat transfer coefficient. Using this, the total heat flux, including the transfer by diffusion, advection and latent phase change, can be written as:
\begin{align*}
  \jq \plus r_{\,12} \, \Bigl(\, \jd{\,,\,1} \plus \ja{\,,\,1} \,\Bigr) & \egal \alpha_{\,q} \, \Bigl(\, T \moins T^{\,\infty} \, \Bigr) \cdot \boldsymbol{n} \plus r_{\,12} \, \alpha_{\,m} \, \Bigl(\, \Pv \moins \Pv^{\,\infty} \, \Bigr) \cdot \boldsymbol{n} \plus r_{\,12} \; \boldsymbol{g}_{\,\infty}\,,
\end{align*}

As the initial condition, the temperature and vapor pressure distributions within the material are considered to be uniform:
\begin{align*}
  \Pv \egal \Pv^{\,i} \,, && T \egal T^{\,i} \,.
\end{align*}


\subsection{Dimensionless representation}

While performing a mathematical and numerical analysis of a given practical problem, it is of capital importance to obtain a unitless formulation of governing equations, due to a number of good reasons. First of all, it enables to determine important scaling parameters (\textsc{Biot}'s numbers for instance). Henceforth, solving one dimensionless problem is equivalent to solve a whole class of dimensional problems sharing the same scaling parameters. Then, dimensionless equations allow to estimate the relative magnitude of various terms, and thus, eventually to simplify the problem using asymptotic methods \cite{Nayfeh2000}. Finally, the floating point arithmetics is designed such as the rounding errors are minimal if computer manipulates the numbers of the same magnitude \cite{Kahan1979}. Moreover, the floating point numbers have the highest density in the interval $(\,0\,,\,1\,)$ and their density decreases when we move further away from this interval. So, it is always better to manipulate numerically the quantities of the order of $\O\,(1)$ to avoid severe round-off errors and to likely improve the conditioning of the problem in hands.

In this way, we define following dimensionless quantities for the temperature and vapor pressure fields:
\begin{align*}
  & u \egal \dfrac{\Pv}{\Pv^{\,\circ}} \,,
  && \uinf \egal \dfrac{\Pv^{\,\infty}}{\Pv^{\,\circ}} \,,
  && v \egal \dfrac{T}{T^{\,\circ}} \,,
  && \vinf \egal \dfrac{T^{\,\infty}}{T^{\,\circ}} \,,
\end{align*}
where $\Pv^{\,\circ}$ and $T^{\,\circ}$ are the reference values of the fields. The time and space domains are also scaled with characteristic values: 
\begin{align*}
  & x^{\,\star} \egal \dfrac{x}{L} \,, 
  && t^{\,\star} \egal \dfrac{t}{t^{\,\circ}} \,, 
\end{align*}
where $L$ is the length of the material sample. All the material thermo-physical properties are scaled considering a reference value, denoted by the super script $^{\,\circ}$ for each parameter:
\begin{align*}
  & c_{\,q}^{\,\star} \egal \dfrac{c_{\,q}}{c_{\,q}^{\,\circ}} \,,
  && a_{\,q}^{\,\star} \egal \dfrac{a_{\,q}}{a_{\,q}^{\,\circ}} \,, 
  && k_{\,q}^{\,\star} \egal \dfrac{k_{\,q}}{k_{\,q}^{\,\circ}} \,,
  && c_{\,m}^{\,\star} \egal \dfrac{c_{\,m}}{c_{\,m}^{\,\circ}}\,,\\[3pt]
  & a_{\,m}^{\,\star} \egal \dfrac{a_{\,m}}{a_{\,m}^{\,\circ}} \,,
  && k_{\,m}^{\,\star} \egal \dfrac{k_{\,m}}{k_{\,m}^{\,\circ}}\,, 
  && a_{\,qm}^{\,\star} \egal \dfrac{a_{\,qm}}{a_{\,qm}^{\,\circ}} \,,
  && k_{\,qm}^{\,\star} \egal \dfrac{k_{\,qm}}{k_{\,qm}^{\,\circ}} \,.
\end{align*}
Then, dimensionless numbers are introduced. The \textsc{Fourier} number characterizes the importance of the heat and mass transfer through the material:
\begin{align*}
  & \mathrm{Fo}_{\,q} \egal \dfrac{k_{\,q}^{\,\circ} \, t^{\,\circ}}{c_{\,q}^{\,\circ} \, L^{\,2}} \,,
  && \mathrm{Fo}_{\,m} \egal \dfrac{k_{\,m}^{\,\circ} \, t^{\,\circ}}{c_{\,m}^{\,\circ} \, L^{\,2}} \,.
\end{align*}
The \textsc{P\'eclet} number translates the importance of the advection relative to the diffusion in the total transfer:
\begin{align*}
  & \mathrm{Pe}_{\,q} \egal \dfrac{a_{\,q}^{\,\circ} \, L}{k_{\,q}^{\,\circ}} \,,
  && \mathrm{Pe}_{\,m} \egal \dfrac{a_{\,m}^{\,\circ} \, L}{k_{\,m}^{\,\circ}} \,,
  && \mathrm{Pe}_{\,qm} \egal \dfrac{a_{\,qm}^{\,\circ} \, L}{k_{\,qm}^{\,\circ}} \,.
\end{align*}
The parameter $\gamma$ quantifies the coupling effects between moisture and heat transfer:
\begin{align*}
  \gamma \egal \dfrac{k_{\,qm}^{\,\circ} \, \Pv^{\,\circ}}{k_{\,q}^{\,\circ} \, T^{\,\circ}} \,.
\end{align*}

The \textsc{Biot} number appears for the boundary conditions, quantifying the transfer from the ambiant air to the porous material:
\begin{align*}
  & \mathrm{Bi}_{\,m} \egal \dfrac{\alpha_{\,m} \cdot L}{k_{\,m}}  \,,
  && \mathrm{Bi}_{\,q} \egal \dfrac{\alpha_{\,q} \cdot L}{k_{\,q}} \,,
  && \mathrm{Bi}_{\,qm} \egal \dfrac{\alpha_{\,qm} \cdot L}{k_{\,qm}} \,.
\end{align*}

In one space dimension, the unitless system of partial differential equations of heat and mass transfer is therefore formulated as:
\begin{align*}
  c_{\,m}^{\,\star}\;\pd{u}{t^{\,\star}} & \egal \mathrm{Fo}_{\,m} \;\pd{}{x^{\,\star}}\;\Biggl(\, k_{\,m}^{\,\star} \;\pd{u}{x^{\,\star}} \moins \mathrm{Pe}_{\,m} \, a_{\,m}^{\,\star} \, u \,\Biggr) \,, \\[3pt]
  c_{\,q}^{\,\star}\;\pd{v}{t^{\,\star}} & \egal \mathrm{Fo}_{\,q} \;\pd{}{x^{\,\star}}\;\Biggl(\, k_{\,q}^{\,\star} \;\pd{v}{x^{\,\star}} \moins \mathrm{Pe}_{\,q} \, a_{\,q}^{\,\star} \, v \,\Biggr) \plus \mathrm{Fo}_{\,q} \, \gamma \; \pd{}{x^{\,\star}}\;\Biggl(\, k_{\,qm}^{\,\star} \;\pd{u}{x^{\,\star}} \moins \mathrm{Pe}_{\,qm} \, a_{\,qm}^{\,\star} \, u \,\Biggr) \,,
\end{align*}
together with the boundary conditions: 
\begin{align*}
  k_{\,m}^{\,\star} \;\pd{u}{x^{\,\star}} \moins \mathrm{Pe}_{\,m} \, a_{\,m}^{\,\star} \, u & \egal \mathrm{Bi}_{\,m} \, \Bigl(\, u \moins u^{\,\infty}\,\Bigr) \,,\\[3pt]
  k_{\,q}^{\,\star} \;\pd{v}{x^{\,\star}} \moins \mathrm{Pe}_{\,q} \, a_{\,q}^{\,\star} \, v \plus  \gamma \, \Biggl(\, k_{\,qm}^{\,\star} \;\pd{u}{x^{\,\star}} \moins \mathrm{Pe}_{\,qm} \, a_{\,qm}^{\,\star} \, u \,\Biggr) & \egal \mathrm{Bi}_{\,q} \, \Bigl(\, v \moins v^{\,\infty}\,\Bigr) \plus \gamma \, \mathrm{Bi}_{\,qm} \, \Bigl(\, u \moins u^{\,\infty}\,\Bigr) \,.
\end{align*}

It can be noted that the dimensionless coefficients $a^{\,\star}\,$, $d^{\,\star}$ translate the non-linearity (or the distorsion) of the diffusion and advection transfer, relatively to the reference state. Using data from \cite{Abadie2009}, the moisture \textsc{Fourier} and \textsc{Biot} numbers for the spruce and the brick are reported in Table~\ref{tab:Fom_Bim}. The \emph{True Moisture Penetration Depth} (TMPD) from \cite{Abadie2009} and the \emph{Moisture Buffer value} (MBV) from \cite{Rode2007} are also given. The brick has a higher \textsc{Fourier} number than spruce. Therefore, the moisture diffusion through this material is predominant, explaining why the TMPD observed in \cite{Abadie2009} is more important. In addition, the \textsc{Biot} number is higher for the spruce. It implies that under an increase of absolute humidity, moisture will penetrate easier in the spruce than in the brick. Combined with a lower \textsc{Fourier} number, it explains why the MBV value is higher for the spruce \cite{Rode2007}. This analysis highlights that the dimensionless numbers, appearing in the formulation of the equations of heat and moisture transfer, enable to understand the material behavior.

\begin{table}
  \centering
  \begin{tabular}{@{}lcccccc}
  \hline
  \hline
  \textit{Material} & $k_{\,m}^{\,\circ} \quad [\,\mathsf{s} \,]$ & $c_{\,m}^{\,\circ} \quad [\,\mathsf{s^{\,2}/m^{\,2}} \,]$ 
  & $\mathrm{Fo}_{\,m}$ & $\mathrm{Bi}_{\,m}$ 
  & TMPD $[\,\mathsf{cm}\,]$ & MBV $[\,\mathsf{g/m^{\,2}}\,]$ \\
  \hline
  \textit{Brick}  &  $ 2 \e{-11} $ & $6 \e{-4} $
  &  $2.9 \cdot 10^{\,-3} $ & $100$ 
  & $5$ & $0.4$ \\
  \textit{Spruce (vertical fiber)} & $ 5 \e{-12} $
  & $9 \e{-3} $ & $5 \cdot 10^{\,-5} $
  & $400$ & $1.5$ & $1.2$\\
  \hline
  \hline
  \multicolumn{7}{c}{\scriptsize Parameter used for computation: $L \egal 0.2 \ \mathsf{m}\ $, $t^{\,\circ} \egal 1 \ \mathsf{h}\ $, $\alpha_{\,m} \egal 10^{\,-8} \ \mathsf{s/m}\ $, $T \egal 23 \ \mathsf{^{\,\circ}C} $}
  \end{tabular}\bigskip
  \caption{\emph{\small{\textsc{Fourier} and \textsc{Biot} numbers.}}}
  \label{tab:Fom_Bim}
\end{table}


\section{Numerical methods}
\label{sec:NM_scalar_equation}

The material properties varies along the space coordinates (and sometimes with the time) and with moisture contents and temperature. Moreover, the boundary conditions are defined according to climate data driven boundary conditions. Therefore, the use of analytical solution is limited and  numerical approaches are necessary to compute the approximate solution of the problem. It introduces a discretisation of the time and space with a local difference approximation of the derivatives when using the \textsc{Taylor} expansion approach. The important aspects of a numerical scheme are (i) it global error and (ii) the appropriate (qualitative and quantitative) behavior of the solution to represent the physical phenomenon. The former is quantified by the accuracy of the method, related to the order of truncation when approximating the derivatives. The latter is associated to the absolute stability of the scheme. A stable scheme avoids to compute an unbounded solution. Moreover, a numerical scheme converges if and only if it is stable (\textsc{Lax}--\textsc{Richtmyer} theorem). It should be remarked that even with a stable scheme, attention should be paid to the choice of the time and space discretisation parameters. A stable scheme does not necessarily imply that a physically realistic solution will be computed. Some examples of such cases can be found in \cite{Gasparin2017} and \cite{Patankar1980}. Moreover, a critical aspect of a numerical scheme is the CPU time to compute the solution of the given problem. Interested readers are invited to consult \cite{Mendes2017, Hairer2009} for more details.

For the description of the numerical schemes, let's consider for simplicity a uniform discretisation of the interval. The discretisation parameters are denoted with $\dx$ for the space and with $\dt$ for the time. The spatial cell $\mathcal{C} \egal \big[\,x_{\,j\,-\,\half},\, x_{\,j\,+\,\half}\,\big]$ is represented in Figure~\ref{fig_Met:stencil}. The values of function $u\,(x,\,t)$ in discrete nodes will be denoted by $u_{\,j}^{\,n}\ \eqdef\ u\,(x_{\,j},\,t^{\,n}\,)\,$ with $j\ \in\ \bigl\{\,1,\,\ldots,\,N\,\bigr\}$ and $n\ =\ 0,\,1,\,2,\,\ldots, \, N_{\,t}\,$.

For the sake of simplicity and without losing the generality, the upper-script $\star$ standing for dimensionless parameters, is dropped out. In addition, the numerical schemes are explained for the one-dimensional linear convection equation written in a conservative form as:
\begin{align}\label{eq:conv_diff}
  & \pd{u}{t} \plus \pd{J}{x} \egal 0 \,, & t & \ > \ 0\,, \;&  x & \ \in \ \big[ \, 0, \, 1 \, \big]\,, \\
  & J \egal a \, u \moins d\;\pd{u}{x} \nonumber \,,
\end{align}
where $u\,(x,\,t)\,$, is the field of interest, $d$ the diffusion coefficient and $a$ the advection coefficient, both considered to be constant. The boundary conditions are also written using a simplified notation:
\begin{subequations}\label{eq:conv_diff_BC}
\begin{align}
  d \;\pd{u}{x} \moins a \, u & \egal \mathrm{Bi} \cdot \left( \, u \moins \uinf \, \right) \,, && x \egal 0 \,, \\[3mm]
  - \ d \;\pd{u}{x} \plus a \, u & \egal \mathrm{Bi} \cdot \left( \, u \moins \uinf \, \right)\,, && x \egal 1 \,,
\end{align}
\end{subequations}
where $\uinf$ is the field in the ambient air surrounding the material. It should be noted that since we consider $1-$dimensional transfer, $\pm \ \pd{u}{x}$ plays the role of normal derivatives.


\subsection{The \SG ~scheme}
\label{sec:SG_scheme}

The straightforward discretisation of Eq.~\eqref{eq:conv_diff} yields the following semi-discrete difference relation:
\begin{align*}
  \od{u_{\,j}}{t} \plus \dfrac{1}{\Delta x}\;\Biggl[\,J_{\,j\,+\,\frac{1}{2}}^{\,n}\ -\ J_{\,j\,-\,\frac{1}{2}}^{\,n}\,\Biggr] \egal 0\,.
\end{align*}

\textsc{Scharfetter} and \textsc{Gummel} assumes that the numerical flux is constant on the dual cell $\mathcal{C}^{\,\star} \egal \big[\,x_{\,j}\,,\,x_{\,j\,+\,1}  \, \big]\,$. Thus, it can be computed giving the following boundary-value problem \cite{Scharfetter1969, Gosse2013, Gosse2017}:
\begin{subequations}\label{eq:SP_equation}
\begin{align}
  J_{\,j\,+\,\half}^{\,n} & \egal a \, u \moins d \;\pd{u}{x} \,, & & \forall x \ \in \ \big[ \, x_{\,j}  \,, \, x_{\,j\,+\,1}  \, \big] \,, \qquad \ \forall  j \ \in \ \bigl\{\,2\,,\ldots,\,N\moins 1 \,\bigl\} \,, \\
  u & \egal u_{\,j}^{\,n} \,, && x \egal x_{\,j}  \,,\\
  u  &\egal u_{\,j+1}^{\,n} \,, && x \egal x_{\,j\,+1\,}  \,.
\end{align}
\end{subequations}

Eq.~\eqref{eq:SP_equation} is a first order differential equation with two boundary conditions and the two unknowns $u$ and $J_{\,j\,+\,\half}^{\,n}\,$.
The solution of Eq.~\eqref{eq:SP_equation} corresponds to the expression of the \textsc{Poincar\'e}--\textsc{Steklov} operator, $\mathcal{S}\,:\ (u_{\,j}^{\,n},\,u_{\,j\,+\,1}^{\,n})\ \mapsto\ J_{\,j\,+\,\half}^{\,n}$  and can be written as:
\begin{align}\label{eq:10}
  & J_{\,j\,+\,\half}^n \egal a \; \dfrac{\biggl(\, u_{\,j}^{\,n} \moins u_{\,j\,+\,1}^{\,n} \, \mathrm{e}^{\,\dfrac{a \, \Delta x}{d}} \,\biggr)}{1 \moins \mathrm{e}^{\,\dfrac{a \, \Delta x}{d}}} \,.
\end{align}
The last equation can be rewritten also as:
\begin{align*}
  & J_{\,j\,+\,\half}^{\,n} \egal \dfrac{d}{\Delta x}\;\Biggl[\,-\ \mathcal{B} \bigl(\, \Theta \,\bigr) \, u_{\,j\,+\,1}^{\,n} \plus \mathcal{B} \bigl(\, - \, \Theta \,\bigr) \, u_{\,j}^{\,n}\,\Biggr]\,,
\end{align*}
where the \textsc{Bernoulli} function $\mathcal{B}\,(\,\cdot \,)$ and the ratio $\Theta$ are defined as:
\begin{align*}
  \mathcal{B}\,(\,\Theta \,) \ \eqdef \ \dfrac{\Theta}{\mathrm{e}^{\,\Theta} \moins 1} \,, && \Theta \ \eqdef \  \dfrac{a \, \Delta x}{d}\,.
\end{align*}

\begin{figure}
  \centering
  \includegraphics[width=.5\textwidth]{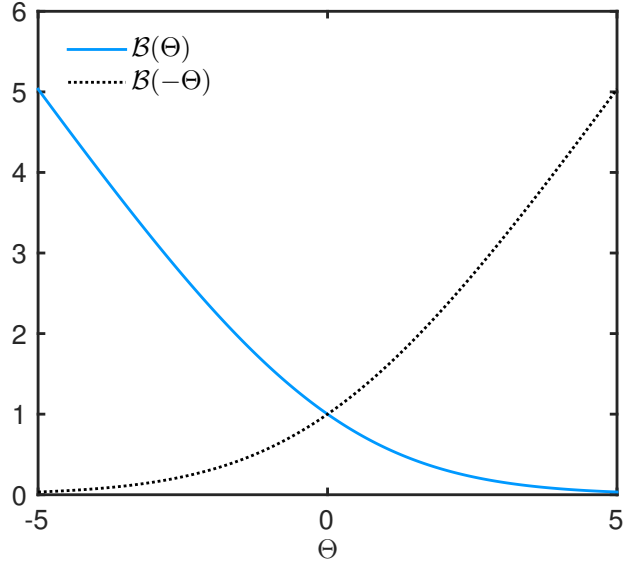}
  \caption{\emph{\small{Plot of the \textsc{Bernoulli} function.}}}
  \label{fig_Met:Bernoulli}
\end{figure}

The behavior of the \textsc{Bernoulli} function is illustrated in Figure~\ref{fig_Met:Bernoulli}. It can be noted that we have the following limiting behavior: 
\begin{subequations}\label{eq:DL_bernoulli}
\begin{align}
  & \lim_{\Theta \ \rightarrow \ 0} \mathcal{B}\,(\,\Theta \,)  \egal 1 \,, 
  && \lim_{\Theta \ \rightarrow \ \infty} \mathcal{B}\,(\,\Theta \,)  \egal 0 \,,\\
  & \lim_{\Theta \ \rightarrow \ 0} \mathcal{B}\,(\,-\Theta \,)  \egal 1 \,, 
  && \lim_{\Theta \ \rightarrow \ \infty} \mathcal{B}\,(\,-\Theta \,)  \egal +\, \infty \,.
\end{align}
\end{subequations}

For the nodes at the boundary surface, $j\ \in\ \bigl\{\,1\,,\,N\,\bigr\}\,$, the flux $J_{\,\half}$ is the solution of 
\begin{align*}
  J_{\,\half}^{\,n} & \egal a \, u \moins d \;\pd{u}{x} \,, & & \forall x \ \in \ \big[ \, 0 \,, \, x_{1}^{\,n} \, \big] \,, \\
  d \;\pd{u}{x} \moins a \, u & \egal \mathrm{Bi} \cdot \left( \, u \moins \uinf \, \right) \,, && x \egal 0 \,,\\
  u  &\egal u_{\,1}^{\,n} \,, && x \egal x_{1}^{\,n} \,.
\end{align*}
and $J_{\,N\,+\,\half}\,$:
\begin{align*}
  J_{\,N\,+\,\half}^{\,n} & \egal a \, u \moins d \;\pd{u}{x} \,, & & \forall x \ \in \ \big[ \, x_{N}^{\,n} \,,\, 1 \, \big] \,, \\
  u &\egal u_{\,N}^{\,n} \,, && x \egal x_{N}^{\,n} \,, \\
  - \, d \;\pd{u}{x} \plus a \, u & \egal \mathrm{Bi} \cdot \left( \, u \moins \uinf \, \right) \,, && x \egal 1 \,.
\end{align*}
Solving these two systems, we get: 
\begin{align*}
  & J_{\,\half}^{\,n} \egal \dfrac{a \, \mathrm{Bi} \,  \biggl(\, \uinf \, \mathrm{e}^{\,\Theta} \moins u_{\,1}^{\,n} \,\biggr)}{\biggl(\, \mathrm{Bi} \, \bigl(\, \mathrm{e}^{\,\Theta} \moins 1\,\bigl) \plus a \, \mathrm{e}^{\,\Theta} \,\biggr)} \,,
  &&
  J_{\,N\,+\,\half}^{\,n} \egal \dfrac{a \, \mathrm{Bi} \,  \biggl(\, \uinf \moins u_{\,N}^{\,n} \, \mathrm{e}^{\,\Theta} \,\biggr)}{\biggl(\, \mathrm{Bi} \, \bigl(\, 1 \moins \mathrm{e}^{\,\Theta} \,\bigl) \moins a \,\biggr)} \,.
\end{align*}

\begin{figure}
  \centering
  \includegraphics[width=.99\textwidth]{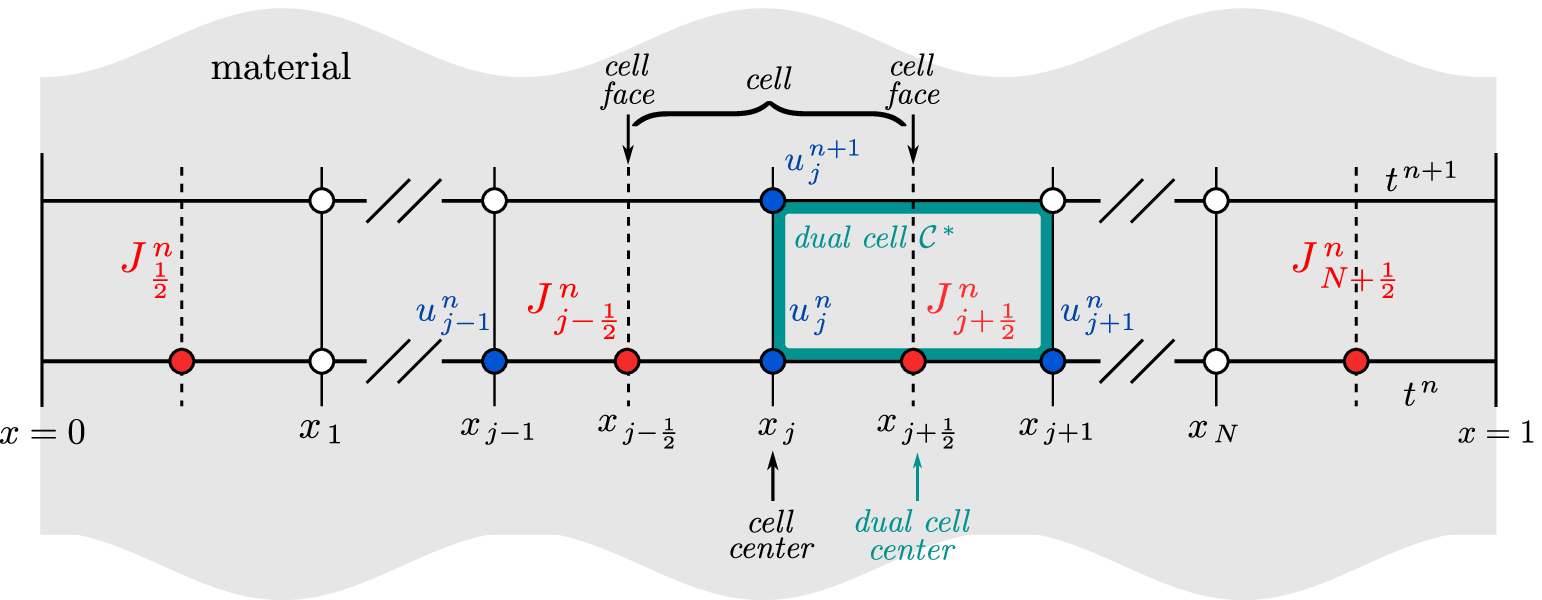}
  \caption{\emph{\small{Stencil of the \SG ~numerical scheme.}}}
  \label{fig_Met:stencil}
\end{figure}

We define $\lambda \ \eqdef \ \dfrac{d \, \dt}{\dx^{\,2}}\,$. When using the \textsc{Euler} explicit approach to approximate the time derivative from Eq.~\eqref{eq:conv_diff}, the \SG ~scheme finally yields to:
\begin{align*}
  u_{\,1}^{\,n+1} & \egal u_{\,1}^{\,n} \plus \dfrac{\dt}{\dx} \, J_{\,\half}^{\,n} \moins \lambda \;\Biggl[ \, -\ \mathcal{B} \bigl(\, \Theta \,\bigr) \, u_{\,2}^{\,n} \plus\mathcal{B} \bigl(\, - \, \Theta \,\bigr) \, u_{\,1}^{\,n} \, \Biggr] \,, &&   \\
  u_{\,j}^{\,n+1} & \egal u_{\,j}^{\,n} \plus \lambda \;\Biggl[ \, \mathcal{B} \bigl(\, \Theta \,\bigr) \, u_{\,j+1}^{\,n} \moins \biggl(\, \mathcal{B} \bigl(\, - \, \Theta \,\bigr) \plus \mathcal{B} \bigl(\, \Theta \,\bigr) \,\biggr) \, u_{\,j}^{\,n} \plus \mathcal{B} \bigl(\, - \, \Theta \,\bigr) \, u_{\,j-1}^{\,n}\, \Biggr]\,, && \\
  &  \quad \ \forall  j \ \in \ \bigl\{\,2\,,\ldots,\,N \moins 1 \,\bigl\} \,, \\
  u_{\,N}^{\,n+1} & \egal u_{\,N}^{\,n} \plus \lambda \;\Biggl[ \, -\ \mathcal{B} \bigl(\, \Theta \,\bigr) \, u_{\,N}^{\,n} \plus \mathcal{B} \bigl(\, - \, \Theta \,\bigr) \, u_{\,N-1}^{\,n} \, \Biggr] \moins \dfrac{\dt}{\dx} \, J_{\,N + \half}^{\,n}\,.
&& 
\end{align*}
The stencil of the scheme is illustrated in Figure~\ref{fig_Met:stencil}. The scheme is first order accurate in time and space $\O\,(\,\dx \plus \dt \,)$. It should be noted that the flux is approximated to the order $\O\,(\,\Delta x\,)\,$ as well \cite{GartlandJr.1993}. This is a remarkable property of the scheme since a derivation usually provokes the loss of one order in accuracy.

\subsection{Specific features of the scheme}
\label{sec:Spec_feat}

The important feature of the \SG~numerical scheme is well balanced as well as asymptotically preserved\footnote{The term \emph{asymptotically preserved} is used nowadays in applied mathematics community. However, the first hystorical term was \emph{uniformly accurate}, which is much clearer to our opinion}. Using the definition of parameter $\Theta \ \eqdef \  \dfrac{a \, \Delta x}{d}\,$, when the advection coefficient is much greater than the diffusion one, $a \ \gg \ d\,$, we have $\Theta \ \rightarrow \ \infty\,$. Inversely, when the advection coefficient is smaller than the diffusion one, $a \ \ll \ d$, we obtain $\Theta \ \rightarrow \ 0\,$. Thus, considering the results from Eq.\eqref{eq:DL_bernoulli}, the limiting behavior of the numerical fluxes is correct independently from grid parameters:
\begin{align*}
  & \lim_{a \ \rightarrow \ 0} J_{\,j\,+\,\half}^{\,n} \egal - \ \dfrac{u_{\,j\,+\,1}^{\,n} \moins u_{\,j}^{\,n}}{\Delta x} \,, 
  && \lim_{d \ \rightarrow \ 0} J_{\,j\,+\,\half}^{\,n} \egal  
  \begin{cases} 
   \ u_{\,j}^{\,n} \,, & \quad  a \ \leqslant \ 0 \,, \\
   \ u_{\,j\,+\,1}^{\,n} \,, & \quad  a \ > \ 0 \,.
\end{cases}
\end{align*}

Furthermore, the computation of $J_{\,j\,+\,\half}^{\,n}$ is exact and it gives an excellent approximation of the physical phenomena. The only hypothesis was done when assuming $J_{\,j\,+\,\half}^{\,n}$ constant in the dual cell $\Bigl[ \, x_{\,j} \,, \, x_{\,j\,+\,1}\, \Bigr]\,$. In addition, when the steady state is reached, the solution computed with the scheme becomes exact \cite{Jerome1991}. Interested readers may consult \cite{Patankar1980, Gosse2013, Gosse2017} as recent works on the \SG~scheme. The approximation of the dispersion relation by the scheme is discussed in Appendix~\ref{sec:annex_disp_relation}. In particular, it reveals that the phase velocity is second-order accurate.

The numerical scheme has other advantages that may be more interesting when applying to physical case studies. First, an explicit form of the solution is obtained. Therefore, no sub-iterations are required to treat the non-linearities of the problem as it is the case when using \CN ~approach for instance. This feature may reduce significantly the CPU time of the algorithm \cite{Berger2017a,Gasparin2017, Gasparin2017b}. When using a fully implicit approach, as for instance in \cite{Simunek2009}, a special iterative approach (\emph{e.g.} the \textsc{Picard} one) is required to treat the non-linearities at each time iteration. Other comments on the advantages of the \SG ~scheme compared to the \CN ~approach are discussed in \cite{Duffy2004}.

It is true that when using explicit approaches, the so-called \textsc{Courant-Friedrichs-Lewy} (CFL) stability condition must be respected. For a classical \textsc{Euler} explicit approach, it is a strong restriction since it implies a fine spatial grid. However, in the linear case, the CFL condition of the \SG ~scheme is given by \cite{Gosse2016}:
\begin{align*}
  \dt \, a \, \tanh \, \Bigl(\, \dfrac{a \, \dx}{2 \, d} \,\Bigr)^{\,-1} \ \leqslant\ \dx \,.
\end{align*}

It can be noted that, if the spatial grid is refined, the \textsc{Taylor} expansion of the hyperbolic tangent gives:
\begin{align*}
  \dfrac{a \, \dx}{2 \, d} \ \rightarrow \ 0 \,, \qquad \tanh\,\Bigl(\, \dfrac{a \, \dx}{2 \, d} \,\Bigr)^{\,-1} \egal \dfrac{2 \, d}{a \, \dx} \plus \dfrac{a \, \dx}{6 \, d} \plus \O(\,\dx^{\,3}\,) \,. 
\end{align*}
Thus, the CFL condition starts to become quadratic $\dt \ \leqslant \ C_{\,1} \cdot \dx^{\,2}\,$. It brings us to the standard CFL condition of the explicit \textsc{Euler} approach. Nevertheless, if the spatial grid is large, $\tanh\,\Bigl(\, \dfrac{a \, \dx}{2 \, d} \,\Bigr) \egal \O\,(1)$ and the CFL condition is improved to $\dt \leqslant C_{\,2} \cdot \dx\,$. The values of $\dx$ have to be in a closed interval, depending on the material properties. It is not necessary to use a fine spatial grid for this approach.

Moreover, a useful point is that, considering Eq.~\eqref{eq:SP_equation}, the exact interpolation of solution $u\,(\,x\,)$ can be computed:
\begin{align}\label{eq:interp_u}
  & u^{\,n}\,(\,x\,) \egal \dfrac{1}{a} \; J_{\,j\,+\,\frac{1}{2}}^{\,n} \plus \dfrac{u_{\,j}^{\,n} \moins u_{\,j+1}^{\,n}}{1 \moins  \mathrm{e}^{\,\Theta}}\cdot\mathrm{exp}\;\biggl(\, \dfrac{a}{d} \, \bigl(\,x \moins x_{\,j} \,\bigr) \,\biggr) \,, && x \ \in \ \big[ \, x_{\,j} \,, \, x_{\,j+1}  \, \big] \,.
\end{align}
Therefore, when using a large spatial grid, one can compute the exact expression of $u\,(\,x\,)$ on the point of interest using Eq.~\eqref{eq:10}. When using the classical methods, an interpolation (\emph{e.g.} linear or cubic) is required. If the solution is steady, Eq.~\eqref{eq:interp_u} provides the exact solution of the problem.

In terms of implementation, it has been highlighted in \cite{Berger2017a} that the \SG ~approach is particularly efficient, in terms of reduction of the CPU cost, when using an adaptive time stepping. In Section~\ref{sec:SG_scheme}, the scheme was presented using an \textsc{Euler} explicit approach for the sake of simplicity. In further sections, the algorithm is implemented using the \textsc{Matlab\;\texttrademark} function \texttt{ode113}, based on the \textsc{Adams}--\textsc{Bashforth}--\textsc{Moulton} approach \cite{Shampine1997}. It is possible to use a \textsc{Runge}--\textsc{Kutta} scheme (function \texttt{ode45} for instance). However, this approach requires intermediate computations between two time iterations. The \textsc{Adams}--\textsc{Bashforth}--\textsc{Moulton} scheme computes directly $u^{\,n\,+\,1}$ as function of the computations at the previous time steps. It is less expensive in terms of computational cost and it was therefore used in the next case studies. The advantages of the \SG ~numerical scheme are synthesized in Table~\ref{tab:synthesis_num_scheme}.

\begin{table}
\centering
\begin{tabular}{@{}m{.75\textwidth}}
  \hline
  \hline
  $\bullet$ Well balanced \\
  $\bullet$ Asymptotic preserving  \\
  $\bullet$ Explicit form of the solution, no sub-iterations required to treat the non-linearities  \\
  $\bullet$ CFL stability condition scaling with $\Delta x$ for large spatial grid  \\
  $\bullet$ Reduced CPU with an adaptive time step algorithm  \\
  $\bullet$ Exact interpolation of the solution $u\,(\,x\,)$   \\
  \hline
  \hline
\end{tabular}\bigskip
\caption{\emph{\small{Synthesis of the \SG ~numerical scheme advantages.}}}
\label{tab:synthesis_num_scheme}
\end{table}


\subsection{Extension to non-linear cases}
\label{sec:ext_NL}

When the coefficients $a$ and $d$ of Eq.~\eqref{eq:conv_diff} are non-linear, \ie ~dependent on $u\,$, we apply the approximation of frozen coefficients on the interval. Hence, the coefficients $a$ and $d$ are constant on the dual cell $\big[ \, x_{\,j}, \, x_{\,j\,+\,1} \, \big]\,$. The flux at the interface is computed using the boundary value problem given in Eq.~\eqref{eq:SP_equation} and the solution yields to:
\begin{align*}
  & J_{\,j\,+\,\half}^{\,n} \egal \dfrac{d_{\,j\,+\,\half}^{\,n}}{\Delta x}\;\Biggl[ \, -\ \mathcal{B} \bigl(\, \Theta_{\,j\,+\,\half}^{\,n} \,\bigr) \, u_{\,j\,+\,1}^{\,n} \plus \mathcal{B} \bigl(\, - \, \Theta_{\,j\,+\,\half}^{\,n} \,\bigr) \, u_{\,j}^{\,n}\, \Biggr]\,,
\end{align*}
where the coefficients are computed according to:
\begin{align*}
  & \Theta_{\,j\,+\,\half}^{\,n} \egal \dfrac{a_{\,j\,+\,\half}^{\,n}}{d_{\,j\,+\,\half}^{\,n}}\;\dx \,,
  && a_{\,j\,+\,\half}^{\,n} \egal a \, \Bigl(\, u_{\,j\,+\,\half}^{\,n}  \,\Bigr) \,, \\[3pt]
  & d_{\,j\,+\,\half}^{\,n} \egal d \, \Bigl(\, u_{\,j\,+\,\half}^{\,n} \,\Bigr) \,,
  && u_{\,j\,+\,\half}^{\,n} \egal \dfrac{1}{2} \, \Bigl(\, u_{\,j}^{\,n} \plus u_{\,j\,+\,1}^{\,n} \,\Bigr) \,.
\end{align*}

When dealing with non-linearities, we reiterate that the \SG ~scheme does not require any sub-iterations at each time iteration $t^{\,n}\,$, since it is explicit. Indeed, the scheme is written in an explicit way, enabling to compute \emph{directly} the coefficients $a$ and $d\,$. On the other hand, the CFL condition of the scheme has to be respected, which is given by \cite{Gosse2016}:
\begin{align*}
  \dt \, \max_{j}\;\Biggl[\, a_{\,j\,+\,\half}\,\tanh \, \Biggl(\, \dfrac{a_{\,j\,+\,\half} \, \dx}{2 \, d_{\,j\,+\,\half}} \,\Biggr)^{\,-1} \,\Biggr] \ \leqslant\ \dx \,.
\end{align*}


\subsection{Comparison of the numerical solution}

To compare and validate the scheme implementation, the error between the solution $u^{\, \mathrm{num}}\,(\,x\,,t\,)\,$, obtained by the numerical method, and a reference solution $u^{\, \mathrm{ref}}\, (\,x \,, t \,)\,$, is computed as a function of $x$ by the following discrete $\ell_{\,2}$ formulation:
\begin{align*}
  \varepsilon_{\,2}\, (\, x\,)\ &\eqdef\ \sqrt{\,\dfrac{1}{N_{\,t}} \;\sum_{j\, =\, 1}^{N_{\,t}} \, \left( \, u_{\, j}^{\, \mathrm{num}}\, (\,x \,, t \,) \moins u_{\, j}^{\mathrm{\, ref}}\, (\,x \,, t \,) \, \right)^{\,2}}\,,
\end{align*}
where $N_{\,t}$ is the number of temporal steps. The global uniform error $\varepsilon_{\,\infty}$ is given by the maximum value of $\varepsilon_{\,2}\, (\, x\,)\,$: 
\begin{align*}
  \varepsilon_{\, \infty}\ &\eqdef\ \sup_{x \ \in \ \bigl[\, 0 \,,\, L \,\bigr]} \, \varepsilon_{\,2}\, (\, x\,) \,.
\end{align*}

As detailed in further Sections, the reference solution $u^{\, \mathrm{ref}}\, (\,x \,, t \,)$ can be given by an analytical solution in exceptional cases, by a numerical pseudo--spectral solution obtained with the \textsc{Matlab\;\texttrademark} open source toolbox \texttt{Chebfun} \cite{Driscoll2014} or even by experimental data. The pseudo--spectral solution employs the function \texttt{pde23t} of \texttt{Chebfun} to compute a numerical solution of a partial derivative equation based on the \textsc{Chebyshev} polynomials representation.


\subsection{Numerical validation}

\begin{figure}
  \centering
  \subfigure[a][\label{fig_AN:scheme_12}]{\includegraphics[width=.7\textwidth]{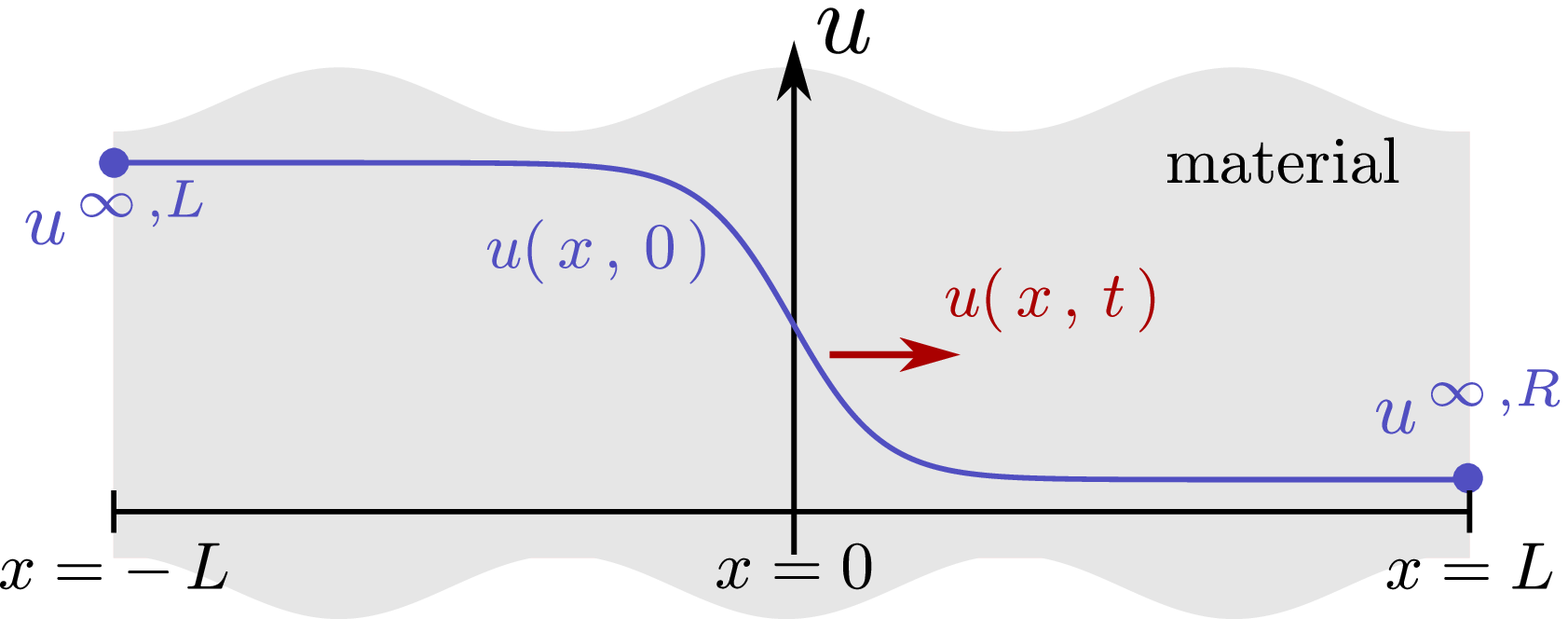}}
  \subfigure[b][\label{fig_AN:scheme_3}]{\includegraphics[width=.7\textwidth]{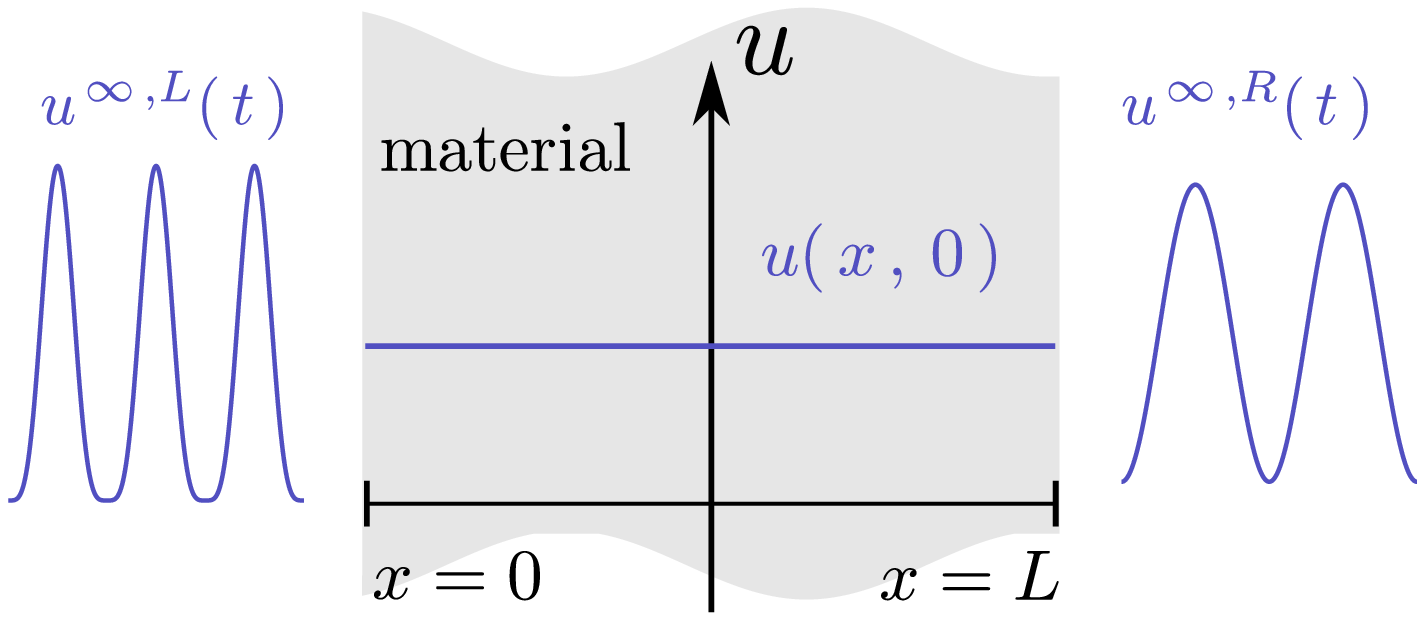}}
  \caption{\emph{\small{Illustration of the case studies $1$ and $2$ (a) and of the case $3$ (b).}}}
\end{figure}

In this Section, the validation of the numerical scheme for a single advective--diffusive equation is proposed. For this purpose, Eq.~\eqref{eq:conv_diff} is written in the form: 
\begin{subequations}\label{eq:conv_diff_NA}
\begin{align}
  & \pd{u}{t} \plus \pd{J}{x} \egal 0 \,, & t & \ > \ 0\,, \;&  x & \ \in \ \Omega_{\,x} \,, \\
  & J \egal a\,(\,u\,) \, u \moins d\,(\,u\,) \;\pd{u}{x} \,, \\
  & a\,(\,u\,) \egal a_{\,0} \plus a_{\,1} \, u \plus a_{\,2} \, u^{\,2} \,,\\
  & d\,(\,u\,) \egal d_{\,0} \plus d_{\,1} \, u \plus d_{\,2} \, u^{\,2} \,.
\end{align}
\end{subequations}
The functions for diffusion and advection coefficients have been exclusively used for the validation of the numerical algorithm and the analysis of the accuracy of the \SG ~nuerical solution. They may not be appropriate for material coefficients experimentally determined. The boundary conditions and the numerical values are specified for each of the three cases. The first two compare the numerical solution with an analytical one. As illustrated in Figure~\ref{fig_AN:scheme_12}, they consider a material with an initial profile $u\,(\,x\,,\,0\,)$ in the material. A \textsc{Dirichlet} conditions is imposed at the boundaries of the material. For the last case, the reference is the \texttt{Chebfun} pseudo--spectral solution. As illustrated in Figure~\ref{fig_AN:scheme_3}, a uniform initial condition is considered in the material with time variable boundary conditions in the ambient air surrounding the material.

\subsubsection{Case 1}

First, we set $a_{\,0} \egal d_{\,0} \egal d_{\,2} \egal 0\,$. In this case, an analytical solution to Eq.~\eqref{eq:conv_diff_NA} can be obtained by direct substitution:
\begin{align*}
  u\,(\,x\,,\,t\,) \egal - \, \dfrac{1}{2} \; A \, \tanh\,\biggl(\, k \,\bigl(\,x \plus x_{\,0} \moins c \,t \,\bigr)  \,\biggr) \moins \dfrac{1}{2} \; \dfrac{a_{\,1}}{a_{\,2}} \,,
\end{align*}
where
\begin{align*}
  & m \ \eqdef \ 2 \; \dfrac{d_{\,1}}{a_{\,2}} \; C_{\,1} \,,
  && k \ \eqdef \ \dfrac{a_{\,2}}{2 \, d_{\,1}} \; A  \,,
  && c \ \eqdef \ - \, \dfrac{1}{4 \, a_{\,2}} \; \biggl(\, a_{\,1}^{\,2} \moins a_{\,2}^{\,2} \, A^{\,2}  \,\biggr) \,,
\end{align*}
and $C_{\,1}$ is an arbitrary constants and $x_{\,0}$ defines the initial position of the front.

The asymptotic values of $u\,(\,x\,,\,t = 0\,)$ provide the \textsc{Dirichlet} type boundary conditions that will be used to compute the numerical solution:
\begin{align*}
  \lim_{x \ \rightarrow \ + \,\infty} u\,(\,x\,,\,0\,) \egal  u^{\,\infty,\, R}  & \egal - \, \dfrac{1}{2 \, a_{\,2} }\;\bigl(\, a_{\,2} \, A \plus a_{\,1} \,\bigr) \,,\\[3pt]
  \lim_{x \ \rightarrow \ - \, \infty} u\,(\,x\,,\,0\,) \egal  u^{\,\infty,\, L} & \egal \dfrac{1}{2 \, a_{\,2} }\;\bigl(\, a_{\,2} \, A \moins a_{\,1} \,\bigr) \,.
\end{align*}

For the numerical application, we take into account the following values:
\begin{align*}
  & x_{\,0} \egal 0 \,, && 
  C_{\,1} \egal 1 \,, &&
  a_{\,1} \egal -1.4 \,, &&
  a_{\,2} \egal 0.2 \,, &&
  d_{\,1} \egal 0.5 \,.
\end{align*}

And as consequences, we have: 
\begin{align*}
  u^{\,\infty,\,R} \egal 1 \,, && 
  u^{\,\infty,\,L} \egal 6 \,.
\end{align*}
The domains are defined as $x \ \in \ \big[ \, -10, \, 10 \, \big]$ and $t \ \in \ \big[ \, 0, \, 5 \, \big]\,$. The numerical solution is computed using the \SG ~scheme, with following spatial discretisation $\dx \egal 0.01$ and an adaptive time step $\dt$ using \textsc{Matlab\;\texttrademark} function \texttt{ode113} \cite{Shampine1997} with an aboslute and relative tolerances set to $10^{\,-5}\,$.

\begin{figure}
  \centering
  \subfigure[a][\label{fig_AN1:profil_u}]{\includegraphics[width=.45\textwidth]{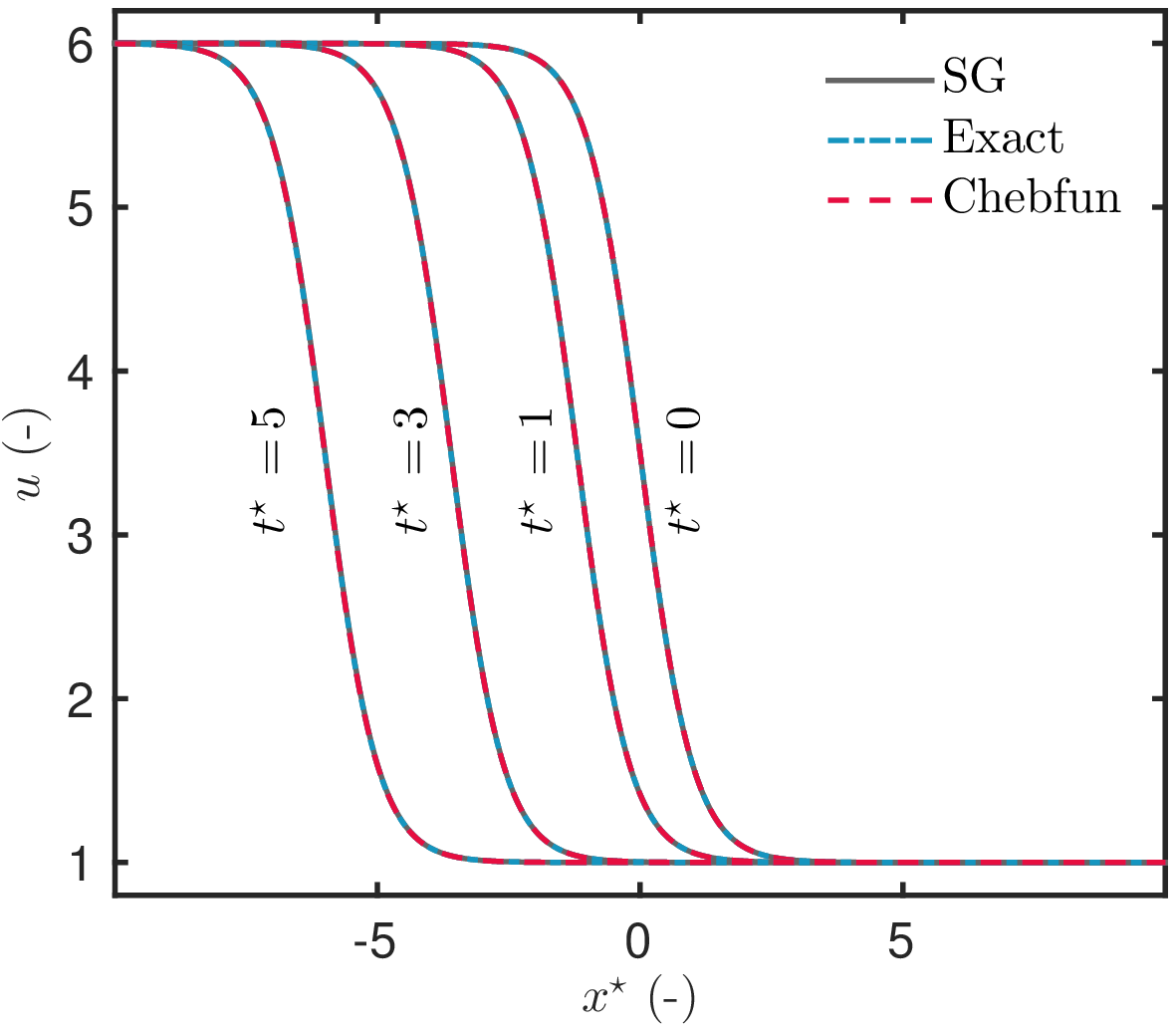}}
  \subfigure[b][\label{fig_AN1:time_u}]{\includegraphics[width=.47\textwidth]{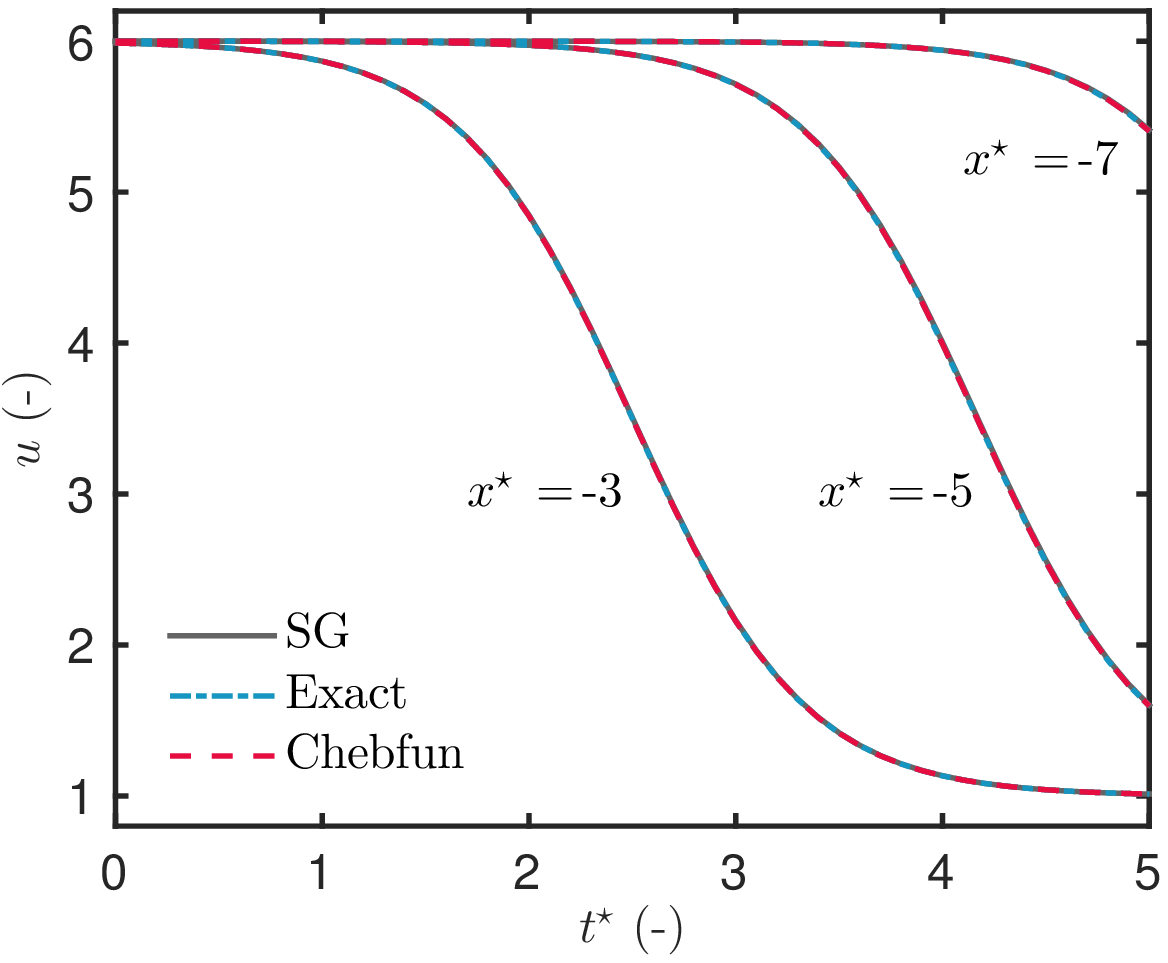}}
  \subfigure[c][\label{fig_AN1:uL2_fx}]{\includegraphics[width=.45\textwidth]{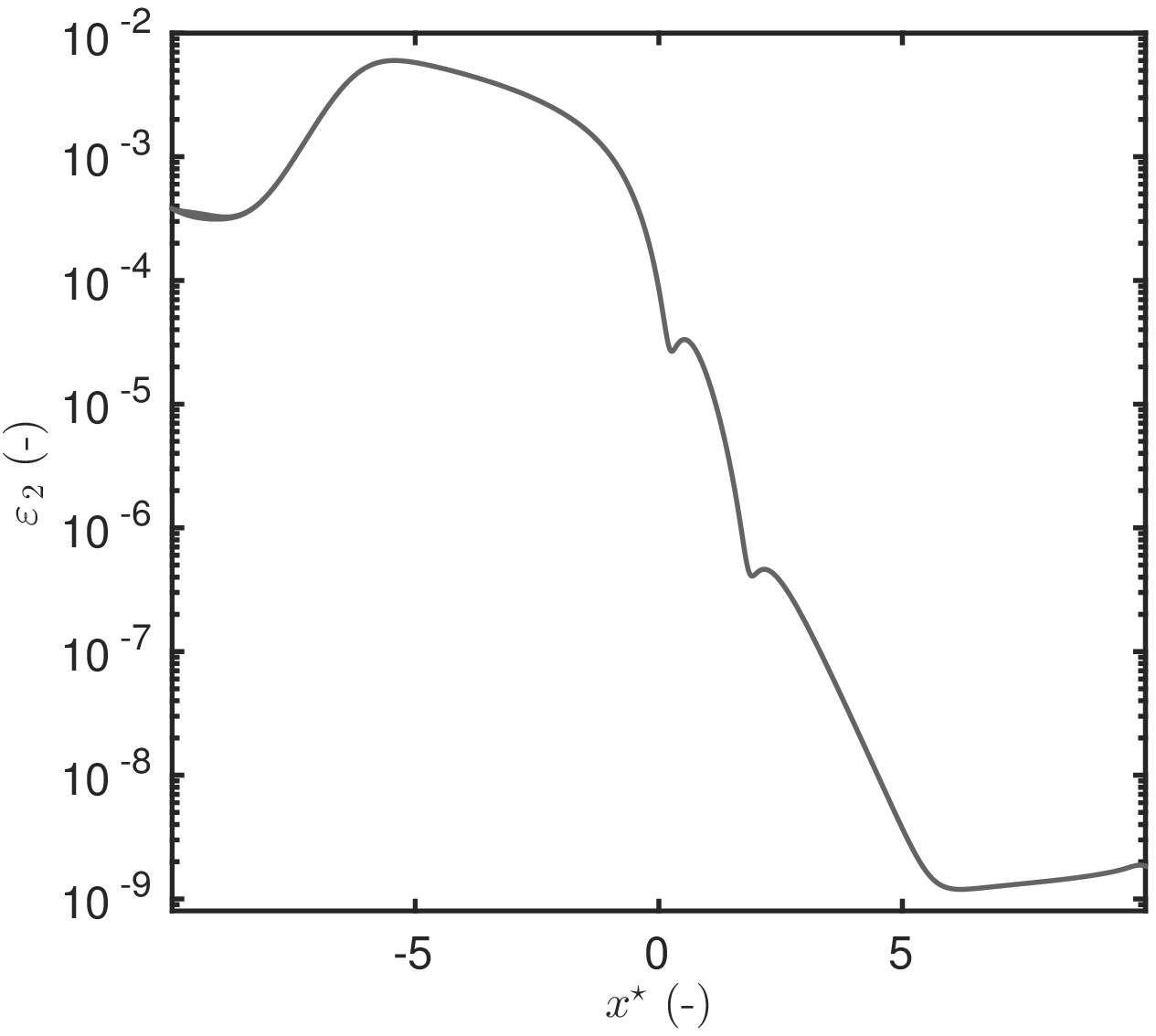}}
  \caption{\emph{\small{Variation of the field $u$ as a function of $x$ (a) and $t$ (b). $\ell_{\,2}$ error as a function of $x$ (c).}}}
\end{figure}

Figures~\ref{fig_AN1:profil_u} and \ref{fig_AN1:time_u} give the variation of the field as a function of time and space. A very good agreement can be noticed among the \SG, \texttt{Chebfun} pseudo--spectral and analytical solutions. As shown in Figure~\ref{fig_AN1:uL2_fx}, the $\ell_{\,2}$ error is lower than $\O(\,10^{\,-2}\,)\,$, which is consistent with the scheme accuracy $\O\,(\dx)\,$. 

\subsubsection{Case 2}

We set as null the coefficients $a_{\,2} \egal d_{\,1} \egal d_{\,2} \egal 0\,$. For these conditions, by direct substitution in Eq.~\eqref{eq:conv_diff_NA}, one can check that an analytical solution can be expressed:
\begin{align*}
  u\,(\,x\,,\,t\,) \egal - \, \dfrac{1}{2} \; A \, \tanh\,\biggl(\, k \,\bigl(\,x \plus x_{\,0} \moins c \,t \,\bigr) \,\biggr) \moins \dfrac{1}{2 \, C_{\,1} \, a_{\,1} } \; \biggl(\, C_{\,1} \, a_{\,0} \plus C_{\,2} \,\biggr) \,,
\end{align*}
where
\begin{align*}
  & A \ \eqdef \ 2 \; \dfrac{d_{\,0}}{a_{\,1}} \; C_{\,1} \,,
  && k \ \eqdef \  C_{\,1}  \,,
  && c \ \eqdef \ - \, \dfrac{C_{\,2}}{ C_{\,1}} \,,
\end{align*}
where $C_{\,1}$ and  $C_{\,2}$ are arbitrary constants and $x_{\,0}$ is the initial position of the front.

The asymptotic values of $u\,(\,x\,,\,t = 0\,)$ give the \textsc{Dirichlet} type boundary conditions that will be used to compute the numerical solution:
\begin{align*}
  \lim_{x \ \rightarrow \ +\,\infty} \,(\,x\,,\,0\,) \egal  u^{\,\infty,\,R}  & \egal \dfrac{1}{2 \, a_{\,1} \, C_{\,1} }\;\bigl(\,-\, 2 \, C_{\,1}^{\,2} \, d_{\,0} - C_{\,1} \, a_{\,0} \moins C_{\,2} \,\bigr) \,,\\[3pt]
  \lim_{x \ \rightarrow \ - \, \infty} u\,(\,x\,,\,0\,) \egal  u^{\,\infty,\,L} & \egal \dfrac{1}{2 \, a_{\,1} \, C_{\,1} }\;\bigl(\, 2 \, C_{\,1}^{\,2} \, d_{\,0} - C_{\,1} \, a_{\,0} \moins C_{\,2} \,\bigr)  \,.
\end{align*}

The following values are used for numerical applications:
\begin{align*}
  & C_{\,1} \egal 0 \,, && 
  C_{\,2} \egal 1 \,, &&
  C_{\,3} \egal -2 \,, &&
  a_{\,0} \egal 0.1 \,, &&
  a_{\,1} \egal 0.3 \,, &&
  d_{\,0} \egal 0.2 \,.
\end{align*}
And therefore,
\begin{align*}
  u^{\,\infty,\,R} \egal 2.5 \,, && 
  u^{\,\infty,\,L} \egal 3.83 \,.
\end{align*}
The time domain is defined as $t \ \in \ \big[ \, 0, \, 3 \, \big]\,$. For the numerical solution, the space domain needs to be defined. Since, $\forall \ t \ \in \bigl[\,0 \,,\, 3 \,\bigr]\,$, $ \bigl|u\,(\, -\,10 \,,\,t\,) \moins u^{\,\infty\,,\,L} \, \bigr| \ \leqslant 3 \e{\,-9}$ and $ \bigl|u\,(\, 10 \,,\,t\,) \moins u^{\,\infty\,,\,R} \, \bigr| \ \leqslant 3 \e{\,-9} \,$, the space domain is set as $x \ \in \ \big[ \, -10, \, 10 \, \big]\,$. As for the previous case, the solution is computed using the \SG ~scheme, with a spatial discretisation $\dx \egal 10^{\,-2} $ and an adaptive time step $\dt \,$ with tolerances set to $10^{\,-5}\,$.

\begin{figure}
  \centering
  \subfigure[a][\label{fig_AN2:profil_u}]{\includegraphics[width=.48\textwidth]{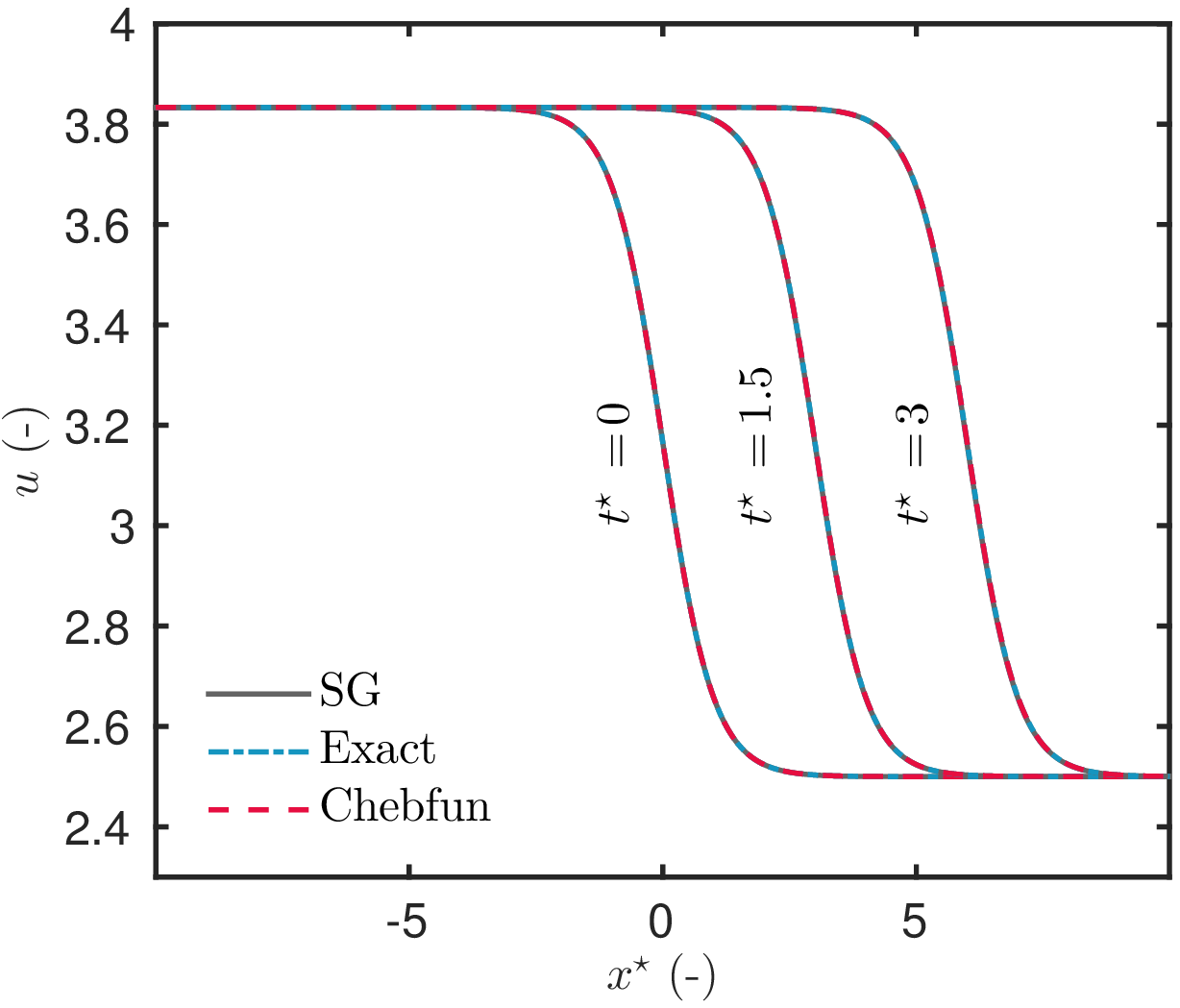}} 
  \subfigure[b][\label{fig_AN2:time_u}]{\includegraphics[width=.48\textwidth]{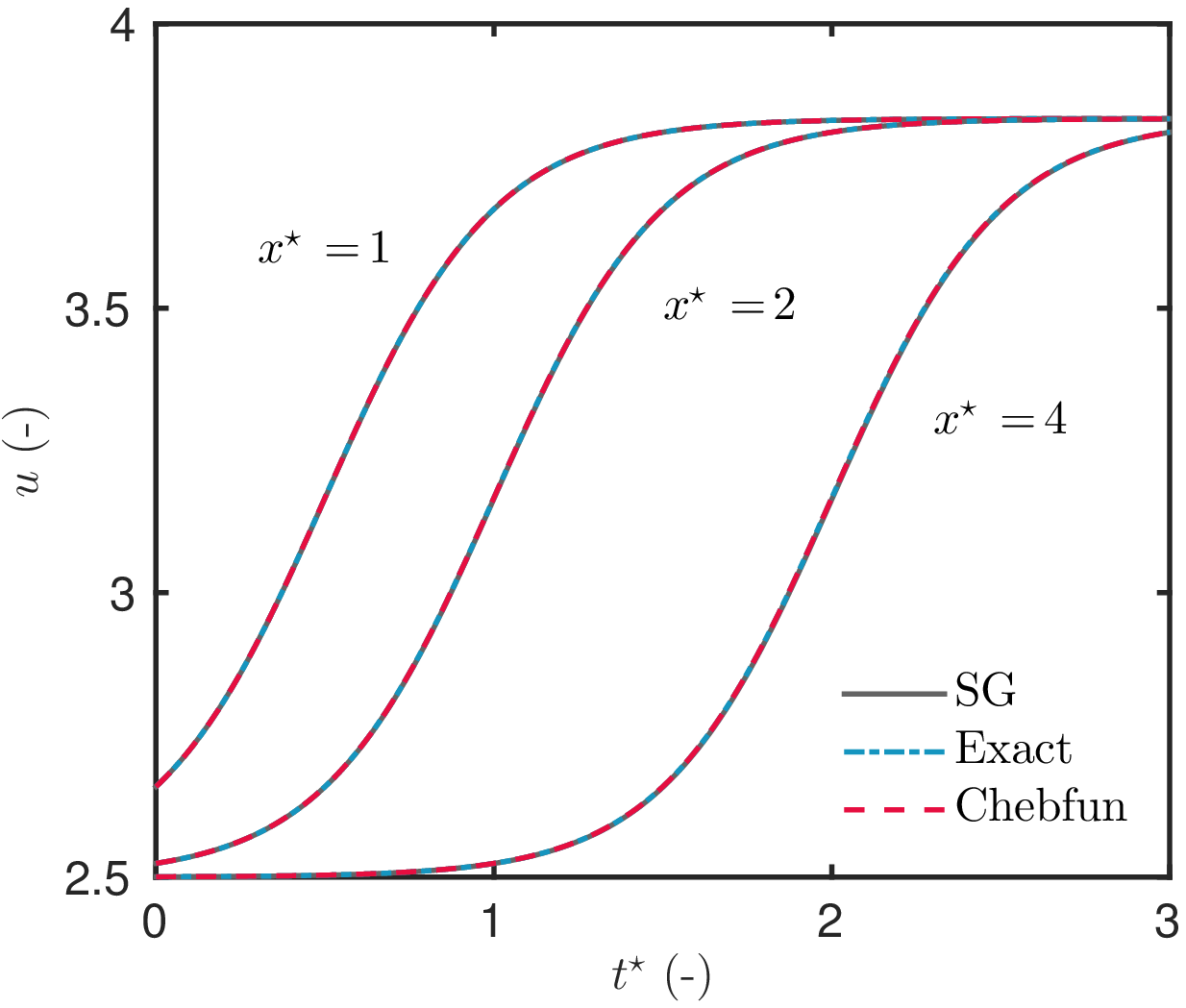}}
  \subfigure[c][\label{fig_AN2:uL2_fx}]{\includegraphics[width=.48\textwidth]{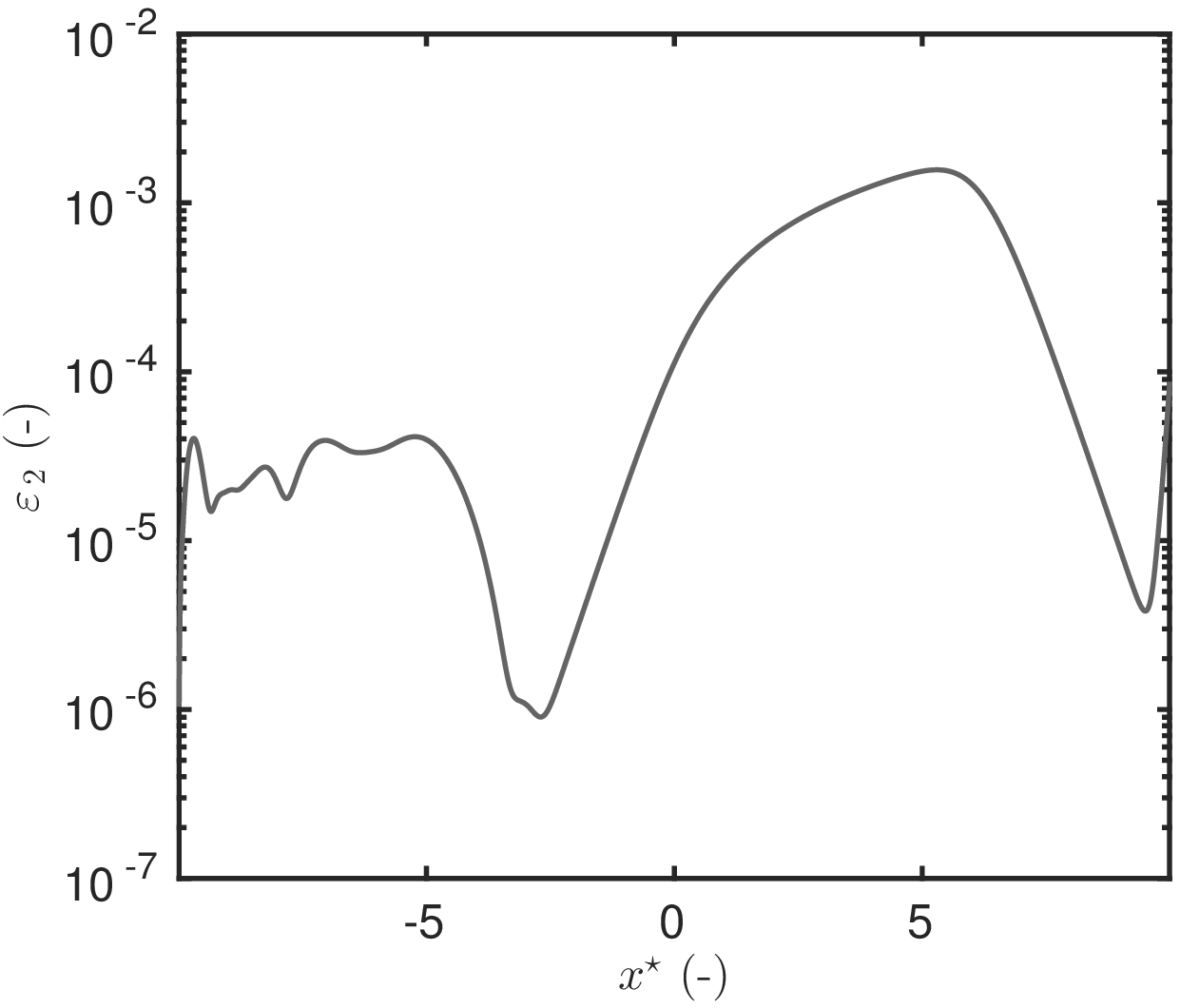}} 
  \caption{\emph{\small{Variation of the field $u$ as a function of $x$ (a) and $t$ (b). $\ell_{\,2}$ error $\varepsilon$ as a function of $x$ (c).}}}
\end{figure}

Results are shown in Figures~\ref{fig_AN2:profil_u}, \ref{fig_AN2:time_u} and \ref{fig_AN2:uL2_fx}. An accurate agreement is observed between the three solutions to represent the physical phenomena. The $\ell_{\,2}$ error is of the order $\O\,(\,10^{\,-3}\,)\,$, highlighting high accuracy of the solution computed with the \SG ~scheme.

\bigskip
\paragraph*{Remark on the analytical solution}

Other analytical solutions can be derived. By direct substitution in Eq.~\eqref{eq:conv_diff_NA}, one can check that, for the particular case $a_{\,1} \egal d_{\,0} \egal d_{\,2} \egal 0 \,$, we have: 
\begin{align*}
  u\,(\,x\,,\,t\,) \egal - \, \dfrac{1}{2} \; A \, \tanh\,\biggl(\, k \,\bigl(\,x \plus x_{\,0} \moins c \,t \,\bigr) \,\biggr) \,,
\end{align*}
where
\begin{align*}
  & A \ \eqdef \ 2 \; \dfrac{d_{\,1}}{a_{\,2}} \; C_{\,1} \,,
  && k \ \eqdef \  \dfrac{1}{2} \; \dfrac{a_{\,2}}{d_{\,1}} \; A  \,,
  && c \ \eqdef \ - \, \dfrac{4}{a_{\,2} \, A^{\,2} \plus 4\,a_{\,0}} \,,
\end{align*}
and $C_{\,1}$ is arbitrary real constants. The asymptotic values are: 
\begin{align*}
  \lim_{x \ \rightarrow \ \infty} u\,(\,x\,,\,t\,) \egal  u^{\,\infty\,,\,R}  & \egal - \dfrac{1}{2} \;A \,,\\[3pt]
  \lim_{x \ \rightarrow \ - \, \infty} u\,(\,x\,,\,t\,) \egal  u^{\,\infty\,,\,L} & \egal \dfrac{1}{2}\;A  \,.
\end{align*}

However, due to the functions $\tanh \,(\,k\,x\,) \,$, the last solution $u$ takes positive and negative values in the interval $x \ \in \ \R\,$. It can be noted that $u^{\,\infty\,,\,R} \ < \ 0$ and $u^{\,\infty\,,\,L} \ > \ 0\,$, for $A \ > \ 0$ (or vice versa). It implies positive and negative values for the diffusion coefficients $d \,$, that has no physical meaning. For this reason, this analytical solution was not used for comparison with the numerical one. However, it can be used to validate the solvers.

In the particular case $a_{\,0} \egal a_{\,2} \egal d_{\,0} \egal d_{\,2} \egal 0 \,$, another two analytical solutions can be derived:
\begin{align*}
  u_{\,1}\,(\,x\,,\,t\,) \egal 4 \, d_{\,1} \, C_{\,2}\;\Biggl[\,a_{\,1}^{\,2} \,\Biggl(\,\tanh\,\biggl(\,  C_{\,2} \, t \moins \dfrac{a_{\,1}}{2 \, d_{\,1}} \; x \plus C_{\,1} \,\biggr) \plus 1 \,\Biggr)\,\Biggr]^{\,-1} \,, \\
  u_{\,2}\,(\,x\,,\,t\,) \egal 4 \, d_{\,1} \, C_{\,2}\;\Biggl[\,a_{\,1}^{\,2} \, \Biggl(\,\tanh\,\biggl(\,  C_{\,2} \, t \plus \dfrac{a_{\,1}}{2 \, d_{\,1}} \;x \plus C_{\,1} \,\biggr) \moins 1 \,\Biggr) \,\Biggr]^{\,-1} \,, 
\end{align*}
and $C_{\,1}$ and $C_{\,2}$ are arbitrary constants. It can be pointed that the asymptotic values are: 
\begin{align*}
  \lim_{x \ \rightarrow \ +\,\infty} u_{\,1}\,(\,x\,,\,t\,) \egal  u_{\,1}^{\,\infty\,,\,R}  & \egal - \infty \,, 
  && \lim_{x \ \rightarrow \ +\,\infty} u_{\,2}\,(\,x\,,\,t\,) \egal  u_{\,2}^{\,\infty\,,\,R} \egal \infty \,, \\[3pt]
  \lim_{x \ \rightarrow \ - \, \infty} u_{\,1}\,(\,x\,,\,t\,) \egal  u_{\,1}^{\,\infty\,,\,L} & \egal 2 \; \dfrac{d_{\,1}}{a_{\,1}^{\,2}} \; C_{\,2}  \,, 
  && \lim_{x \ \rightarrow \ - \, \infty} u_{\,2}\,(\,x\,,\,t\,) \egal  u_{\,2}^{\,\infty\,,\,L} \egal - \, 2 \; \dfrac{d_{\,1}}{a_{\,1}^{\,2}} \; C_{\,2}\,.
\end{align*}
One asymptotic value of both analytical solutions tends to infinity, having no physical meaning. This solution can be used for validation purpose only on a finite domain.


\subsubsection{Case 3}

The previous comparison cases considered only \textsc{Dirichlet} boundary conditions. Here, a non-linear case of transfer with \textsc{Robin} boundary conditions is investigated. The time and space domains are defined as $x \ \in \ \big[ \, 0, \, 1 \, \big]$ and $t \ \in \ \big[ \, 0, \, 6 \, \big] \,$, respectively. The material properties are fixed to:
\begin{align*}
  & 
  a_{\,0} \egal 0.5 \,, &&
  a_{\,1} \egal 0.3 \,, &&
  a_{\,2} \egal 0 \,, &&
  d_{\,0} \egal 0.9 \,, &&
  d_{\,1} \egal 0.1 \,, &&
  d_{\,2} \egal 0 \,.
\end{align*}

The initial condition is set to $u\,(\,x\,,\,t\egal0\,) \egal 0\,$. Moreover, in accordance with Eq.~\eqref{eq:conv_diff_BC}, we have: 
\begin{align*}
  &
  \text{At} \quad x \egal 1 \,: &&
  \mathrm{Bi}^{\,R} \egal 1.3 \,, &&
  u^{\,\infty\,,\,R} (\,t\,) \egal 1.9 \, \sin \Bigl(\, 2 \, \pi\;\dfrac{t}{6}\;\Bigr)^{\,2} \,. \\
  &
  \text{At} \quad x \egal 0 \,: &&
  \mathrm{Bi}^{\,L} \egal 0.5 \,, &&
  u^{\,\infty\,,\,L} (\,t\,) \egal 0.3 \, \Bigl(\, 1 \moins \cos \bigl(\, \pi \, t \,\bigr) \, \Bigr)^{\,2}  \,.
\end{align*}

\begin{figure}
  \centering
  \subfigure[a][\label{fig_AN3:profil_u}]{\includegraphics[width=.48\textwidth]{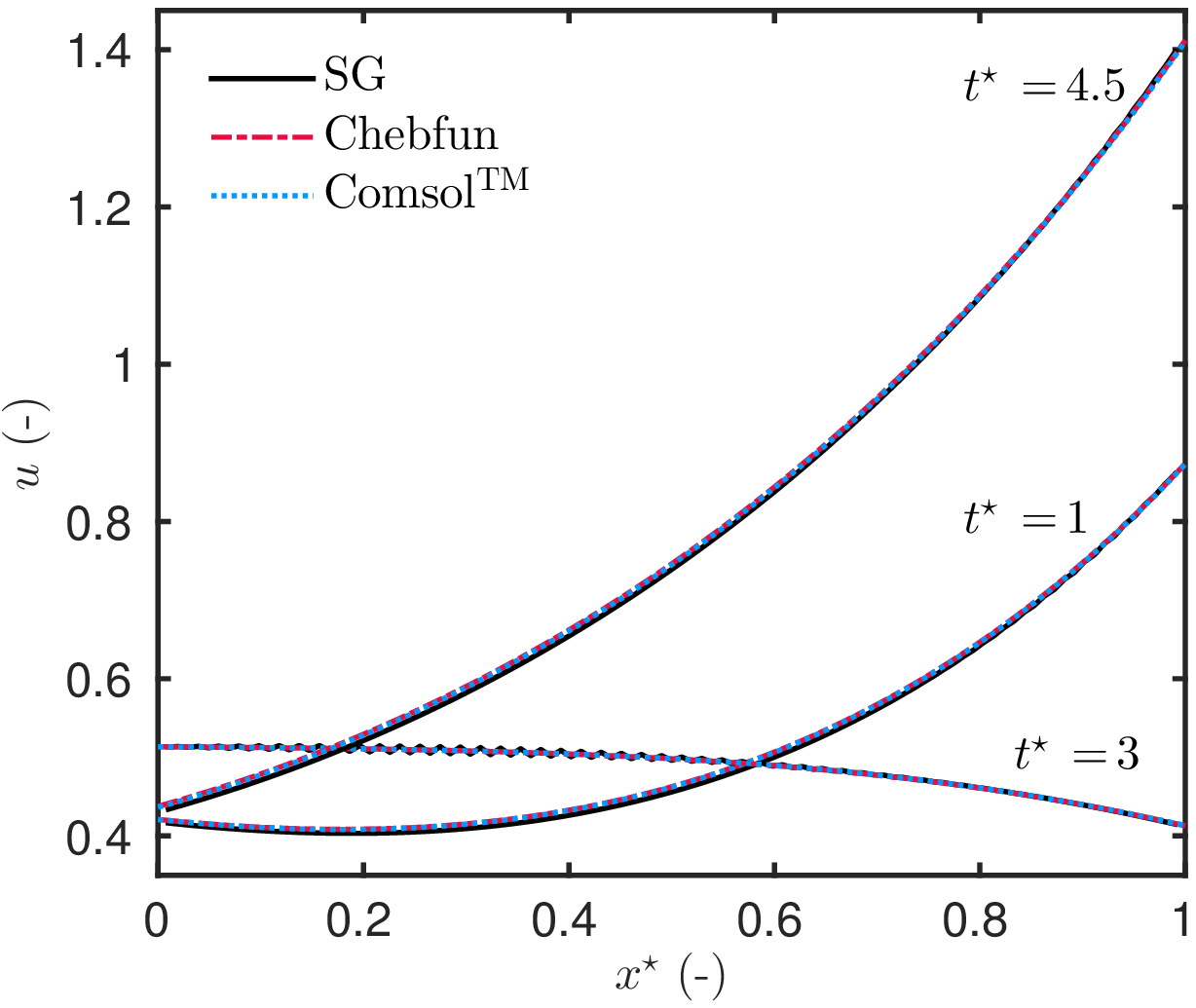}} 
  \subfigure[b][\label{fig_AN3:time_u}]{\includegraphics[width=.48\textwidth]{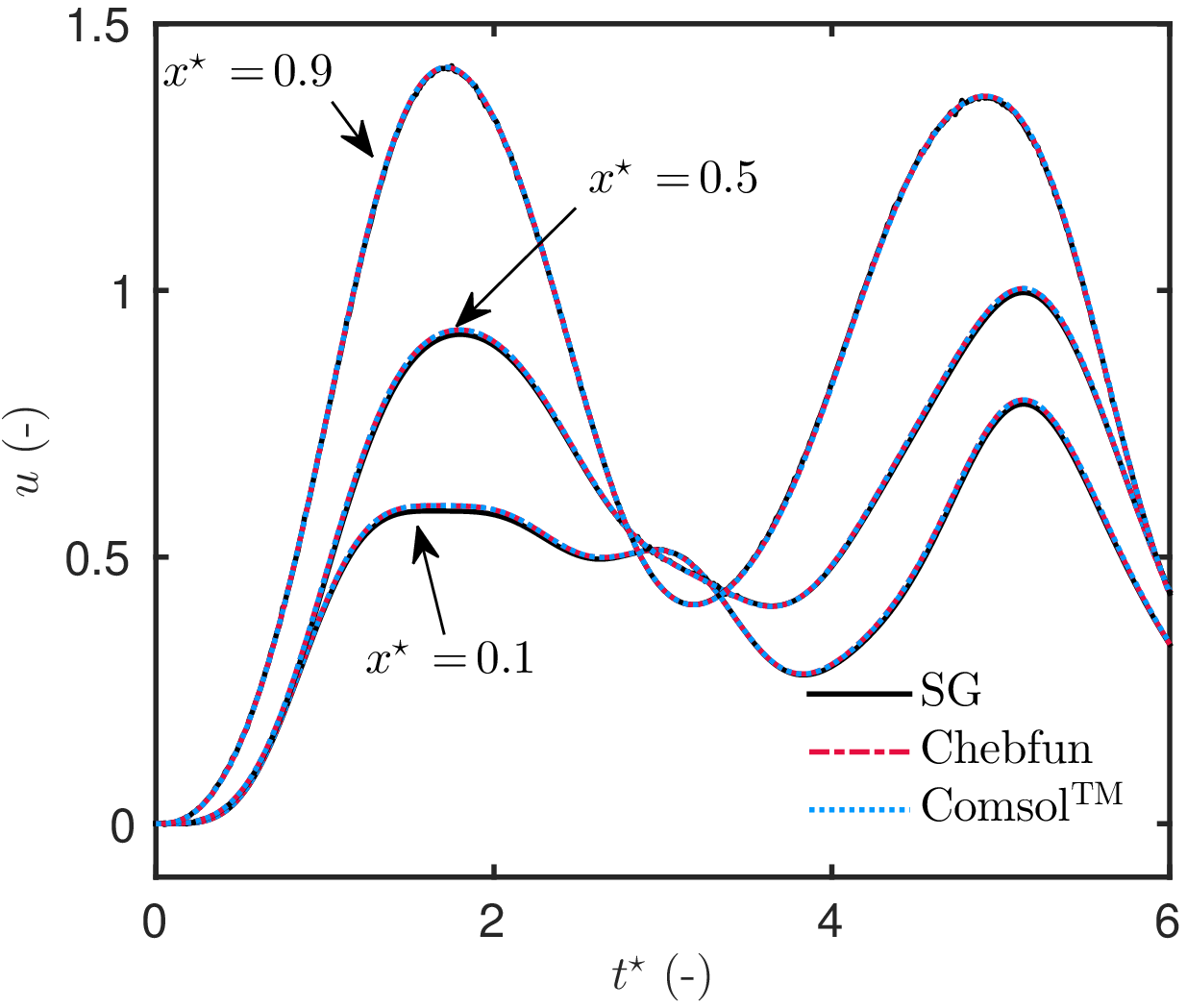}} \\
  \subfigure[c][\label{fig_AN3:uL2_fx}]{\includegraphics[width=.48\textwidth]{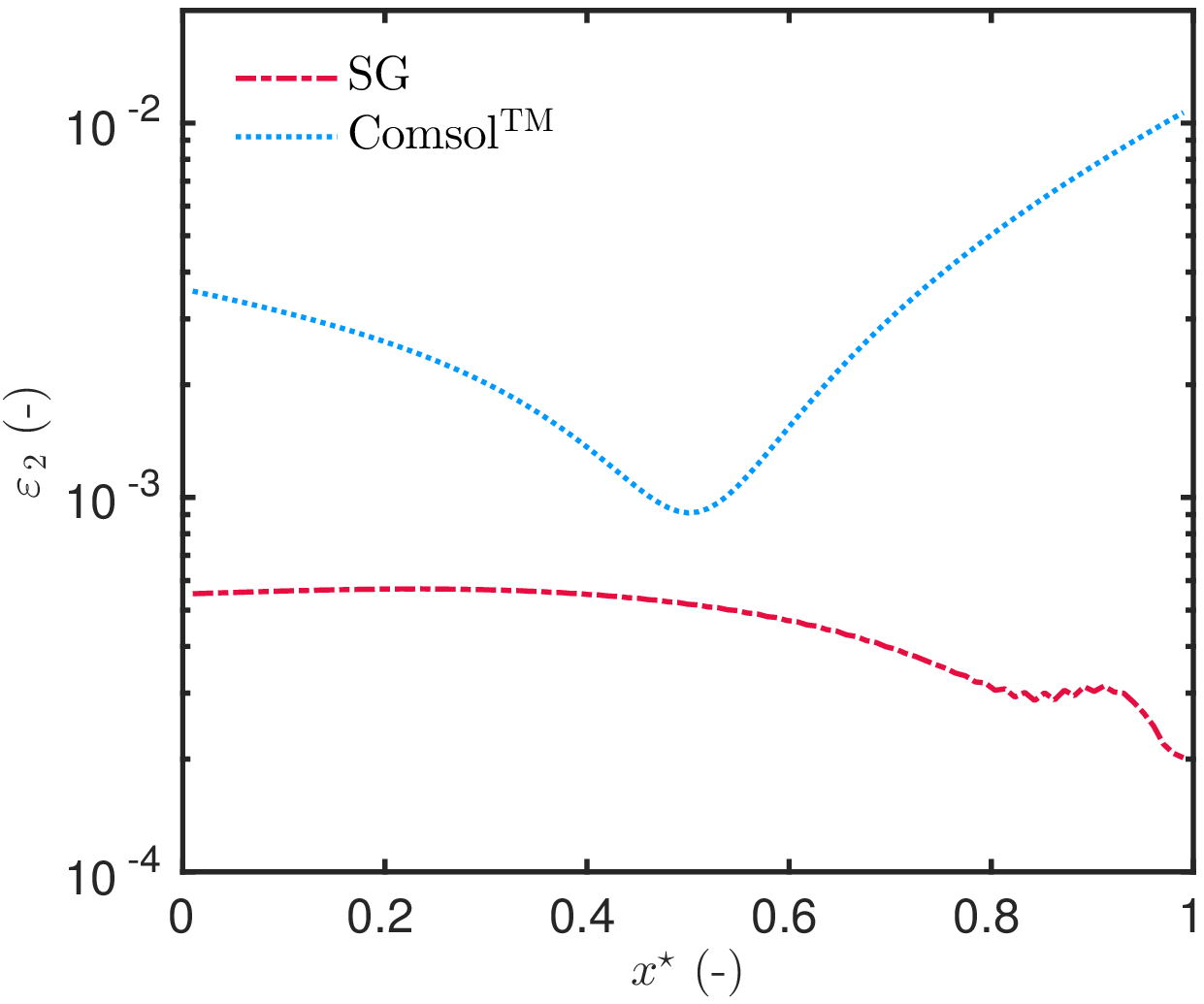}} 
  \subfigure[d][\label{fig_AN3:Res_ft}]{\includegraphics[width=.48\textwidth]{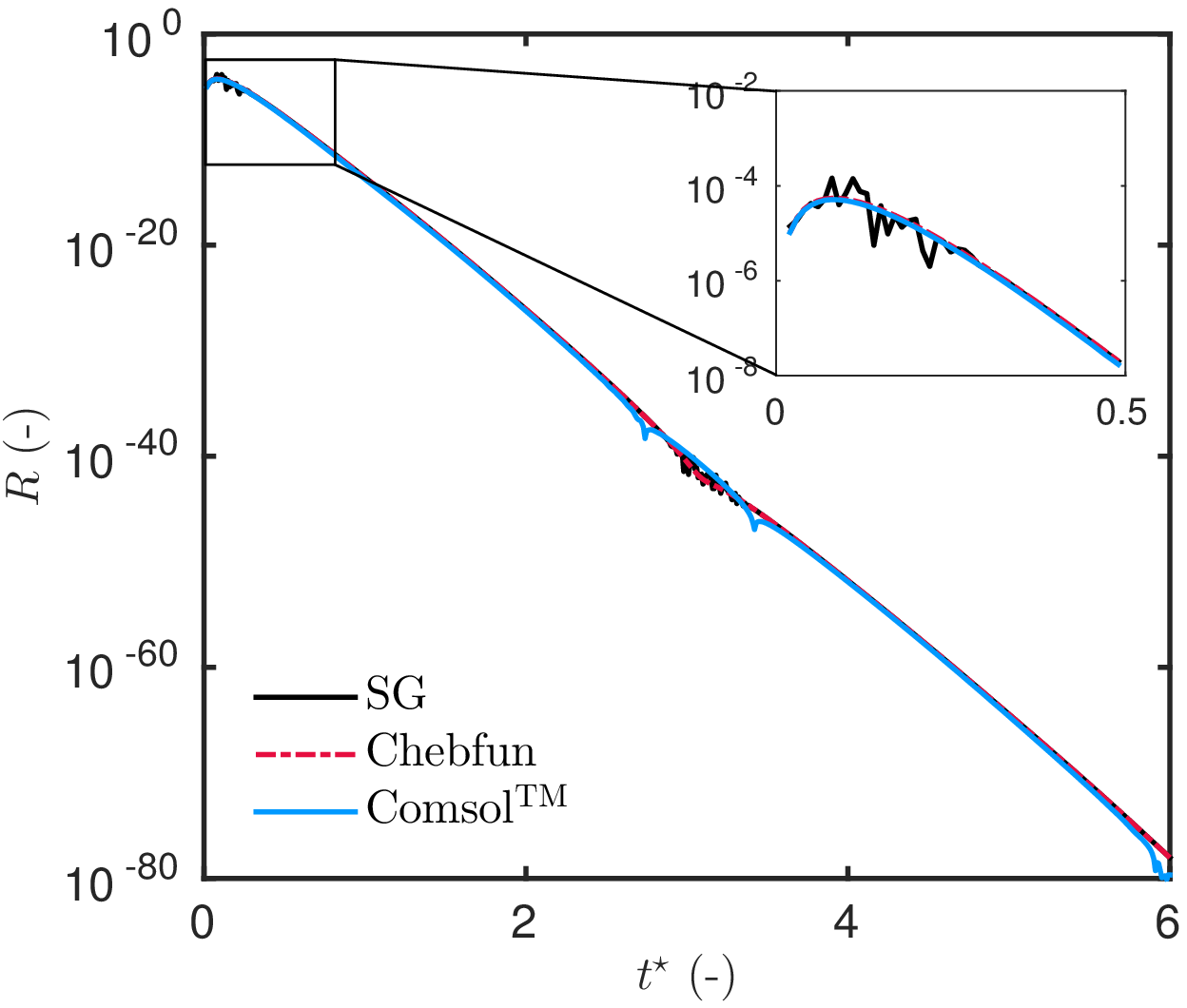}} 
  \caption{\emph{\small{Variation of the field $u$ as a function of $x$ (a) and $t$ (b). $\ell_{\,2}$ error $\varepsilon$ as a function of $x$ (c). Time variation of the residual $R \,(\,t\,)$ of the conservation law Eq.~\eqref{eq:cons_law}.}}}
\end{figure}

The \SG ~numerical solution is computed for a spatial discretisation $\dx \egal 10^{\,-2}\,$ and an adaptive time step with all tolerances set to $10^{\,-5}\,$. Moreover, in order to confirm the analysis of the studied scheme accuracy, the commercial software \COMSOL ~is used to compute the solution. It is based on a finite-element approach with a backward implicit time discretisation. The same spatial mesh $\dx \egal 10^{\,-2}\,$ is used. As no analytical solution was found, the reference solution is the one computed with the \texttt{Chebfun} package. Figures~\ref{fig_AN3:profil_u} and \ref{fig_AN3:time_u} show that the physical phenomena are perfectly represented by the \SG ~numerical solution. The field $u$ follows the variation of the boundary conditions. Once again, a very good agreement is observed between the \COMSOL, the \SG ~and the \texttt{Chebfun} solutions. The $\ell_{\,2}$ error is lower than $\O\,(\,10^{\,-2}\,) \,$ for both \COMSOL ~and \SG ~solutions. It is important to note that the evaluation of the CPU time is not accomplished since the \SG ~and \COMSOL ~algorithms are developed in different languages and environments.

Another possibility to verify the accuracy of the solution is to verify the conservation law. In this particular case ($a_{\,2} \egal d_{\,2} \egal 0 $), a non-trivial conservation law can be derived:
\begin{multline}\label{eq:cons_law}
  \pd{}{t}\;\Biggl[\, u \cdot \mathrm{e}^{\,-\,\dfrac{2 \, a_{\,1}}{d_{\,1}} \; \bigr(\,x \moins \dfrac{\Delta}{d_{\,1}} \; t \,\bigl)} \,\Biggr] \plus\\ 
  \pd{}{x}\;\Biggl[\,\Biggl(\, \dfrac{\Delta}{d_{\,1}} \; u \moins \bigl(\,d_{\,0} \plus d_{\,1} \, u \,\bigr)\;\pd{u}{x} \,\Biggr)\cdot \mathrm{e}^{\,-\,\dfrac{2 \, a_{\,1}}{d_{\,1}} \; \bigr(\,x \moins \dfrac{\Delta}{d_{\,1}} \; t \,\bigl)}\,\Biggr] \egal 0 \,,
\end{multline}
with $\Delta \ \eqdef \ a_{\,0} \, d_{\,1} \moins 2 \, a_{\,1} \, d_{\,0} \,$. In order to verify the accuracy of the numerical solution computed, the residual $R$ is calculated according to:
\begin{align*}
  R (\,t\,) \egal \int_{0}^1 \,  & \Biggl\{ \,\pd{}{t}\;\Biggl[\, u \cdot \mathrm{e}^{\,-\,\dfrac{2 \, a_{\,1}}{d_{\,1}} \; \bigr(\,x \moins \dfrac{\Delta}{d_{\,1}} \; t \,\bigl)} \,\Biggr] \\
  & \plus \pd{}{x}\;\Biggl[\,\Biggl(\, \dfrac{\Delta}{d_{\,1}} \; u \moins \bigl(\,d_{\,0} \plus d_{\,1} \, u \,\bigr) \;\pd{u}{x} \,\Biggr)\cdot \mathrm{e}^{\,-\,\dfrac{2 \, a_{\,1}}{d_{\,1}} \; \bigr(\,x \moins \dfrac{\Delta}{d_{\,1}} \; t \,\bigl)}\,\Biggr] \, \Biggr\} \, \mathrm{d}x\,,
\end{align*}
that can be rewritten in the following form:
\begin{align*}
  R (\,t\,)  \egal & \pd{}{t}\;\Biggl[\, \int_{\,0}^{\,1} \,  u \cdot \mathrm{e}^{\,-\,\dfrac{2 \, a_{\,1}}{d_{\,1}} \; \bigr(\,x \moins \dfrac{\Delta}{d_{\,1}} \; t \,\bigl)} \, \mathrm{d}x \,\Biggr] \\
  & \plus \Biggl[\,\Biggl(\, \dfrac{\Delta}{d_{\,1}} \; u \moins \bigl(\,d_{\,0} \plus d_{\,1} \, u \,\bigr) \;\pd{u}{x} \,\Biggr)\cdot \mathrm{e}^{\,-\,\dfrac{2 \, a_{\,1}}{d_{\,1}} \; \bigr(\,x \moins \dfrac{\Delta}{d_{\,1}} \; t \,\bigl)}\,\Biggr]_{\,x \,=\,1} \\
  & \moins \Biggl[\,\Biggl(\, \dfrac{\Delta}{d_{\,1}}\;u \moins \bigl(\,d_{\,0} \plus d_{\,1} \, u \,\bigr) \;\pd{u}{x} \,\Biggr)\cdot \mathrm{e}^{\,-\,\dfrac{2 \, a_{\,1}}{d_{\,1}} \; \bigr(\,x \moins \dfrac{\Delta}{d_{\,1}} \; t \,\bigl)}\,\Biggr]_{\,x \,=\, 0} \,.
\end{align*}

\begin{figure}
\centering
\includegraphics[width=.65\textwidth]{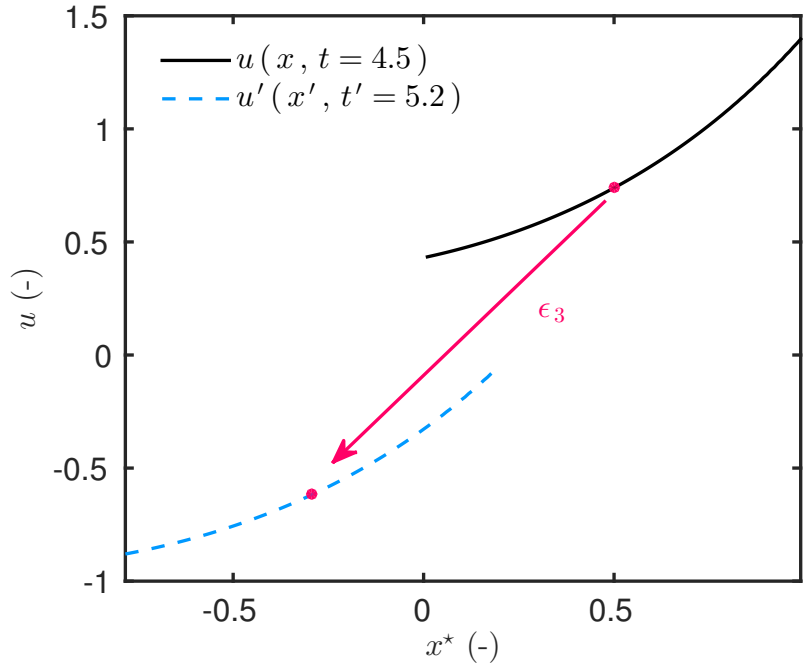}
\caption{\emph{\small{Illustration of the use of the scaling symmetry to compute $u\,(\,x^{\,\prime} \,,\,t^{\,\prime}\,)$ using $u\,(\,x \,,\,t\,)$ and the scaling parameter $\epsilon_{\,3} \egal -1.5 \,$.}}}
\label{fig_AN3:sym_u}
\end{figure}

Figure~\ref{fig_AN3:Res_ft} shows the time variation of the residual of the conservation law. The conservation law is verified by numerical solutions indicating a satisfying accuracy of the numerical methods. Another remark concerns the symmetry point transformation. These symmetries enables to translate physical observations (as space or time translations) and can be used in further studies when exploring the equation solutions. Three symmetries have been identified:
\begin{align*}
  1. \ \text{Space translation:} &
  \left.
  \begin{cases} 
  &\ x^{\,\prime} \egal x \plus \epsilon_{\,1} \\
  &\ t^{\,\prime} \egal t \\
  &\ u^{\,\prime} \egal u  \\
  \end{cases}
  \ \right.  \\
  2. \ \text{Time translation:} &
  \left.
  \begin{cases} 
  &\ x^{\,\prime} \egal x  \\
  &\ t^{\,\prime} \egal t \plus \epsilon_{\,2}\\
  &\ u^{\,\prime} \egal u  \\
  \end{cases}
  \ \right. \\
  3. \ \text{Scaling symmetry:} \ &
  \left.
  \begin{cases} 
  &\ x^{\,\prime} \egal x \moins \dfrac{\Delta}{d_{\,1}} \; \biggl(\, 1 \moins \mathrm{e}^{\,-\,d_{\,1} \, \epsilon_{\,3} } \,\biggr) \\
  &\ t^{\,\prime} \egal t \, \mathrm{e}^{\,- \,d_{\,1} \, \epsilon_{\,3} } \\
  &\ u^{\,\prime} \egal u \, \mathrm{e}^{\,d_{\,1} \, \epsilon_{\,3} } \moins \dfrac{d_{\,0}}{d_{\,1}} \; \biggl(\, 1 \moins \mathrm{e}^{\,d_{\,1} \, \epsilon_{\,3} } \,\biggr)  \\
  \end{cases}
  \ \right. 
\end{align*}
The symmetry can be used when exploring the equation solutions. Knowing a solution and a particular symmetry, an infinity of solution can be computed. An example has been included in the manuscript for the scaling symmetry $3\,$. Figure~\ref{fig_AN3:sym_u} shows the solution $u\,(\,x \,,\, t\,)$ obtained using the \SG ~numerical scheme. Adopting the scaling parameter $\epsilon_{\,3} \egal -1.5 \,$, the solution $u^{\,\prime} \, (\,x^{\,\prime} \,,\, t^{\,\prime}\,)$ can be obtained by applying the symmetry.

\section{Numerical method for coupled equations}
\label{sec:NM_coupled_equations}

For the sake of simplicity and without losing the generality, the numerical schemes are described for the system of two coupled linear advection--diffusion equations, written as: 
\begin{subequations}\label{eq:sys_conv_diff}
\begin{align}
  & \pd{u}{t} \plus \pd{f}{x} \egal 0 \,, & t & \ > \ 0\,, \;&  x & \ \in \ \big[ \, 0, \, 1 \, \big]\,, \\
  & \pd{v}{t} \plus \pd{g}{x} \egal 0 \,, & t & \ > \ 0\,, \;&  x & \ \in \ \big[ \, 0, \, 1 \, \big]\,, \\
  & f \egal \ai{11} \, u \moins \di{11} \;\pd{u}{x} \,, \\[3pt]
  & g \egal \ai{22} \, v \moins \di{22} \;\pd{v}{x} \plus \ai{21} \, u \moins \di{21} \;\pd{u}{x} \,,
\end{align}
\end{subequations}
where $u\,(x,\,t)$ and $v\,(x,\,t)\,$, $x\ \in\ \Omega_{\,x}\,$, $t\ >\ 0\,$, are the fields of interest, $d_{\,i\,j}\,$, the diffusion coefficients and, $a_{\,i\,j}\,$, the advection coefficients, both considered as constants (in this section). The frozen coefficients are used here only to generalize the results obtained for the case of a scalar linear advection--diffusion equation. At $x \egal 0\,$, the boundary conditions are:
\begin{subequations}\label{eq:sys_conv_diff_BC}
\begin{align}
  \di{11} \;\pd{u}{x} \moins \ai{11} \, u & \egal \mathrm{Bi}_{\,11} \cdot \left( \, u \moins \uinf \, \right) \,, \\[3mm]
  \di{22} \;\pd{v}{x} \moins \ai{22} \, v \plus \di{21} \;\pd{u}{x} \moins \ai{21} \, u & \egal \mathrm{Bi}_{\,22} \cdot \left( \, v \moins \vinf \, \right) \plus \mathrm{Bi}_{\,21} \cdot \left( \, u \moins \uinf \, \right)\,,
\end{align}
\end{subequations}
and at $x \egal 1\,$:
\begin{subequations}\label{eq:sys_conv_diff_BC}
\begin{align}
  \di{11} \;\pd{u}{x} \moins \ai{11} \, u & \egal -\,\mathrm{Bi}_{\,11} \cdot \left( \, u \moins \uinf \, \right) \,, \\[3mm]
  \di{22} \;\pd{v}{x} \moins \ai{22} \, v \plus \di{21} \;\pd{u}{x} \moins \ai{21} \, u & \egal -\,\mathrm{Bi}_{\,22} \cdot \left( \, v \moins \vinf \, \right) \moins \mathrm{Bi}_{\,21} \cdot \left( \, u \moins \uinf \, \right)\,,
\end{align}
\end{subequations}
where $\uinf\,(\,t\,)$ and $\vinf\,(\,t\,)$ are the field variables in the ambient air surrounding the material.


\subsection{Extension of the \SG ~scheme to the case of a system}

The discretisation of Eqs.~\eqref{eq:sys_conv_diff} yields to the following semi-discrete difference equations:
\begin{align*}
  \od{u_{\,j}}{t} \plus \dfrac{1}{\Delta x}\;\Biggl[\, f_{\,j\,+\,\frac{1}{2}}^{\,n}\ -\ f_{\,j\,-\,\frac{1}{2}}^{\,n}   \, \Biggr] \egal 0 \,, \\
  \od{v_{\,j}}{t} \plus \dfrac{1}{\Delta x}\;\Biggl[\, g_{\,j\,+\,\frac{1}{2}}^{\,n}\ -\ g_{\,j\,-\,\frac{1}{2}}^{\,n}   \, \Biggr] \egal 0 \,.
\end{align*}

As for the scalar case, the \SG ~scheme assumes the fluxes $f_{\,j\,+\,\half}^{\,n}$ and $g_{\,j\,+\,\half}^{\,n}$ to be constant on the dual cell $\bigl[\,x_{\,j} \,, \, x_{\,j\,+\,1} \, \bigr]\,$. The flux $f_{\,j\,+\,\half}^{\,n}$ is the solution of the following boundary-value problem:
\begin{subequations}\label{eq:SP_equation_u}
\begin{align}
  f_{\,j\,+\,\half}^{\,n} & \egal \ai{11} \, u \moins \di{11} \;\pd{u}{x} \,, & & \forall x \ \in \ \big[ \, x_{\,j}  \,, \, x_{\,j\,+\,1} \, \big] \,, \qquad \ \forall  j \ \in \ \bigl\{\,2\,,\ldots,\,N\moins 1 \,\bigl\} \,, \\
  u & \egal u_{\,j}^{\,n} \,, && x \egal x_{\,j}  \,,\\
  u  &\egal u_{\,j\,+\,1}^{\,n} \,, && x \egal x_{\,j\,+\,1}\,.
\end{align}
\end{subequations}
While $g_{\,j\,+\,\half}^{\,n}\,$, $\forall  j \ \in \ \bigl\{\,2\,,\ldots,\,N \moins 1 \,\bigl\}$ is given as the solution of:
\begin{subequations}\label{eq:SP_equation_v}
\begin{align}
  g_{\,j\,+\,\half}^{\,n} & \egal \ai{22} \, v \moins \di{22} \;\pd{v}{x} \plus \ai{21} \, u \moins \di{21} \;\pd{u}{x} \,, & & \forall x \ \in \ \big[\,x_{\,j}  \,, \, x_{\,j\,+\,1}  \, \big]\,, \\
  v & \egal v_{\,j}^{\,n} \,, && x \egal x_{\,j}  \,,\\
  v  &\egal v_{\,j\,+\,1}^{\,n} \,, && x \egal x_{\,j\,+\,1}  \,.
\end{align}
\end{subequations}
We define $\Thi{\,k\,l} \ \eqdef \ \dfrac{a_{\,k\,l} \, \Delta x}{d_{\,k\,l}} \,$. Then, the computation of $f_{\,j\,+\,\half}^{\,n}$ from Eq.~\eqref{eq:SP_equation_u} is straightforward:
\begin{align*}
  & f_{\,j\,+\,\half}^{\,n} \egal \dfrac{\di{1\,1}}{\Delta x}\;\Biggl[\,-\ \mathcal{B}\bigl(\, \Thi{1\,1} \,\bigr) \, u_{\,j\,+\,1}^{\,n} \plus \mathcal{B} \bigl(\, - \, \Thi{1\,1} \,\bigr)\,u_{\,j}^{\,n}\, \Biggr] \,,
\end{align*}
Using results from Eq.~\eqref{eq:interp_u}, the solution $u\,(\,x\,)$ can also be computed exactly as well:
\begin{align}\label{eq:SP_solution_u}
  & u^{\,n}\,(\,x\,) \egal \dfrac{1}{\ai{1\,1}} \;f_{\,j\,+\,\frac{1}{2}}^{\,n} \plus \dfrac{\Bigr( \, u_{\,j}^{\,n} \moins u_{\,j\,+\,1}^{\,n} \, \Bigr)}{1 \moins  \mathrm{e}^{\,\Theta}}\,\mathrm{exp} \biggl(\, \dfrac{\ai{1\,1}}{\di{1\,1}} \; \bigl(\,x \moins x_{\,j} \,\bigr) \,\biggr) \,, 
  && x \ \in \ \big[ \, x_{\,j} \,, \, x_{\,j\,+\,1}  \, \big]\,.
\end{align}

For the computation of $g_{\,j\,+\,\frac{1}{2}}^{\,n}$ from Eq.~\eqref{eq:SP_equation_v}, the solution $u\,(\,x\,)$ from Eq.~\eqref{eq:SP_solution_u} is used. For the sake of notation compactness, the expression of the flux $g_{\,j\,+\,\frac{1}{2}}^{\,n}$ is provided in the \texttt{Maple} sheet provided as a supplementary material.

All specific features of the \SG ~scheme mentioned in Section~\ref{sec:Spec_feat} are still valid for the system of differential equations~\eqref{eq:sys_conv_diff_BC}. The scheme is well-balanced and asymptotic preserving. An exact computation of solutions $u$ and $v$ can be computed in the interval $\Bigl[\,x_{\,j} \,, \, x_{\,j\,+\,1} \, \Bigr]\,$. The CFL condition is extended as:
\begin{align*}
  \dt \, \max_{1 \, \leqslant \, k \,,\, l \, \leqslant \, 2} \ d_{\,kl} \ \max_{1 \, \leqslant \, j \, \leqslant \, N}\;\Biggl[\, \dfrac{a_{\,kl}}{d_{\,kl}} \; \tanh \, \Biggl(\, \dfrac{a_{\,kl} \, \dx}{2 \, d_{\,kl}} \,\Biggr)^{\,-1} \,\Biggr] \ \leqslant\ \dx \,.
\end{align*}

The treatment of non-linear problems, where coefficients $\ai{kl}$ and $\di{kl}$ depend on $u$ and $v\,$, is completely analogous to Section~\ref{sec:ext_NL} with the approach of the frozen coefficients on the dual cell. The problem of heat and mass transfer formulated in Section~\ref{sec:HM_transfer} corresponds to a weakly coupled system of differential equations. When considering highly coupled equations, the \SG ~approach can be applied for each equation by assuming the flux as constant and computing the latter by solving the associated boundary value problem.


\subsection{Numerical validation}

To validate the numerical \SG ~scheme for a system of coupled differential equations, two cases are considered. The first one, which considers constant material properties, is used to undertake a convergence study on the discretisation parameter $\dx$ and $\dt\,$. The second case, with material properties depending on the fields, will highlight the accuracy of the scheme to treat a non-linear problem.


\subsubsection{Case 1}

In this case, the material properties do not depend on the fields and they are fixed to:
\begin{align*}
  &  \ai{11} \egal 0.02 \,, 
  && \di{11} \egal 0.09 \,, 
  && \ai{22} \egal 0.03 \,, 
  && \di{22} \egal 0.07 \,, 
  && \ai{21} \egal 0.01 \,, 
  && \di{21} \egal 0.03 \,. 
\end{align*}
The initial condition is set to $u \egal v \egal 0\,$. For the \textsc{Robin} type boundary conditions, the \textsc{Biot} numbers are equal to: 
\begin{align*}
  & x \egal 0 \,: 
  && \mathrm{Bi}_{\,11} \egal 1.5 \,, 
  && \mathrm{Bi}_{\,22} \egal 0.6 \,, 
  && \mathrm{Bi}_{\,21} \egal 0.2 \,, \\
  & x \egal 1 \,:
  && \mathrm{Bi}_{\,11} \egal 1.3 \,, 
  && \mathrm{Bi}_{\,22} \egal 1.1 \,, 
  && \mathrm{Bi}_{\,21} \egal 0.8 \,. 
\end{align*}
In the ambient air, the fields vary according to sinusoidal variations:
\begin{align*}
  & x \egal 0 \,: 
  && \uinf (\,t\,) \egal  0.2 \, \sin^{\,2} \bigl(\, \pi \, t \,\bigr) \,, 
  && \vinf (\,t\,) \egal  0.6 \, \sin^{\,2} \biggl(\, \dfrac{2}{5} \; \pi \, t \,\biggr) \,,  \\[3pt]
  & x \egal 1 \,:
  && \uinf (\,t\,) \egal  0.9 \, \sin^{\,2} \biggl(\, \dfrac{2}{6} \;\pi \, t \,\biggr) \,, 
  && \vinf (\,t\,) \egal  0.5 \, \sin^{\,2} \biggl(\, \dfrac{2}{3} \; \pi \, t \,\biggr) \,.
\end{align*}

\begin{figure}
  \centering
  \subfigure[a][\label{fig_AN4:profil_u}]{\includegraphics[width=.48\textwidth]{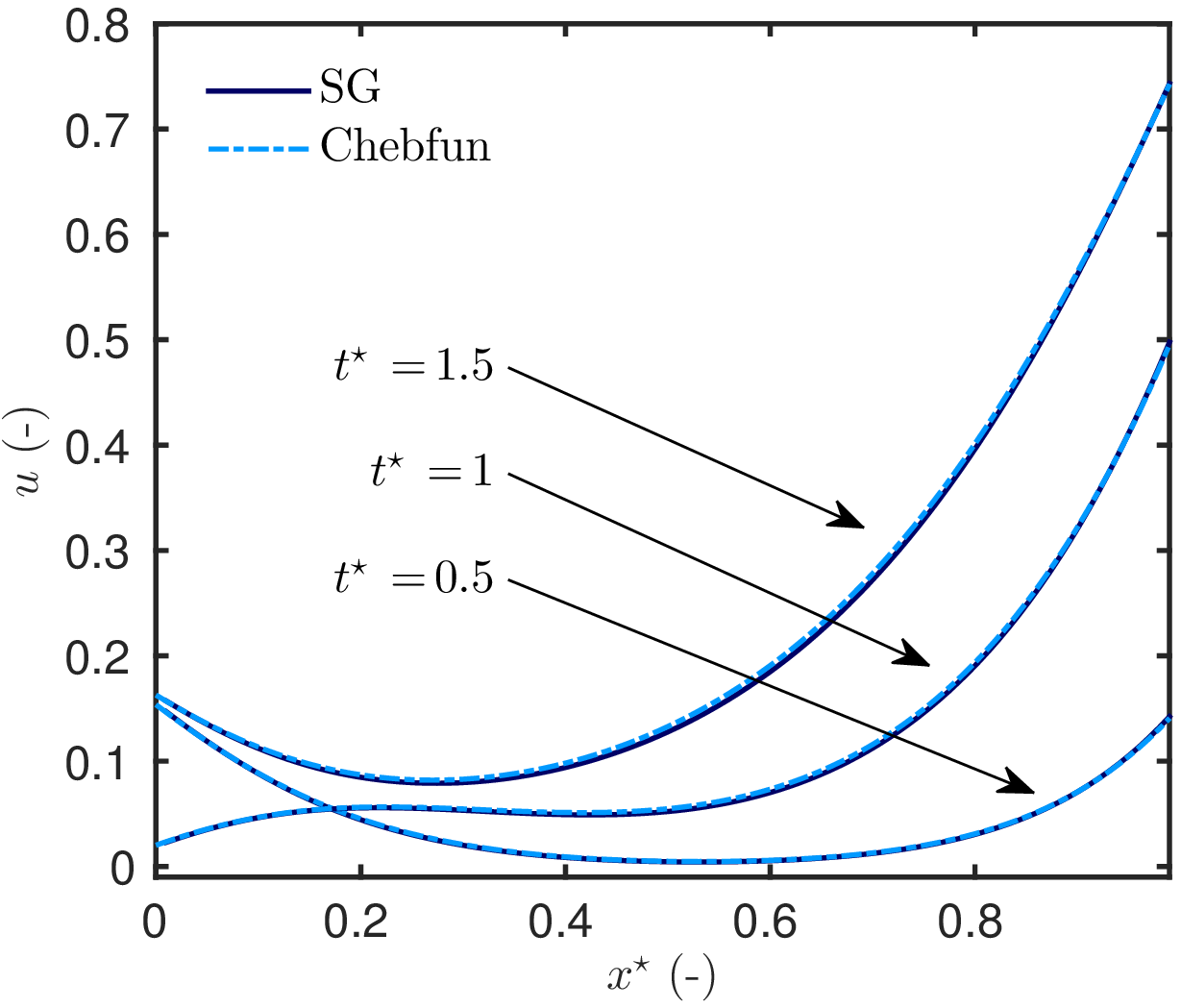}} 
  \subfigure[b][\label{fig_AN4:profil_v}]{\includegraphics[width=.48\textwidth]{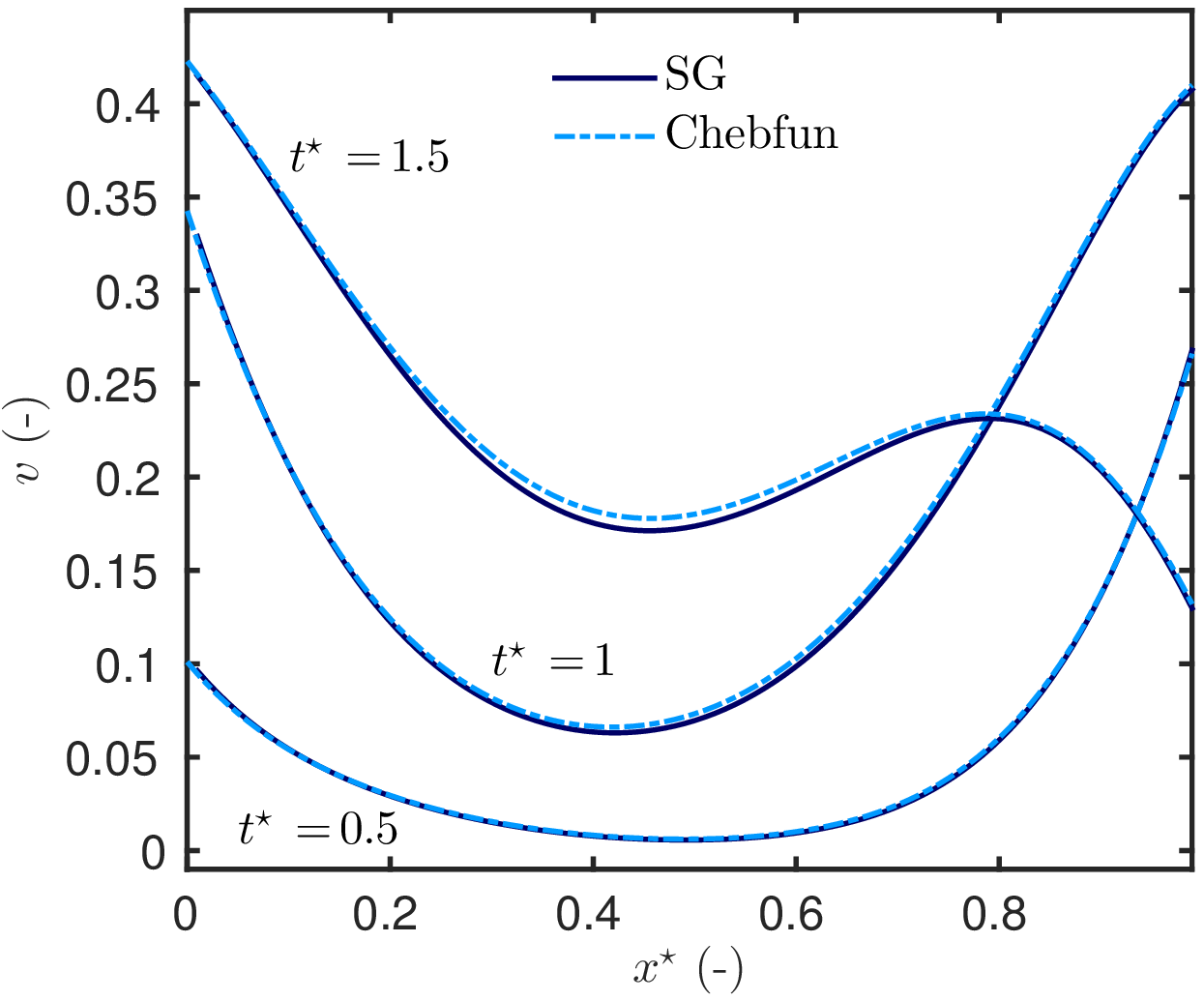}}
  \subfigure[c][\label{fig_AN4:time_u}]{\includegraphics[width=.48\textwidth]{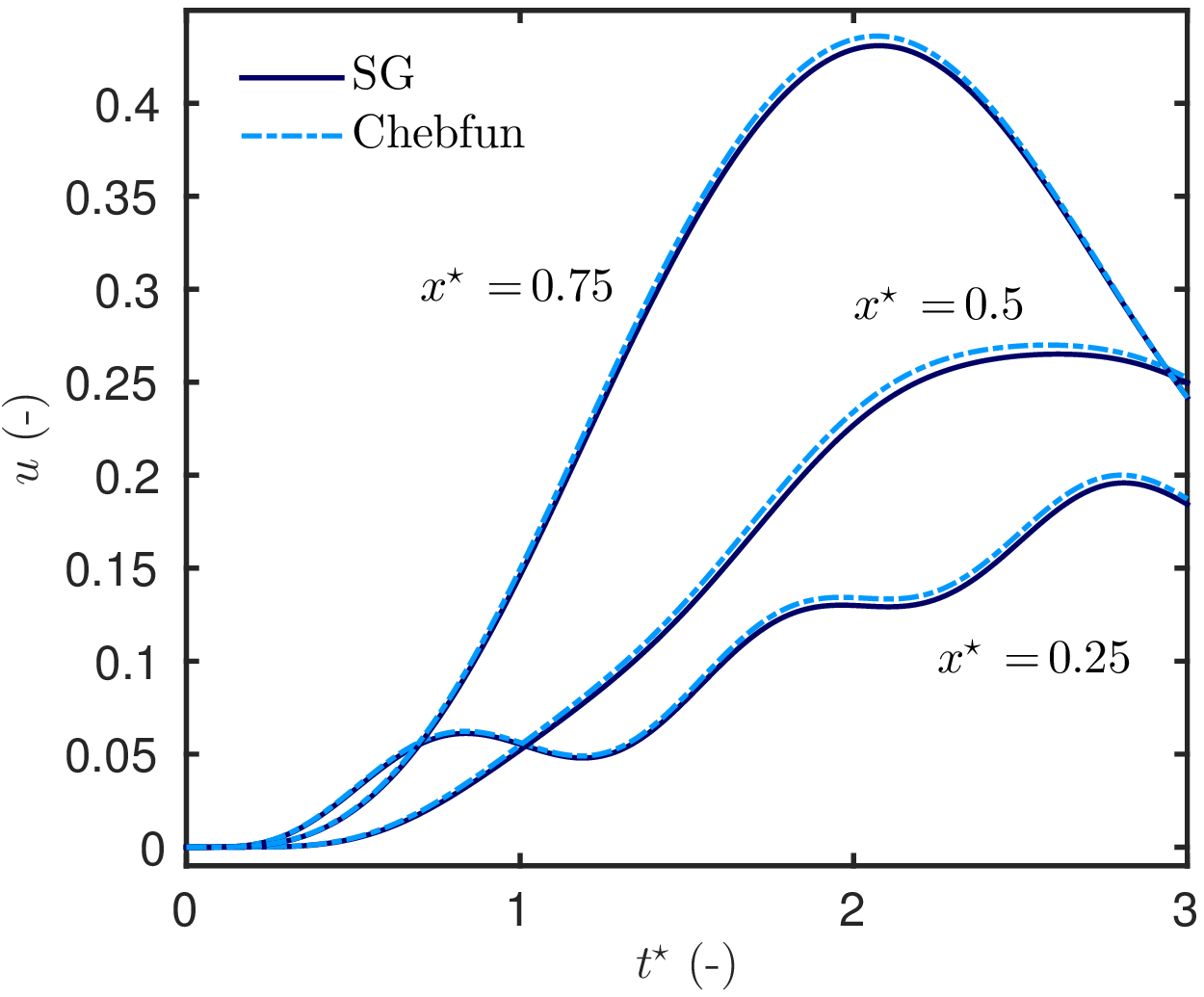}} 
  \subfigure[d][\label{fig_AN4:time_v}]{\includegraphics[width=.48\textwidth]{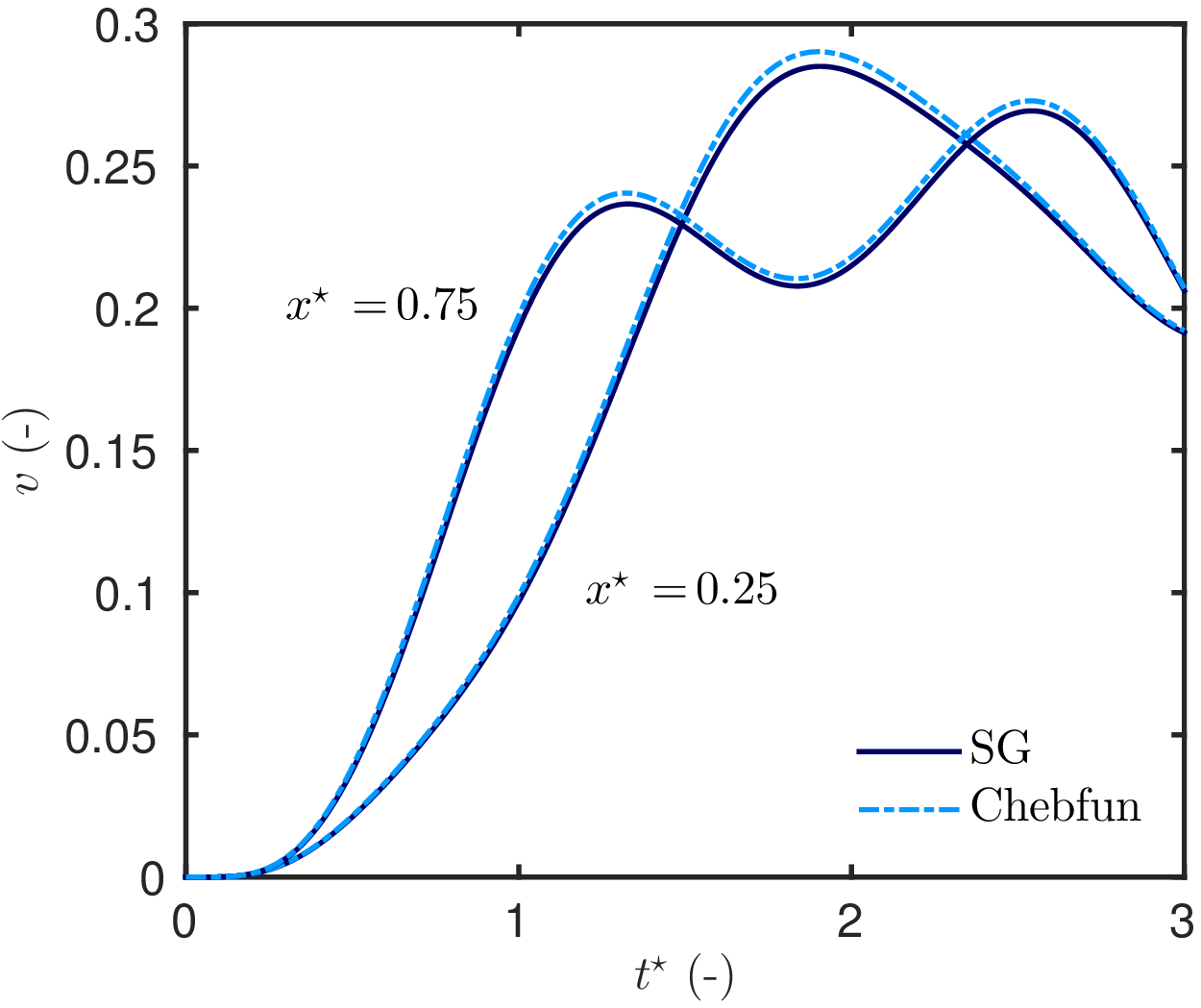}} 
  \caption{\emph{\small{Variation of the fields as a function of $x$ (a,b) and $t$ (c,d).}}}
\end{figure}

\begin{figure}
  \centering
  \includegraphics[width=.55\textwidth]{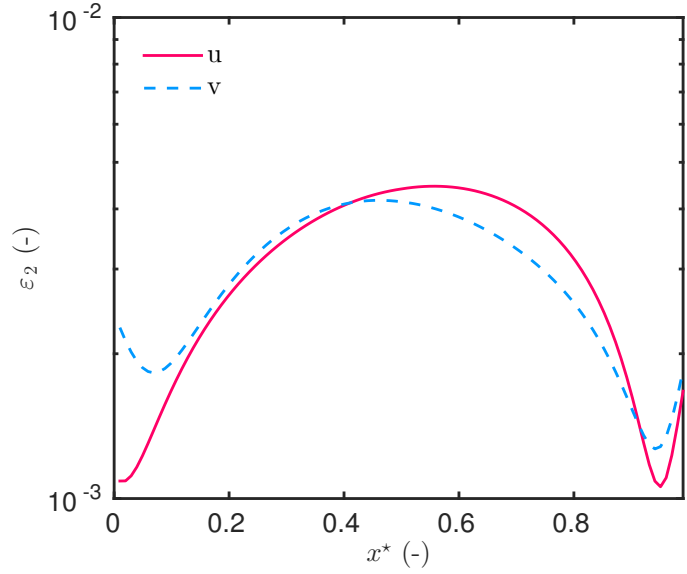}
  \caption{\emph{\small{$\ell_{\,2}$ error as a function of $x\,$.}}}
  \label{fig_AN4:L2_fx}
\end{figure}

\begin{figure}
  \centering 
  \subfigure[][\label{fig_AN4:SG_L2_fdx}]{\includegraphics[width=.48\textwidth]{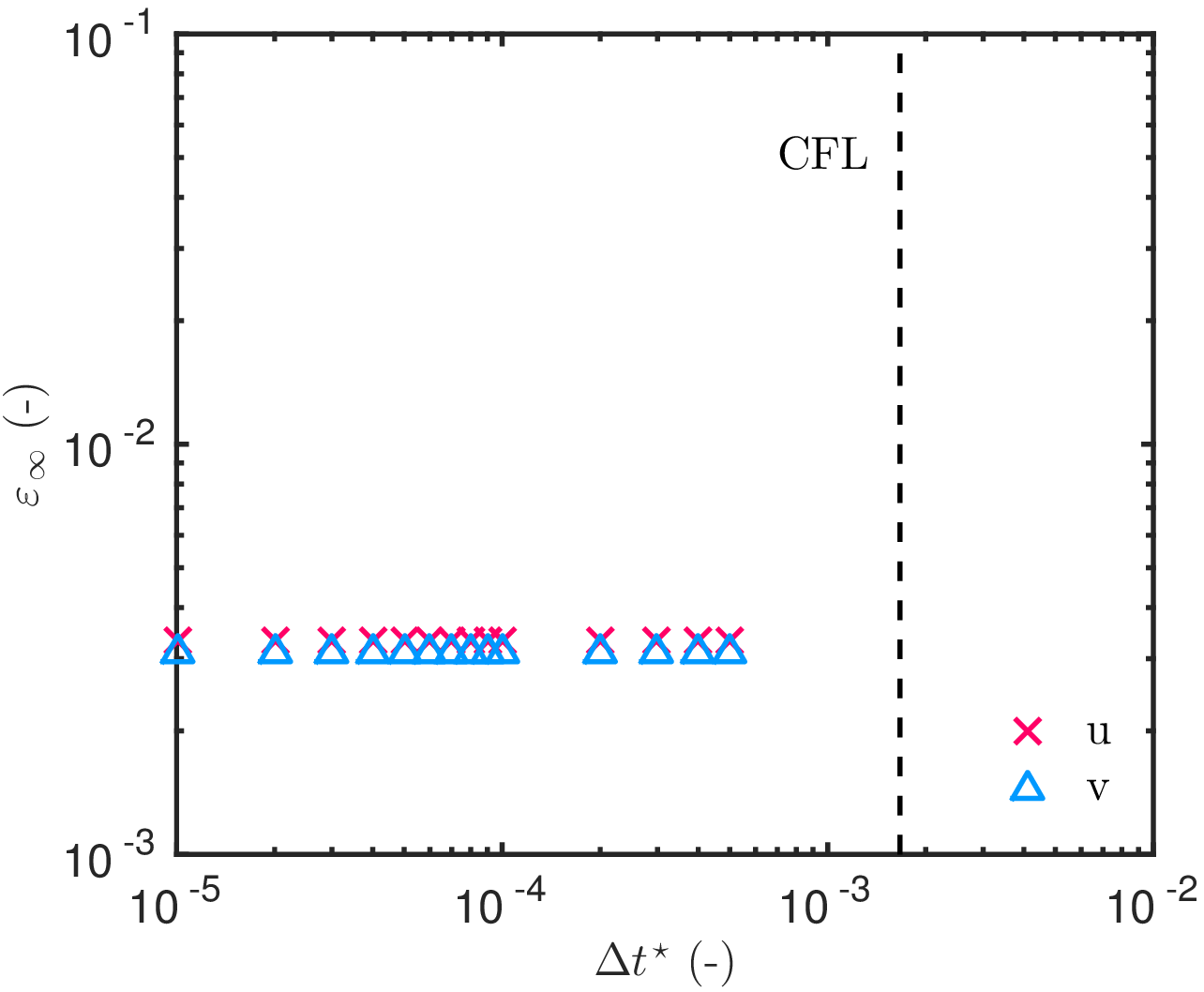}}
  \subfigure[][\label{fig_AN4:SG_L2_fdt}]{\includegraphics[width=.48\textwidth]{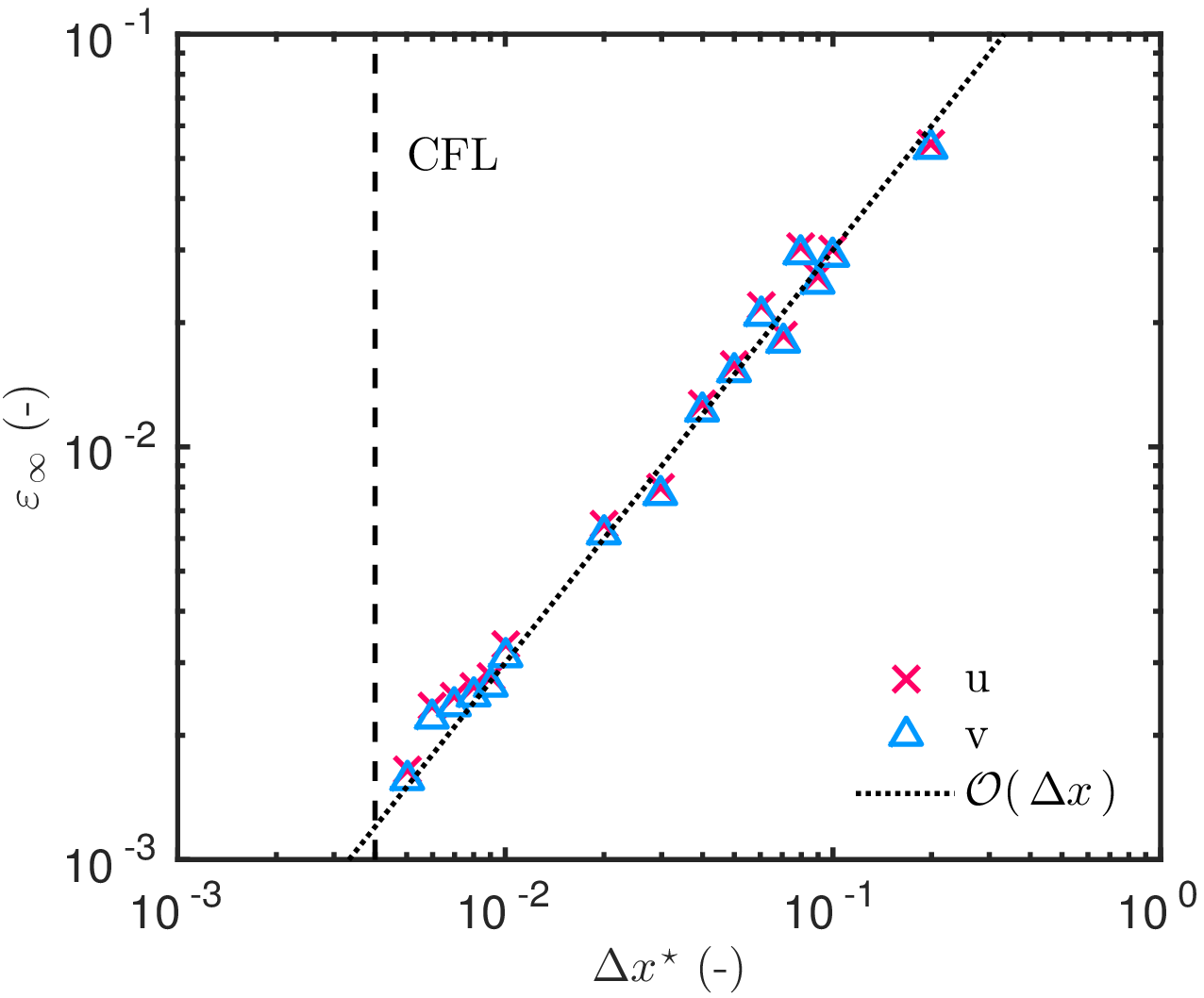}}
  \caption{\emph{\small{Variation of the $\ell_{\,2}$ error as a function of $\dx$ ($\dt \egal 10^{\,-4}$) (a) and $\dt$ ($\dx \egal 10^{\,-2}$) (b) using an explicit \SG ~scheme.}}}
\end{figure}

\begin{figure}
  \centering 
  \subfigure[c][\label{fig_AN4:SGoed_L2_fdx}]{\includegraphics[width=.48\textwidth]{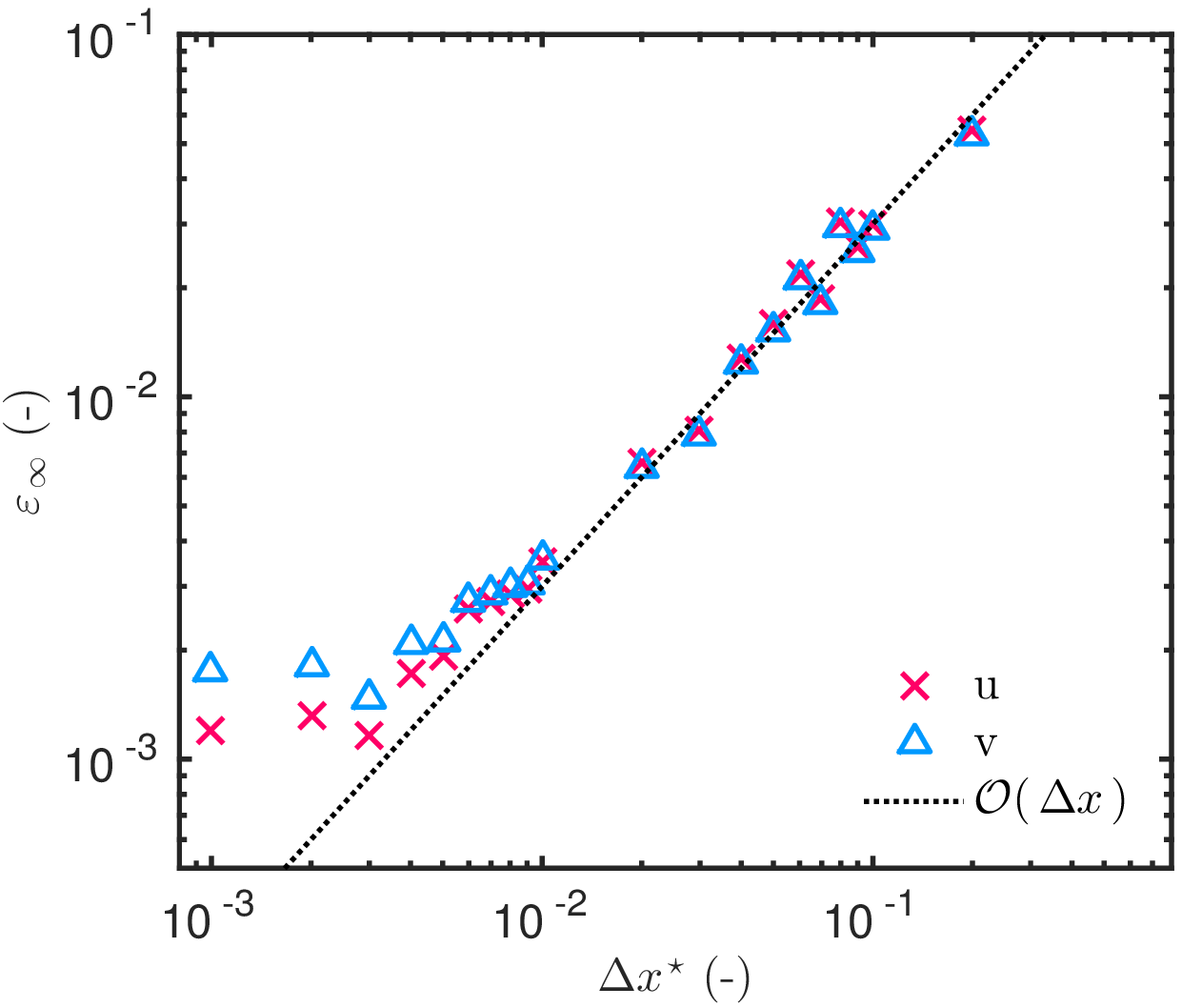}}
  \subfigure[][\label{fig_AN4:SG_CPU_fdx}]{\includegraphics[width=.48\textwidth]{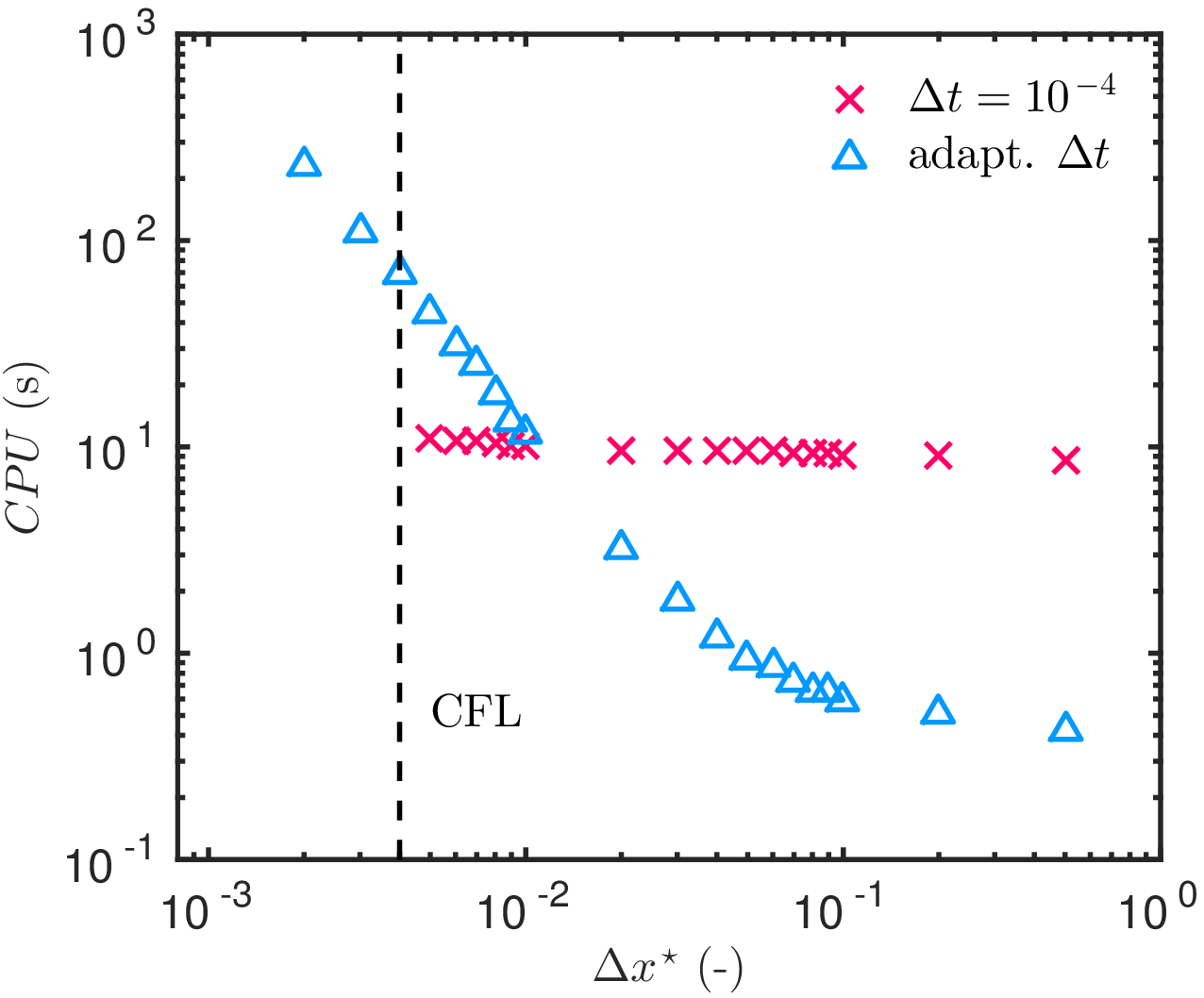}}
  \caption{\emph{\small{Variation of the $\ell_{\,2}$ error as a function of $\dx$ using an adaptive in time \SG ~scheme (a) and variation of the CPU time as a function of $\dx$ (b).}}}
\end{figure}

The simulation final time is $t \egal 3\,$. The discretisation parameters used for the computation are $\dx \egal 10^{\,-2}$ and $\dt \egal 10^{\,-4}\,$. These parameters respect the CFL conditions: $\dt \, \leqslant \, 5 \cdot 10^{\,-4}\,$. The variation of the fields $u\,(\,x\,,\,t\,)$ and $v\,(\,x\,,\,t\,)$ as a function of time and space is illustrated in Figures~\ref{fig_AN4:profil_u}--\ref{fig_AN4:time_v}. It follows the variation of boundary conditions and physical phenomena, which are well reflected. Moreover, a very good agreement can be noticed between the solution computed with the \SG ~scheme and the reference one. For both fields, the $\ell_{\,2}$ error is less than $5 \cdot 10^{\,-3}$ as shown in Figure~\ref{fig_AN4:L2_fx}. A convergence study has been carried out by varying $\dt$ or $\dx$ and fixing the other one. Figure~\ref{fig_AN4:SG_L2_fdt} shows the variation of the error as a function of $ \dt $ for a fixed spatial discretisation $\dx \egal 10^{\,-2}\,$. The error is invariant and equals to the absolute error of the scheme for the range of $\dt$ considered. The scheme is not able to compute a solution when the CFL condition is not respected. Figure~\ref{fig_AN4:SG_L2_fdx} gives the error $\varepsilon_{\,2}$ as a function of $\dx$ for a fixed $\dt \egal 10^{\,-4} \,$. It can be noted that the error $\ell_{\,2}$ as a similar behavior for both fields. In addition, the \SG ~scheme is first-order accurate in space $\O\,(\,\dx\,)\,$. For this parametric study, the computational time of the scheme has been compared for two approaches: (i) with a fixed  time step $\dt \egal 10^{\,-4}$ and (ii) with an adaptive time step using the \textsc{Matlab\;\texttrademark} function \texttt{ode113} and two tolerances set to $10^{\,-5}\,$. As shown in Figure~\ref{fig_AN4:SG_CPU_fdx}, using an adaptive time step enables an important reduction of the computation time when $\dx$ is relatively large without losing any accuracy. Figure~\ref{fig_AN4:SGoed_L2_fdx} gives the variation of the error as a function of the discretisation parameter $\dx\,$. Thanks to the time adaptive feature of the algorithm, it enables to respect the CFL condition for any value of space discretisation parameter $\dx \,$.

\subsubsection{Case 2}

The material properties are now depending on the fields $u$ and $v$:
\begin{align*}
  &  \ai{11} \egal 0.02 \plus 0.3 \, u \plus 0.6 \, v^{\,2} \,, 
  && \di{11} \egal 0.09 \plus 0.5 \, u^{\,2} \plus 0.5 \, v \,, \\
  & \ai{22} \egal 0.03 \plus 0.2 \, u \plus 0.1 \, v \,, 
  && \di{22} \egal 0.07 \plus 0.6 \, u^{\,2} \plus 0.5 \, v^{\,2} \,, \\
  & \ai{21} \egal 0.01 \plus 0.3 \, u \plus 0.5 \, v \,, 
  && \di{21} \egal 0.03 \plus 0.1 \, u \plus 0.3 \, u^{\,2} \plus 0.5 \, v \,. 
\end{align*}
The initial condition and \textsc{Biot} numbers are similar to the ones from the previous case. The boundary conditions are: 
\begin{align*}
  & x \egal 0 \,, 
  && \uinf\,(\,t\,) \egal  0.2 \, \biggl(\, 1 \moins \cos^{\,2}\,\bigl(\, \pi \, t \,\bigr) \,\biggr) \,, 
  && \vinf\,(\,t\,) \egal  0.6 \, \sin^{\,2}\,\bigl(\, \pi \, t \,\bigr) \,,  \\
  & x \egal 1 \,,
  && \uinf\,(\,t\,) \egal  0.9 \, \sin^{\,2}\,\biggl(\, \dfrac{2}{6} \;\pi \, t \,\biggr) \,, 
  && \vinf\,(\,t\,) \egal  0.5 \, \sin^{\,2}\,\biggl(\, \dfrac{2}{3} \; \pi \, t \,\biggr) \,.
\end{align*}

\begin{figure}
  \centering
  \subfigure[a][\label{fig_AN5:profil_u}]{\includegraphics[width=.48\textwidth]{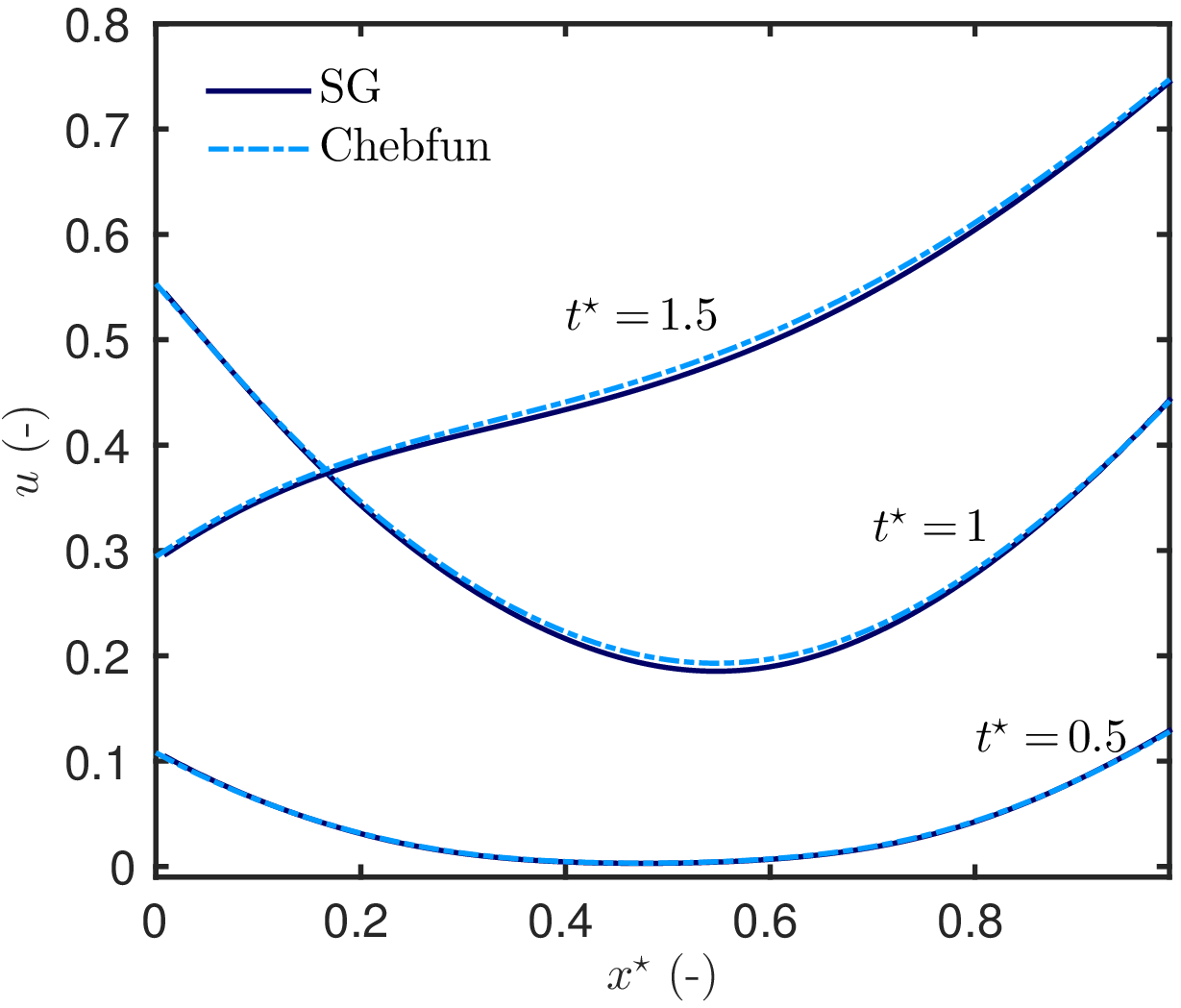}} 
  \subfigure[b][\label{fig_AN5:profil_v}]{\includegraphics[width=.48\textwidth]{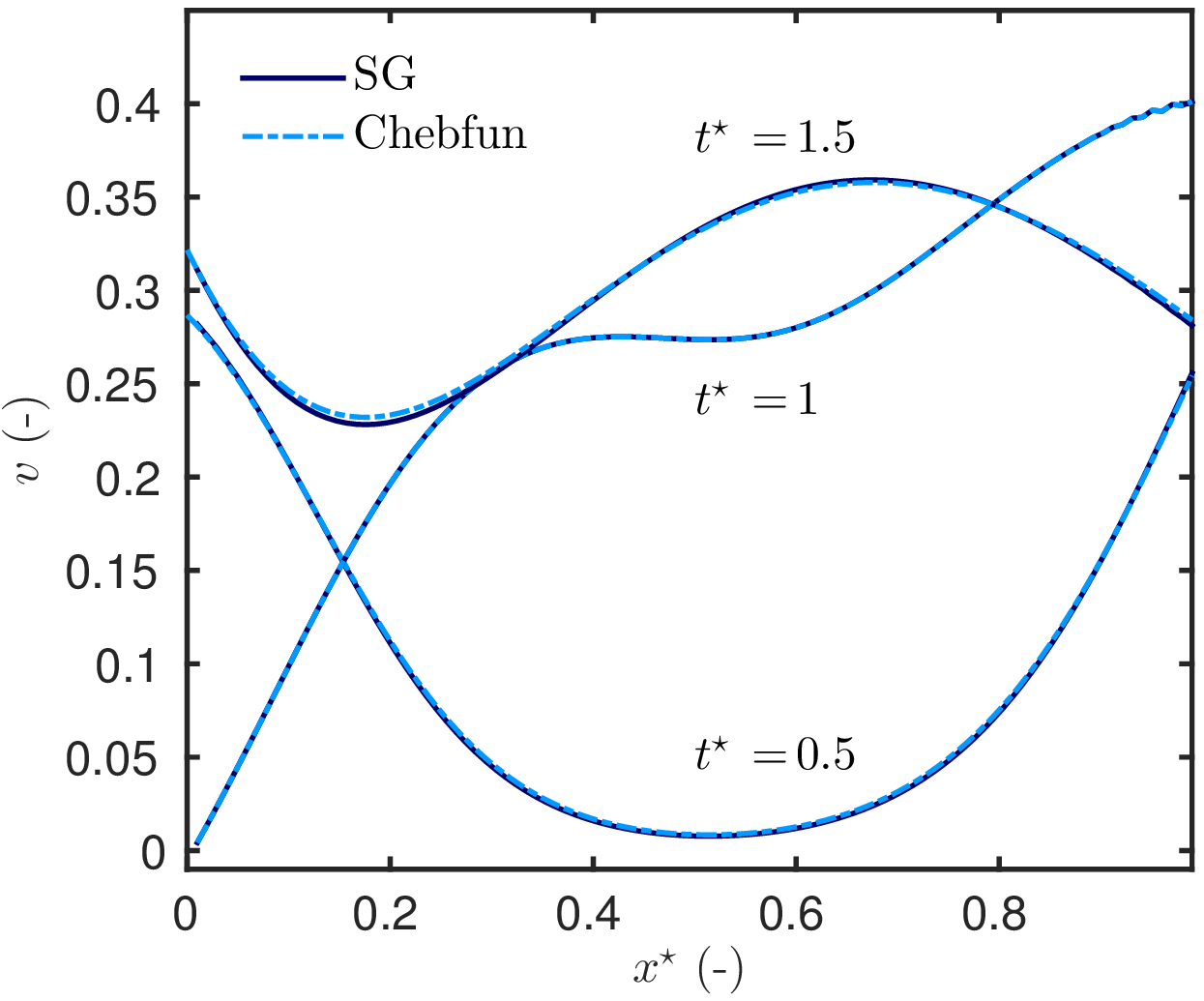}}
  \subfigure[c][\label{fig_AN5:time_u}]{\includegraphics[width=.48\textwidth]{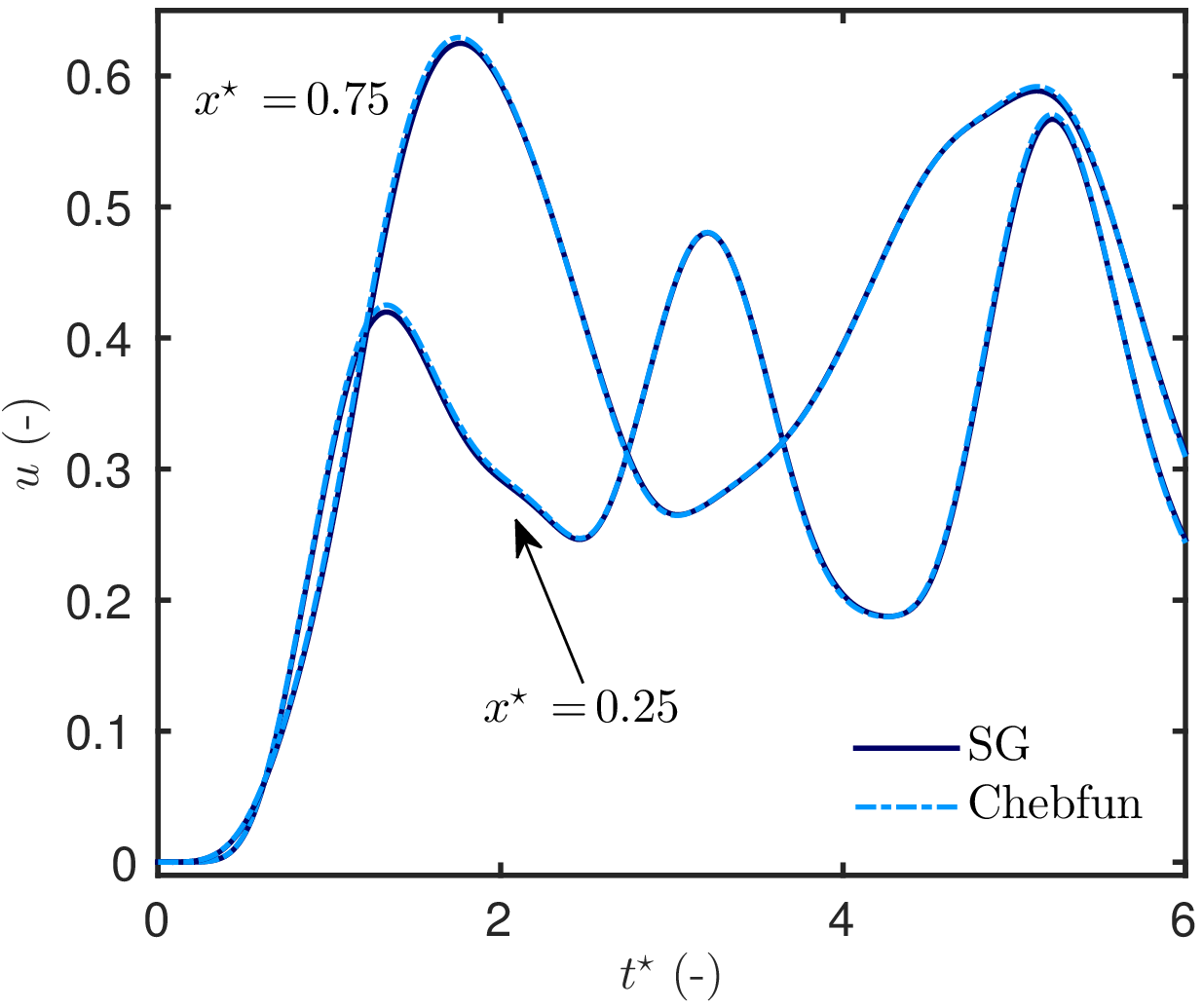}}
  \subfigure[d][\label{fig_AN5:time_v}]{\includegraphics[width=.48\textwidth]{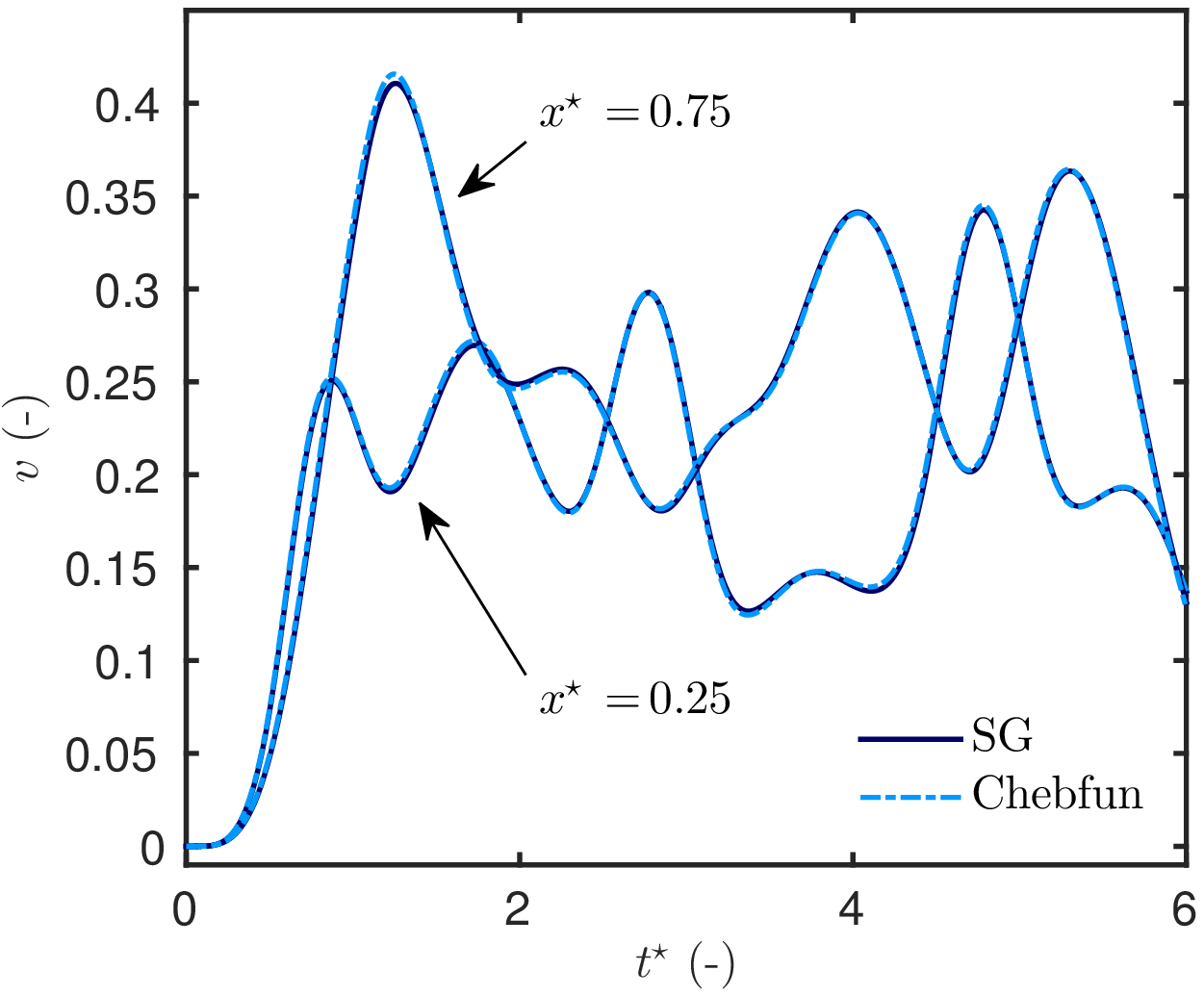}}
  \caption{\emph{\small{Variation of the fields as a function of $x$ (a,b) and $t$ (c,d).}}}
\end{figure}

\begin{figure}
  \centering
  \includegraphics[width=.5\textwidth]{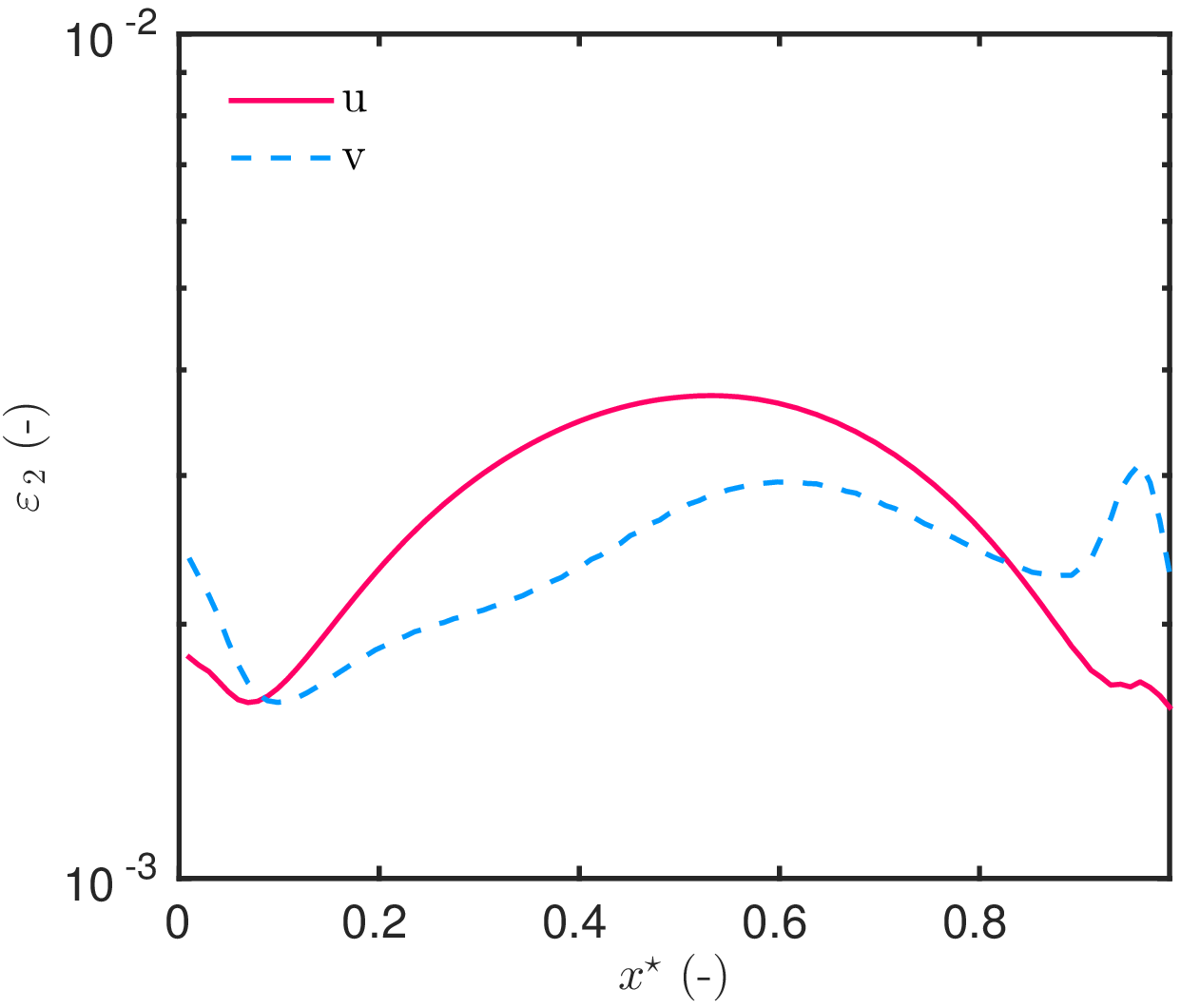}
  \caption{\emph{\small{$\ell_{\,2}$ error for the two fields $u\,(\,x\,,\,t\,)$ and $v\,(\,x\,,\,t\,)\,$.}}}
  \label{fig_AN5:L2_fx}
\end{figure}

The simulation final time is $t \egal 6 \,$ and the solution is computed with $\dx \egal 0.01$ along with an adaptive time step with both tolerances set to $10^{\,-5}\,$. A perfect agreement between the reference and \SG ~solutions can be seen in Figures~\ref{fig_AN5:profil_u}--\ref{fig_AN5:time_v}. The absolute error is lower than $4 \cdot 10^{\,-3}$ as shown in Figure~\ref{fig_AN5:L2_fx}, validating the scheme for this non-linear case.

\newpage
\section{Experimental comparison}
\label{sec:Exp_comp}

\subsection{Description of the case study}

As the advantages of the \SG ~scheme were highlighted in previous test cases, an important step in the validation of a physical model is its capacity to represent the physical phenomena. For this, results from the numerical model are compared with experimental data from \cite{James2010}, which enables to investigate both advective and diffusive effects on the moisture front. A gypsum board, of length $L \egal 37.5 \ \mathsf{mm}$ and initially conditioned at the relative humidity $\phi \egal 0.3 \,$, is submitted to an adsoprtion-desorption cycle (30--72--30) for $48 \ \mathsf{h}$. The temperature is maintained almost constant during the whole test at $T \egal 23.5 \ \mathsf{^\circ C}\,$. The constant surface transfer coefficient is equal to $\alpha_{\,q} \egal 3.45 \e{-8} \ \mathsf{W/m^{\,2}/K}$ and $\alpha_{\,m} \egal 2.41 \e{-8} \ \mathsf{s/m}\,$. The material properties are recalled in Figure~\ref{fig_Exp:mat_prop} and can be found in \cite{James2010}. The sorption moisture equilibrium curve with its hysteresis characteristic is reminded and illustrated in Figure~\ref{fig_Exp:sorption}.

\begin{figure}
  \centering
  \includegraphics[width=.69\textwidth]{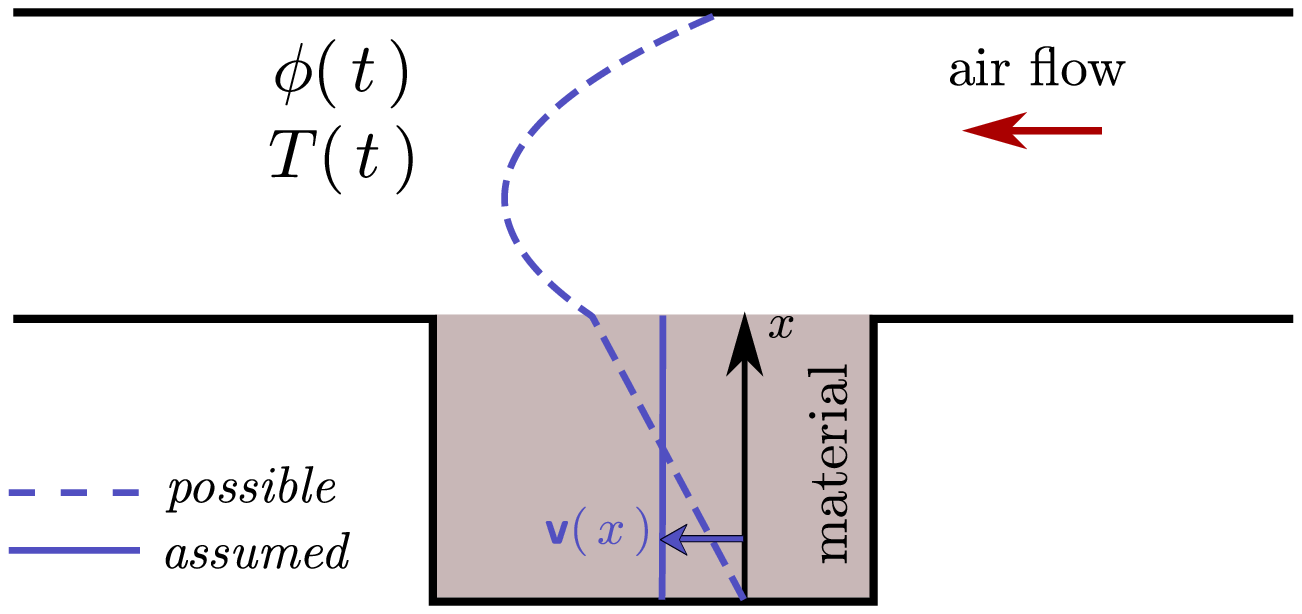}
  \caption{\emph{\small{Illustration experimental facility.}}}
  \label{fig_AN:schema_vit}
\end{figure}

\begin{figure}
  \centering
  \includegraphics[width=.55\textwidth]{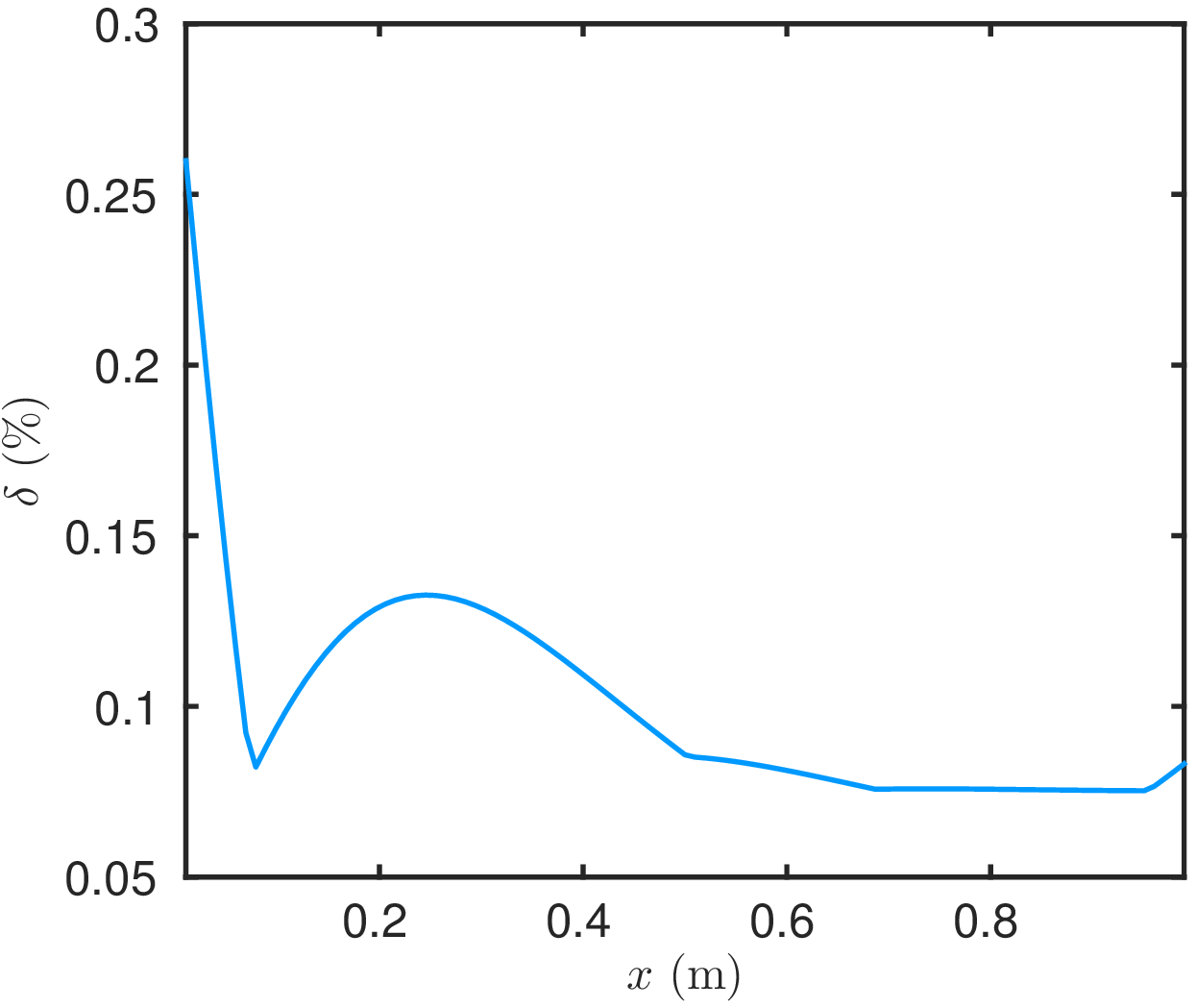}
  \caption{\emph{\small{Verification of the assumption made in the derivation of the equations.}}}
  \label{fig_AN:eps2H}
\end{figure}

The problem is solved with the \SG ~numerical scheme considering a large spatial discretisation parameter $\dx \egal 0.1\,$, an adaptive time step and both tolerances set to $10^{\,-3}\,$. Before analyzing carefully the numerical prediction and the experimental data, it is important to verify the hypothesis that was done in Section~\ref{sec:heat_transfer_sec}. In Eq.~\eqref{eq:heat_transport} , the term $\sum_{i\,=\,1}^{\,2} \grad \bigl(\, c_{\,i} \, T \,\bigr) \, \scal \, \bigl(\, \ja{\,,\,i} \plus \jd{\,,\,i} \,\bigr)$ has been neglected according to the suggestion of \textsc{Luikov} \cite[Chapter~6]{Luikov1966}. The sensitivity of this assumption is verified by evaluating the contribution of this term compared to the others:
\begin{align*}
  \delta \egal \max_{\,t} \, \left| \, \dfrac{\sum_{i\,=\,1}^{\,2} \grad \bigl(\, c_{\,i} \, T \,\bigr) \, \scal \, \bigl(\, \ja{\,,\,i} \plus \jd{\,,\,i} \,\bigr)}{\div \, \jq \plus r_{\,1\,2} \, \div \Bigl(\, \ja{\,,\,1} \plus \jd{\,,\,1} \,\Bigr) \plus \sum_{i\,=\,1}^{\,2} \grad \bigl(\, c_{\,i} \, T \,\bigr) \, \scal \, \bigl(\, \ja{\,,\,i} \plus \jd{\,,\,i} \,\bigr)} \, \right| \,.
\end{align*}
The variation of $\delta$ is given in Figure~\ref{fig_AN:eps2H}. It can be noted that this term contributes to the sum, at most, $0.25 \% \,$. This simplifying hypothesis is therefore acceptable.

\subsection{Results and discussion}

\begin{figure}
  \centering
  \subfigure[a][\label{fig_Exp:Px1DA_ft}]{\includegraphics[width=.48\textwidth]{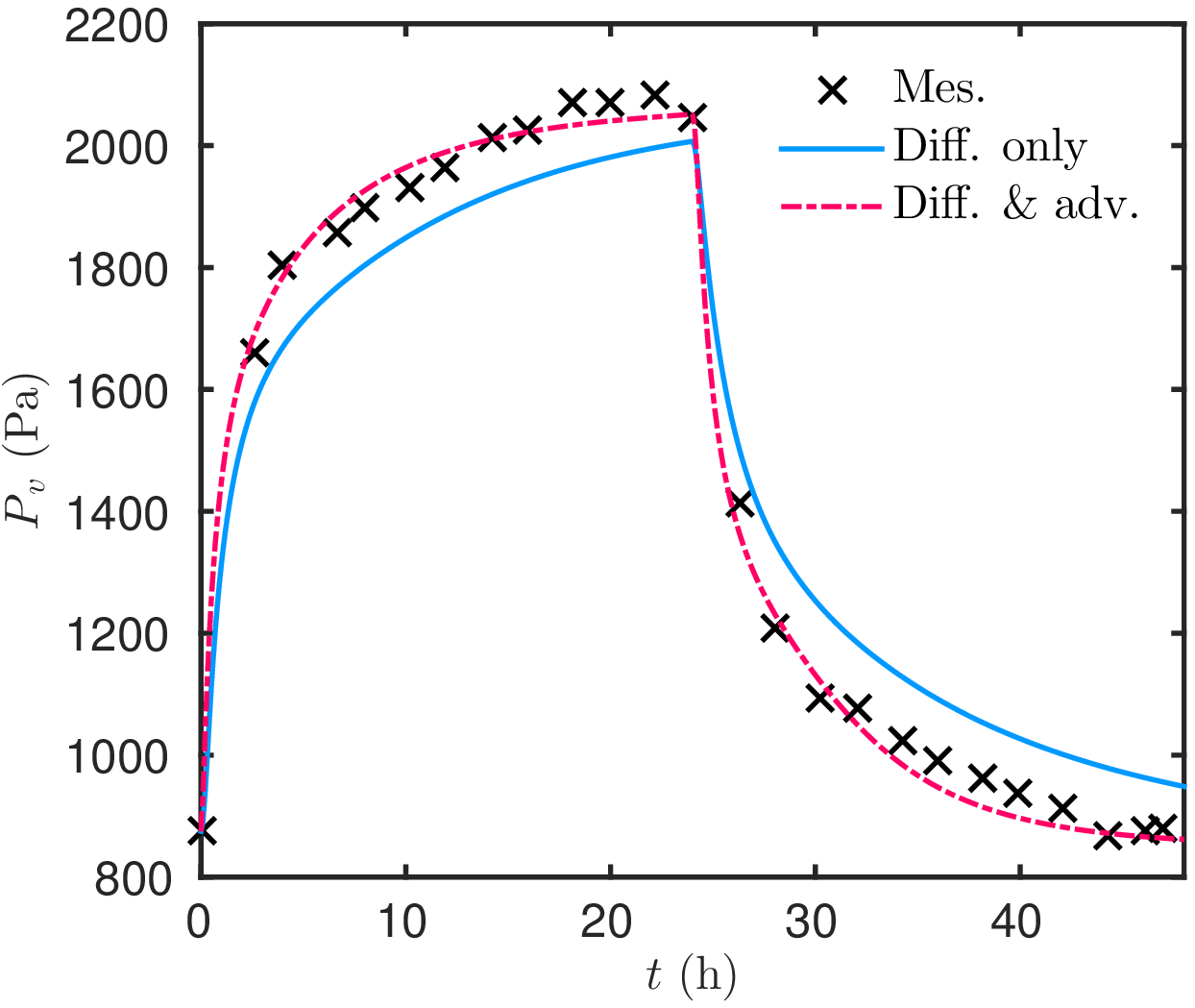}}
  \subfigure[b][\label{fig_Exp:Px2DA_ft}]{\includegraphics[width=.48\textwidth]{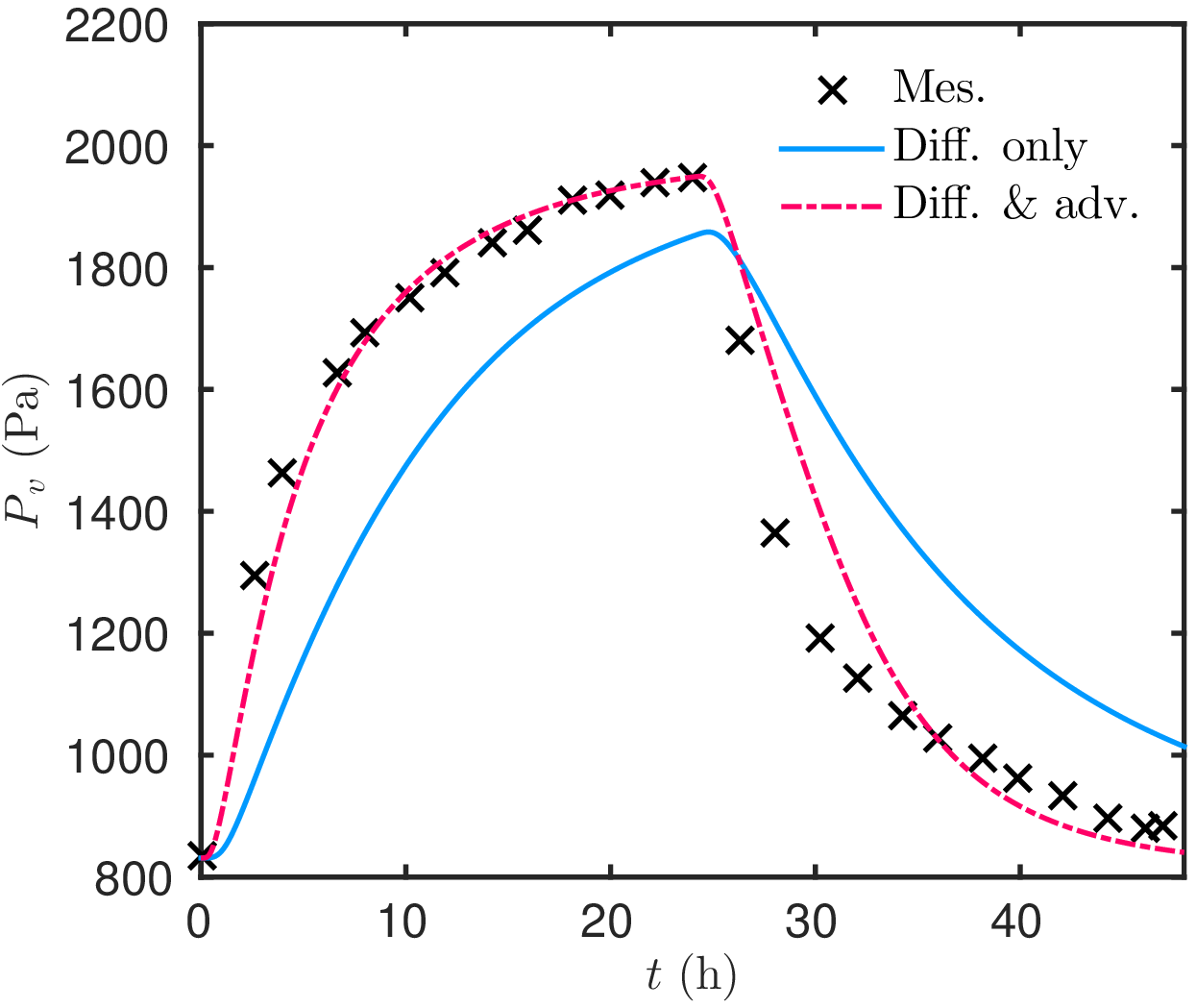}}
  \subfigure[c][\label{fig_Exp:Px1DAH_ft}]{\includegraphics[width=.48\textwidth]{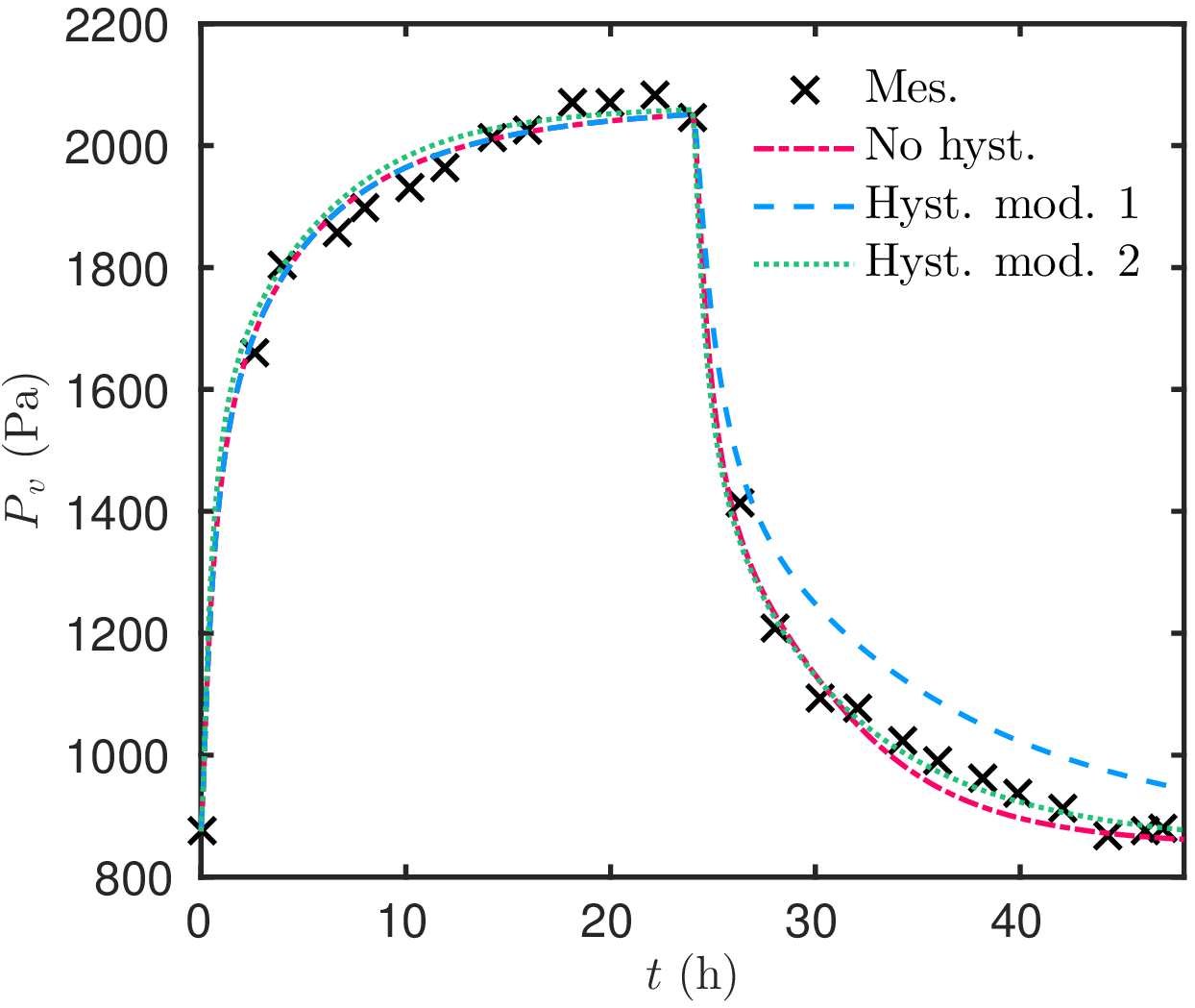}}
  \subfigure[d][\label{fig_Exp:Px2DAH_ft}]{\includegraphics[width=.48\textwidth]{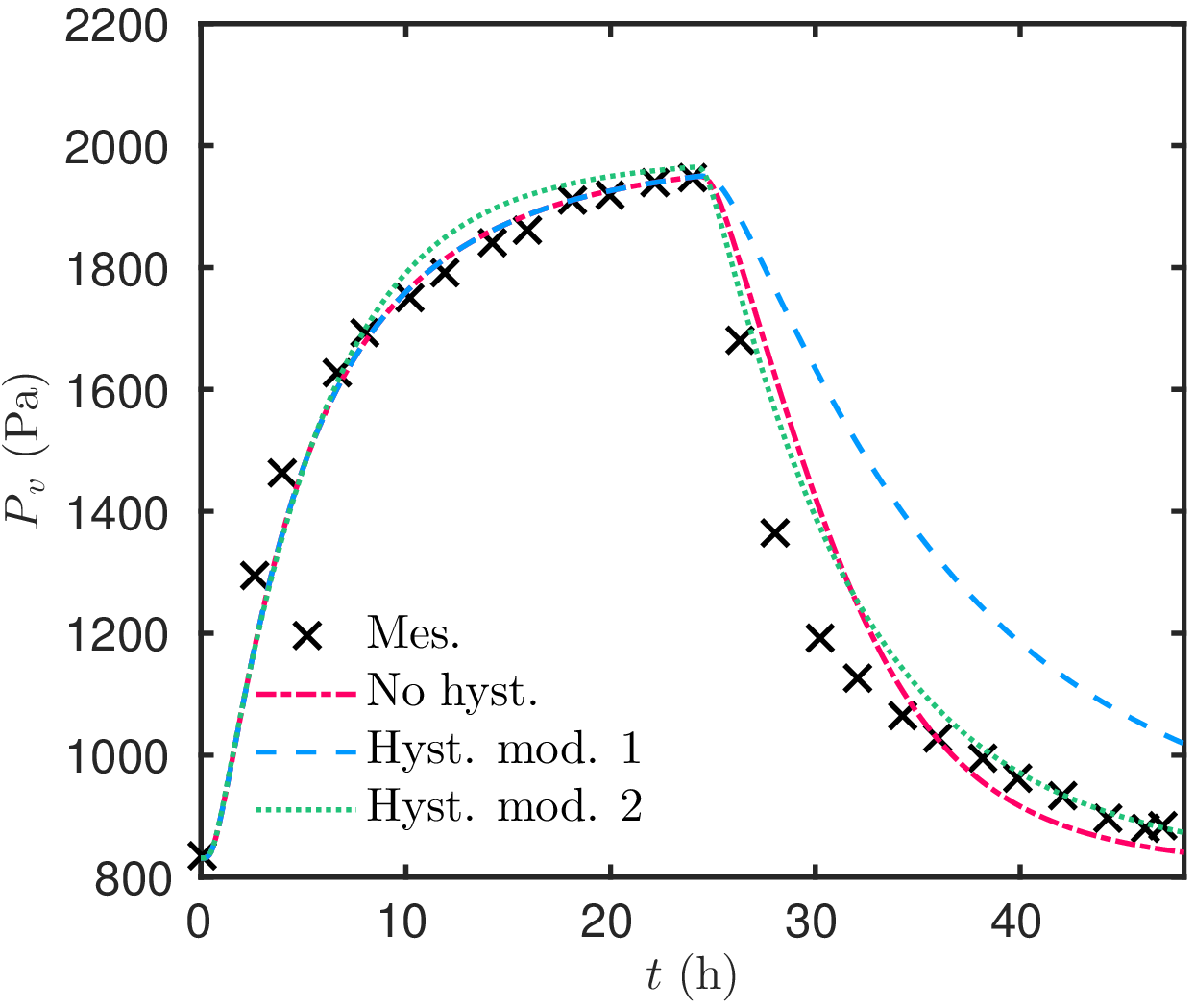}}
  \caption{\emph{\small{Evolution of the vapor pressure at $x \egal 12 \ \mathsf{mm}$ (a,c) and $x \egal 25 \ \mathsf{mm}$ (b,d).}}}
\end{figure}

\begin{figure}
  \centering
  \subfigure[a][\label{fig_Exp:Tx1DA_ft}]{\includegraphics[width=.48\textwidth]{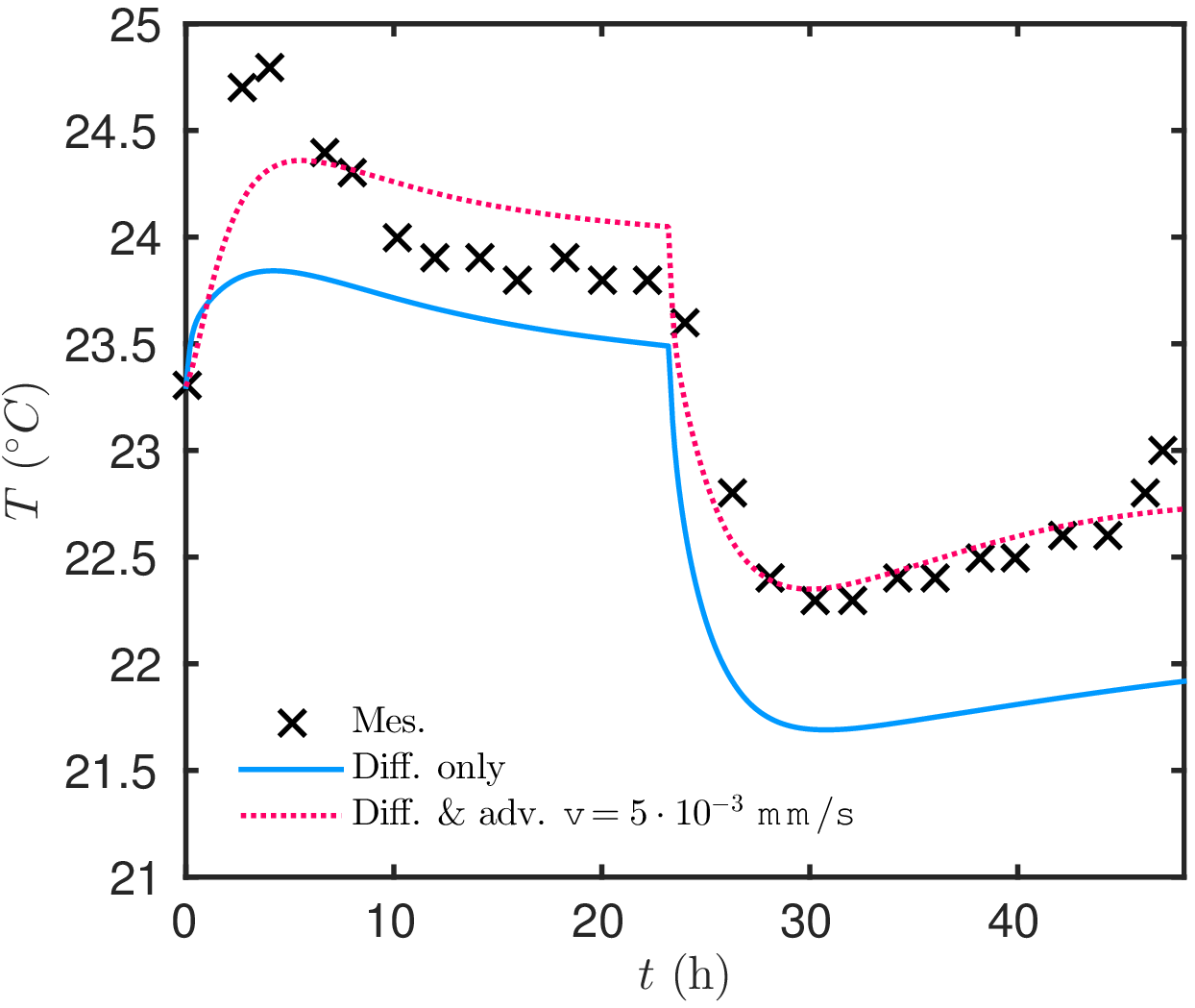}} 
  \subfigure[b][\label{fig_Exp:Tx2DA_ft}]{\includegraphics[width=.48\textwidth]{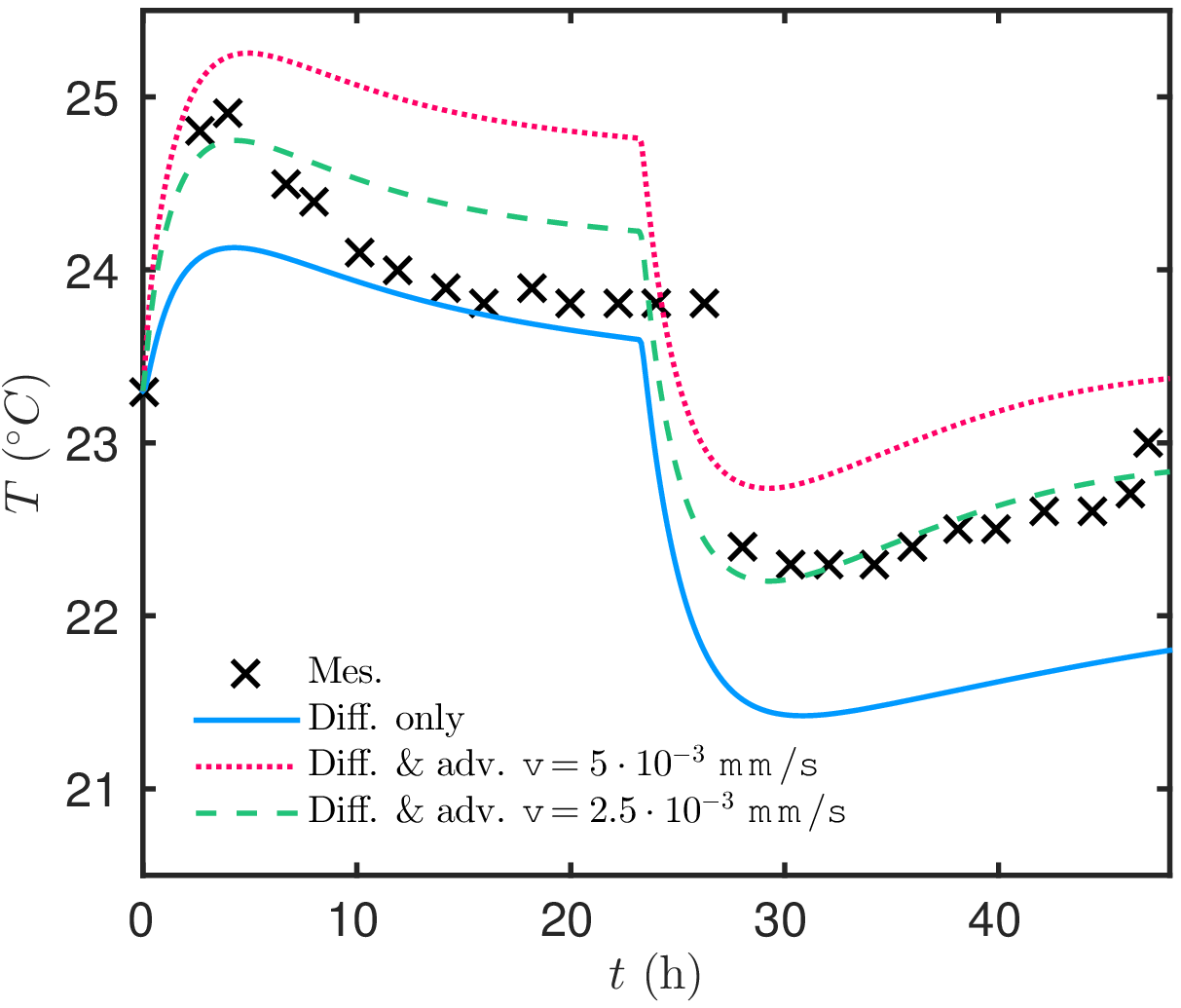}}
  \caption{\emph{\small{Evolution of the temperature at $x \egal 12 \ \mathsf{mm}$ (a) and $x \egal 25 \ \mathsf{mm}$ (b).}}}
\end{figure}

The purpose is now to compare the numerical predictions with the experimental data. The experimental data are given at $x \egal \bigl\{ 12.5 \,,\, 25 \,\bigr\} \ \mathsf{mm}\,$. The numerical solution is obtained at this point using the exact interpolation by Eq.~\eqref{eq:SP_solution_u}, also provided in the \textsc{Maple\;\texttrademark} supplementary file. The experimental facility is illustrated in Figure~\ref{fig_AN:schema_vit}. At the top of the material, an airflow is used to impose the temperature and relative humidity conditions. Due to this imposed airflow, it is supposed that there is an non-null velocity profile within the material. A probable profile of the velocity is shown in Figure~\ref{fig_AN:schema_vit}. However, the physical model does not take into account the momentum equation. Thus, the velocity is supposed to be constant and equal to its spatial average taken along the material height, as a first-order approximation. For each simulation, the velocity is estimated using an interior-point algorithm by minimizing the residual with the experimental data at each measurement point. Results are reported in Table~\ref{tab:res_vit}. Figures~\ref{fig_Exp:Px1DA_ft} and \ref{fig_Exp:Px2DA_ft} illustrate the variation of the vapor pressure at measurement points for a physical model considering only diffusion mechanism and another one taking into account both diffusion and advection phenomena. First, it can be noted that the model with only diffusion underestimates the adsorption phase and overestimates the desorption phase. By considering the advection transfer in the material, there is a better agreement between the experimental data and the numerical results. Similar conclusion can be drawn for the temperature evolution, shown in Figures~\ref{fig_Exp:Tx1DA_ft} and \ref{fig_Exp:Tx2DA_ft}. The model with diffusion and advection slightly overestimates the temperature at $x \egal 25 \ \mathsf{mm}\,$. Using the interior--point optimisation algorithm only for this measurement point, a lower velocity is estimated $\vi \egal 2.5 \e{-3} \ \mathsf{mm/s}\,$. As illustrated in Figure~\ref{fig_Exp:Tx2DA_ft}, the numerical results have a better agreement with the experimental measurements. This analysis illustrates that considering the mass average velocity as constant in space is a first-order approximation as discussed in \cite{Wang2000}. In addition, the velocity may also vary in time. For instance, at $t \egal 10 \ \mathsf{h}\,$, the numerical model overestimates vapor pressure, which might be explained by an overestimation of velocity. It should be remarked that considering this velocity, the \textsc{P\'eclet} number is of order $\O\,(\,10^{\,-2}\,)$ for moisture transport, validating the hypothesis neglecting the dispersion effects in the moisture transport.

\begin{figure}
  \centering
  \subfigure[a][\label{fig_Exp:km_fphi}]{\includegraphics[width=.50\textwidth]{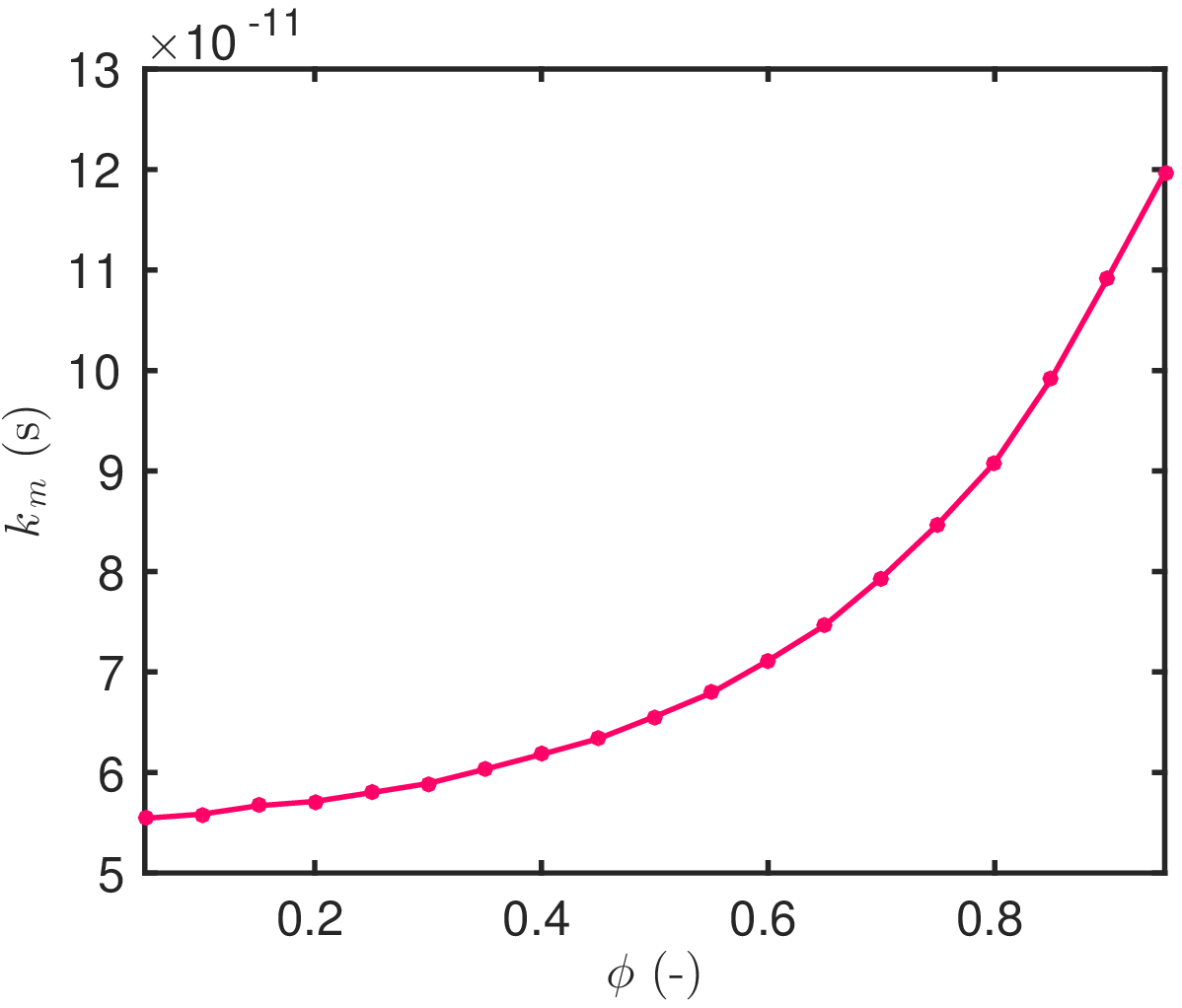}}
  \subfigure[b][\label{fig_Exp:sorption}]{\includegraphics[width=.48\textwidth]{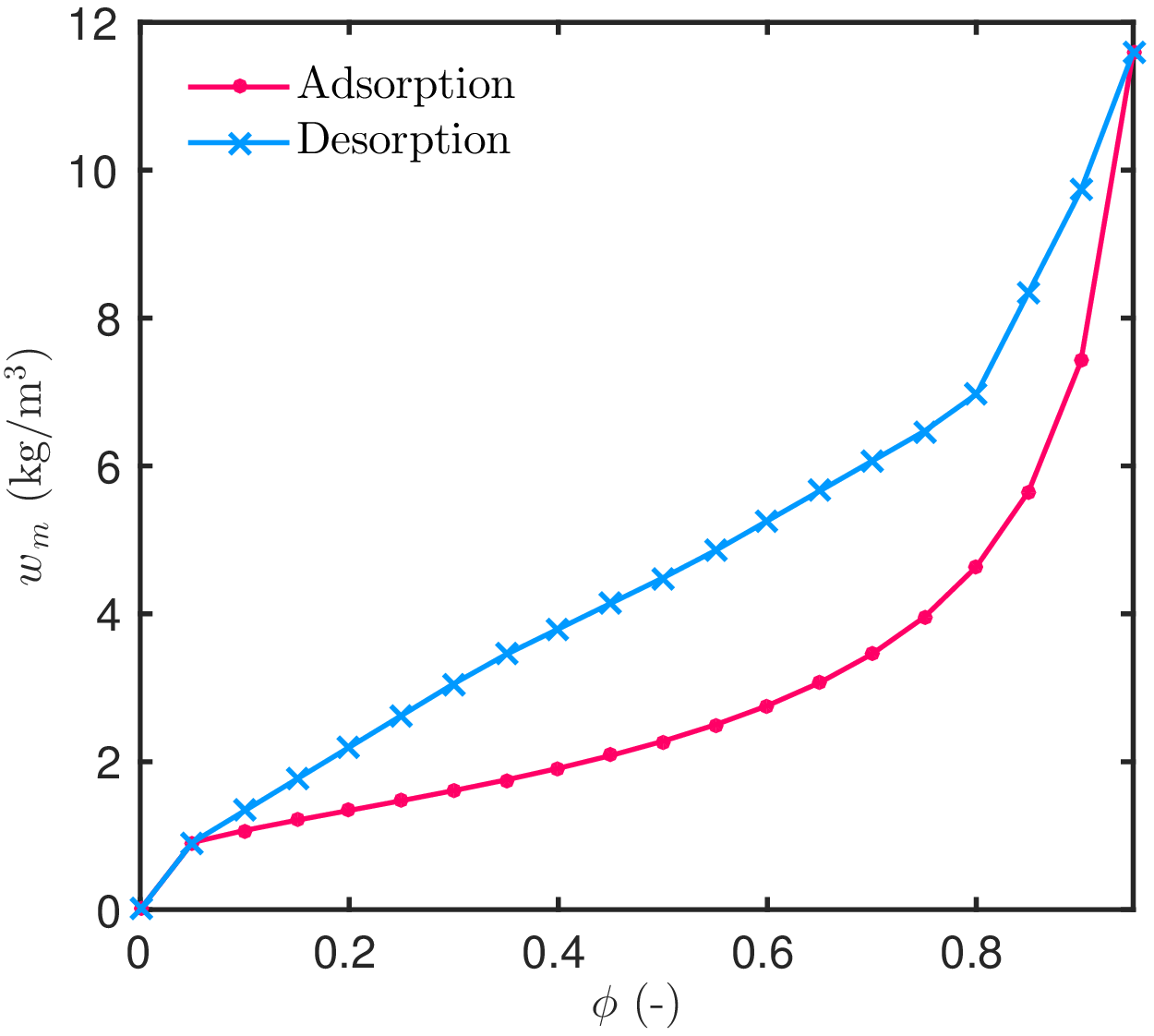}} 
  \caption{\emph{\small{Moisture permeability (a) and, adsorption and desorption curves of the moisture content for the gypsum board (b) (see \cite{James2010} for more details).}}}
  \label{fig_Exp:mat_prop}
\end{figure}

However, some discrepancies still remain for the model considering both diffusion and advection mechanisms, particularly for the measurement point $x \egal 25 \ \mathsf{mm} \,$, for $t \in \bigl[\, 28 \,,\, 48 \,\bigr]\,$. As mentioned in \cite{James2010}, these discrepancies may be due to the hysteresis effect on the moisture sorption curve. Therefore the physical model has been improved by considering the hysteresis effect on the coefficient $c_{\,m}\,$. The first approach considers only the adsorption and desorption curves illustrated in Figure~\ref{fig_Exp:sorption}. In control literature, it is referred as the \emph{bang--bang} model. The second verifies a differential equation that is solved at the same time as the coupled heat and moisture problem and that enables smoother transition between both curves. The computation of the coefficient $c_{\,m}$ for both approaches can be summarized:
\begin{align*}
& \text{Hysteresis model 1:} && c_{\,m} \egal
\begin{cases} 
  \ c_{\,m}^{\,\mathrm{ads.}}  \,, & \quad  \pd{\phi}{t} \ < \ 0 \,, \\
  \ c_{\,m}^{\,\mathrm{des.}}  \,, & \quad  \pd{\phi}{t} \ \geqslant \ 0 \,.
\end{cases} \\[3pt]
& \text{Hysteresis model 2:} && 
\pd{c_{\,m}}{t} \egal \beta \cdot 
\mathrm{sign} \, \biggl(\, \pd{\phi}{t} \,\biggr)
\cdot \biggl(\, c_{\,m} \moins c_{\,m}^{\,\mathrm{ads.}} \,\biggr) \, 
\biggl(\, c_{\,m} \moins c_{\,m}^{\,\mathrm{des.}} \,\biggr) \,, \\
& 
&& \mathrm{sign} \, \bigl(\,X\,\bigr) \egal 
\begin{cases} 
  \ 1 \,, & \quad  X \ > \ 0 \,, \\
  \ 0 \,, & \quad  X \egal 0 \,, \\
  \ -\,1 \,, & \quad  X \ < \ 0 \,.
\end{cases}
\end{align*}
where $c_{\,m}^{\,\mathrm{ads.}}$ and $c_{\,m}^{\,\mathrm{des.}}$ are respectively the adsorption and desorption curves. These curves depend on the relative humidity $\phi$ and are experimentally determined. Analytical functions of the experimental curves provided in \cite{James2010, James2009} are fitted. The coefficient $\beta$ is a numerical parameter which controls the transition velocity between the two curves.

\begin{figure}
  \centering
  \includegraphics[width=.56\textwidth]{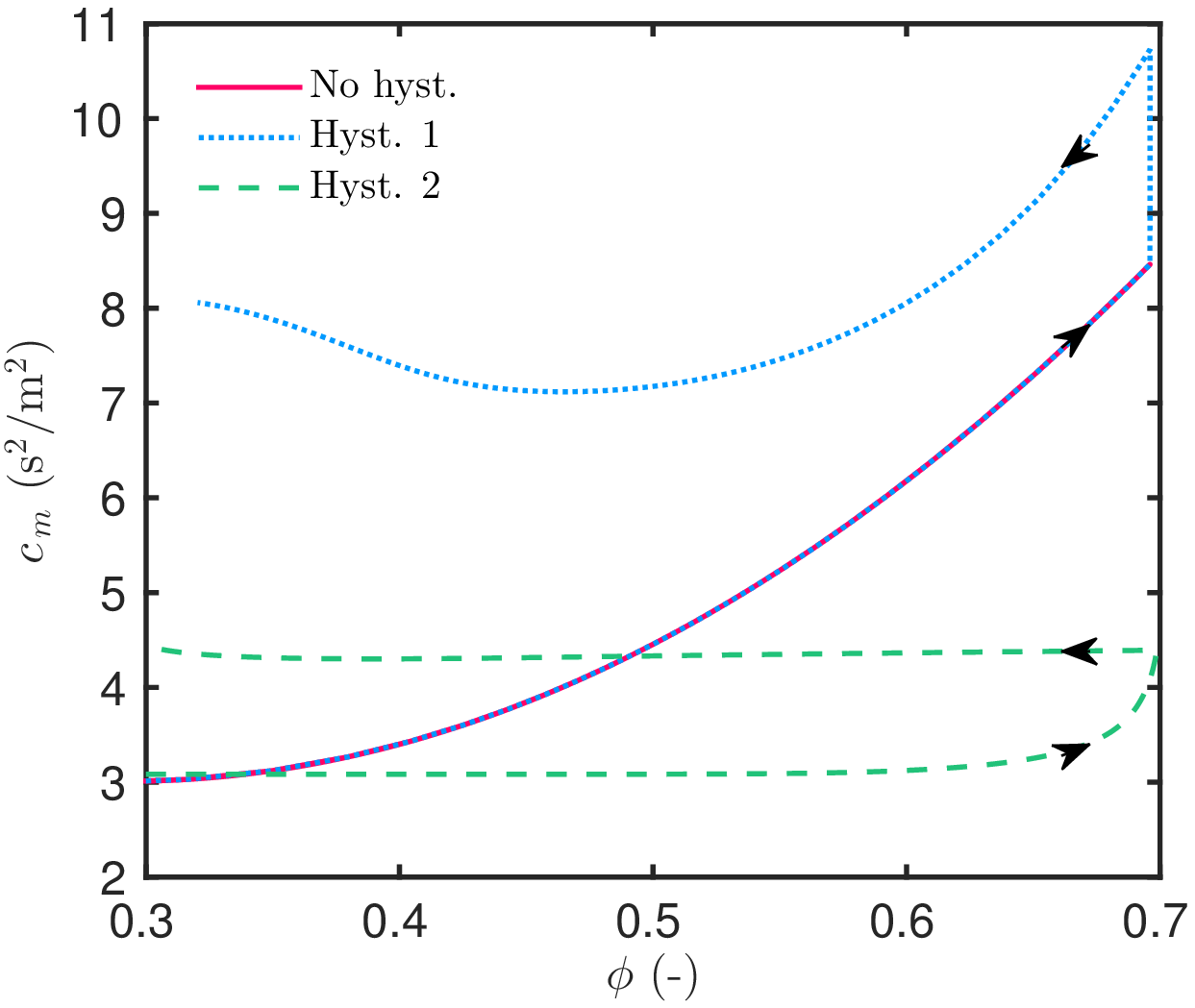}
  \caption{\emph{\small{Evolution of the moisture capacity parameter $c_{\,m}$ for different hysteresis models.}}}
  \label{fig_Exp:cM_fphi}
\end{figure}

The results of the implementation of two hysteresis models are illustrated in Figures~\ref{fig_Exp:Px1DAH_ft} and \ref{fig_Exp:Px2DAH_ft}. The first hysteresis model is not able to reduce the discrepancies. Indeed, the approach considering only the adsorption and desorption curves is too minimalist. The second hysteresis model provides a better agreement, particularly at $x \egal 25 \ \mathsf{mm}\,$. Figure~\ref{fig_Exp:cM_fphi} shows the variation of the coefficients $c_{\,m}$ that have been plotted as a function of the computed relative humidity. For the model without hysteresis, the coefficient varies along only one curve. The hysteresis model $1$ switches between the adsoption and desorption curves without any interpolation and without ensuring the continuity of the physical characteristic. Since, the coefficient $c_{\,m}$ is proportional to the derivative $\pd{w}{\phi}\,$, a discontinuity in the variation of the coefficient is observed at $\phi \egal 0.7\,$. Moreover, the magnitude of the coefficient $c_{\,m}$ in the model $1$ is higher than for the other models, which explains the higher values of the vapor pressure shown in Figures~\ref{fig_Exp:Px1DAH_ft} and \ref{fig_Exp:Px2DAH_ft} for $t\ \in\ \bigl[\, 28 \,,\, 48 \,\bigr]\,$. Oppositely, the variation of the coefficient is continuous for the second hysteresis model, while the derivative is discontinuous. For a numerical parameter $\beta \egal 0.02\,$, the numerical results have a satisfying agreement with the experimental data. The estimated velocity equals to $\vi \egal 4.2 \e{-3} \ \mathsf{mm/s}\,$. The hysteresis effect does not show an important impact on the temperature residual as noticed in Table~\ref{tab:res_vit}.

\subsection{Local sensitivity analysis}

\begin{figure}
  \centering
  \subfigure[a][\label{fig_Exp:Theta_x1}]{\includegraphics[width=.48\textwidth]{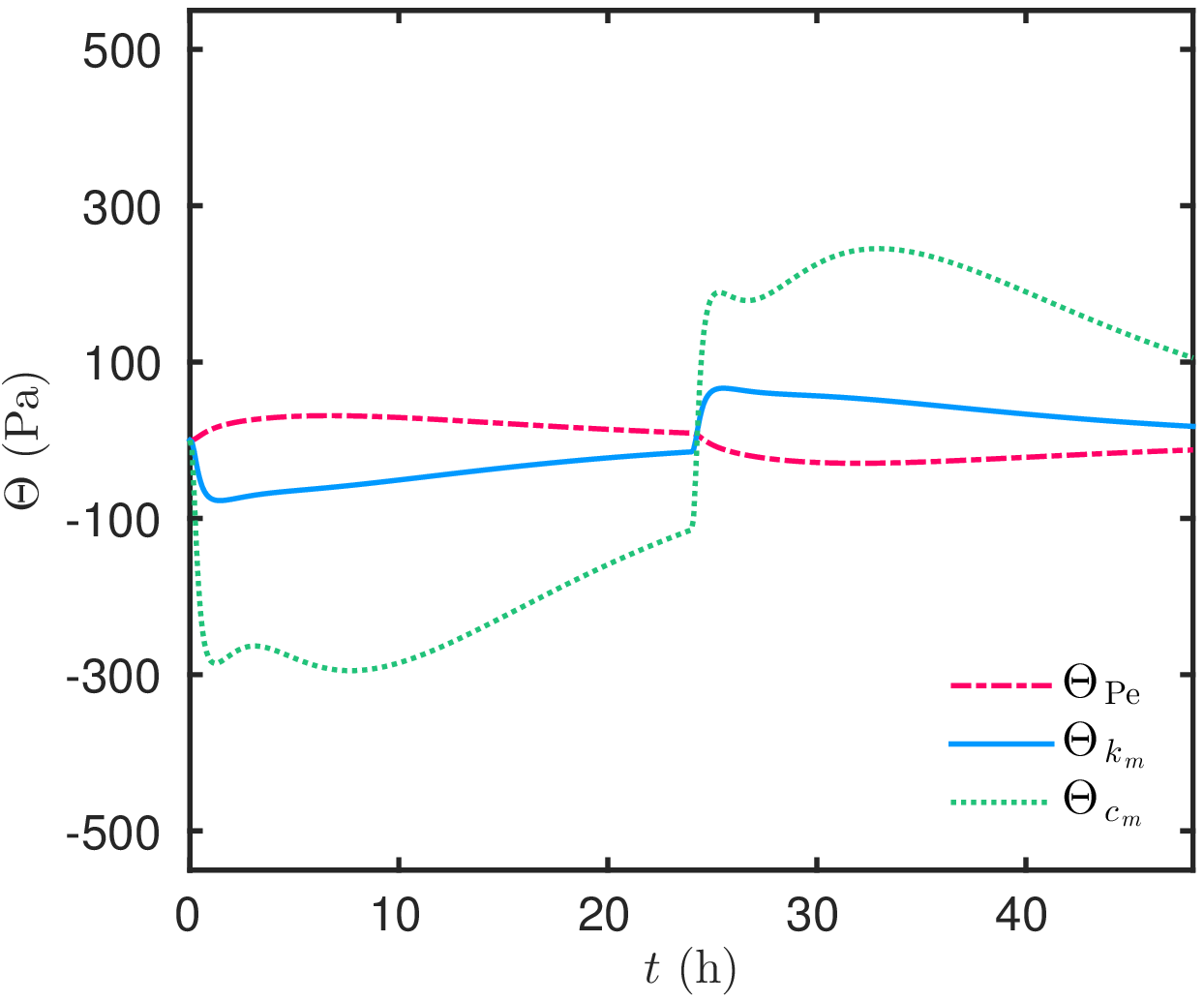}}
  \subfigure[b][\label{fig_Exp:Theta_x2}]{\includegraphics[width=.48\textwidth]{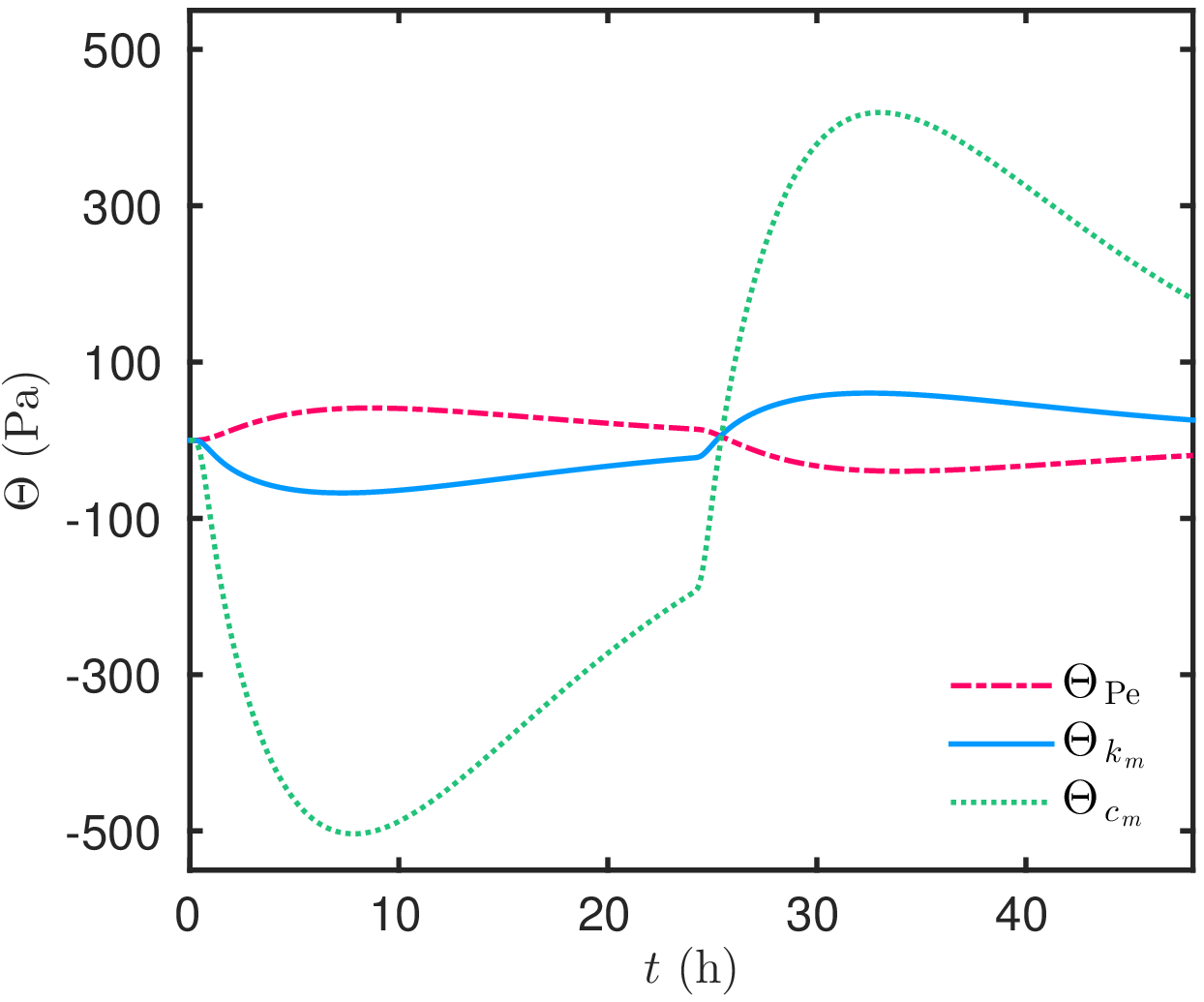}}
  \caption{\emph{\small{Sensitivity coefficients of parameters $k_{\,m} \,$, $c_{\,m}$ and $\mathrm{Pe}_{\,m}$ at $x \egal 12.5$ $\mathsf{mm}$ (a) and $x \egal 25$ $\mathsf{mm}$ (b).}}}
  \label{fig_Exp:Theta}
\end{figure}

To compare the relative importance of each mechanism among moisture advection, diffusion and storage, a brief and local sensitivity analysis is carried out by computing the sensitivity functions $\Theta\,$:
\begin{align*}
  & \Theta_{\,k_{\,m}} \egal k_{\,m} \;\pd{\Pv}{k_{\,m}} \,, 
  && \Theta_{\,c_{\,m}} \egal c_{\,m} \;\pd{\Pv}{c_{\,m}} \,, 
  && \Theta_{\,\mathrm{Pe}_{\,m}} \egal \mathrm{Pe}_{\,m} \;\pd{\Pv}{\mathrm{Pe}_{\,m}} \,.
\end{align*}
The sensitivity function evaluates, as its name clearly indicates, the local sensitivity of the numerically computed vapor pressure field with respect to a change in the parameter. A small magnitude value of $\Theta$ indicates that large changes in the parameter yield to small changes in the field. Here, it has been computed for the first order of material properties. Figures~\ref{fig_Exp:Theta_x1} and \ref{fig_Exp:Theta_x2} show the time evolution of each sensitivity function. For the diffusion and advection parameters, the sensitivity increases during the transient regimes of the simulation and then decreases as the simulation reaches the steady state. It can be noted that both mechanisms have the same order of magnitude of sensitivity. Contrarily, the sensitivity to the moisture capacity parameter $c_{\,m}$ has higher variations. Moreover, the magnitude is higher for the measurement point $x \egal 25 \ \mathsf{mm}\,$. It indicates that the moisture capacity has higher impact on the vapor pressure. It is related to the fact that the simulation performed with the different hysteresis models have more impact on the measurement at this point, as noticed in Figure~\ref{fig_Exp:Px2DAH_ft}. This local sensitivity analysis highlights the importance of each mechanism among the advection  and diffusion transfer, and the moisture storage, for this material and for the range of temperatures and relative humidities used in the experiments.

\begin{table}
  \centering
  \begin{tabular}{@{}lccc}
  \hline\hline
  \textit{Model} & \multicolumn{2}{c}{\textit{Residual} $\varepsilon_{\,2}$} & \textit{Estimated velocity} \\
   & \textit{Vapor pressure} & \textit{Temperature} & $\mathsf{(mm/s)}$ \\
   & $\mathsf{(Pa})$ & $\mathsf{(^{\,\circ}C})$ & \\
  \hline
  \textit{Diffusion} & $0.19$ & $3.4 \e{-3}$ & -\\
  \textit{Diffusion and advection, no hysteresis} & $0.065$ & $1.5 \e{-3}$ & $5 \e{-3}$\\
  \textit{Diffusion and advection, hysteresis model} $1$ & $0.13$ & $1.6 \e{-3}$ & $4 \e{-3}$\\
  \textit{Diffusion and advection, hysteresis model} $2$ & $0.039$ & $1.4 \e{-3}$ & $4.2 \e{-3}$ \\
  \hline\hline
  \end{tabular}
  \bigskip
  \caption{\emph{\small{Residual with experimental data and estimated velocity.}}}
  \label{tab:res_vit}
\end{table}


\section{Conclusion}
\label{sec:conclusion}

When comparing measurements to numerical simulations of moisture transfer through porous materials, discrepancies have been reported in several works from the literature \cite{Busser2018, Berger2017a}. Indeed, the numerical model is built considering only diffusion transfer through porous materials as physical phenomenon. As a result, the simulation underestimates the adsorption process or overestimates the desorption process. One possible explanation is the absence of advection transfer in the governing equations. Therefore, this paper investigated the influence of the advection and diffusion transfer in a heat and moisture coupled model.

To solve efficiently the coupled advection-diffusion differential equations, an innovative numerical scheme, the so-called \SG, has been considered. This scheme has been proposed in 1969, for the first time for data analysis problems, and it is still studied theoretically (with the latest theoretical results from 2016). It has the advantages of being well-balanced and asymptotically preserving. In addition, the interpolation of the solution on any spatial point is given by an exact expression. The numerical efficiency has been first analysed for non-linear cases of a single scalar differential equation. Its accuracy has been validated with two analytical solutions and with a reference solution computed using the \texttt{Chebfun} package. The extension of the scheme for a system of weakly coupled differential equations has been proposed. Consequently, the numerical scheme and its implementation has been validated with a \texttt{Chebfun} reference solution for both linear and non-linear cases. A parametric study of the discretisation parameters $\dt$ and $\dx$ has also been carried out. As expected, the \SG ~scheme has a CFL stability condition. Nevertheless, the approach is particularly interesting when using large spatial discretisation and an adaptive time step to enable important computational savings without losing the accuracy of the solution.

In Section~\ref{sec:Exp_comp}, the numerical results have been compared to experimental data from \cite{James2010}. An adsorption--desorption cycle is performed for a gypsum board material. The temperature and vapor pressure profiles within the material are provided. Comparative results between a purely diffusive and the improved mathematical models have been presented. The purely diffusive model underestimates the sorption phase and overestimates the desorption phase. With the improved advective--diffusive model, there is a better agreement between the numerical results and the experimental data. The momentum equation has not been taken into account in the physical model. Thus, a constant mass average velocity within the material porous structure has been estimated. Despite the inclusion of the advection transfer mechanism provides a better agreement with the experimental data, some of discrepancies still remain, particularly at the end of the desorption cycle, which might be due to the presence of hysteresis effects in the moisture capacity of the material. Thus, the model has been improved by adding also a third differential equation on the moisture capacity, enabling to interpolate between the adsorption and desorption equilibrium curves. This hysteretic diffusive--advective model provided the best results with a residual lower than $0.04$ for the vapor pressure and $1.4 \e{-3}$ for the temperature.

The estimated velocity has been discussed highlighting that the velocity may decrease with space and time. A constant velocity hypothesis remains as a first-order approximation. Further research is needed to include the momentum equation in the physical model to have a better calculation of the mass average velocity and hopefully provide better results in the comparison with the experimental data.


\bigskip

\subsection*{Acknowledgments}
\addcontentsline{toc}{subsection}{Acknowledgments}

The authors acknowledge the \textsc{Brazilian} Agencies CAPES of the Ministry of Education, the CNPQ of the Ministry of Science, Technology and Innovation, for the financial support for the project CAPES--COFECUB Ref.~774/2013. The authors also would like to acknowledge Dr. J.~\textsc{Garnier} (LAMA, UMR 5127 CNRS, Universit\'e \textsc{Savoie Mont Blanc}) for his precious discussions on numerical matters for the hysteresis phenomena, Dr. B.~\textsc{Rysbaiuly} for his valued discussions on writing the physical problem and its mathematical formulation, Dr. A.~\textsc{Agbossou} and Dr. M.~\textsc{Cugnet} for their help with the \COMSOL ~simulations, and Dr. L.~\textsc{Gosse} for his appreciated discussions on the \SG ~numerical scheme.

\bigskip


\section*{Nomenclature}

\begin{tabular*}{0.7\textwidth}{@{\extracolsep{\fill}} |@{} >{\scriptsize} c >{\scriptsize} l >{\scriptsize} l| }
  \hline
  \multicolumn{3}{|c|}{\emph{Latin letters}} \\
  $a_{\,m}$ & moisture advection coefficient & $[\mathsf{s/m}]$ \\
  $a_{\,q}$ & heat advection coefficient & $[\mathsf{J/(K.m^{\,2}.s)}]$ \\
  $a_{\,qm}$ & heat advection coeff. under vap. press. grad. & $[\mathsf{W.s^{\,2}/(kg.m)}]$ \\
  $b$ & volume ratio & $[\mathsf{-}]$ \\
  $c_{\,m}$ & moisture storage capacity & $[\mathsf{kg/(m^3.Pa)}]$ \\
  $c_{\,q}$ & volumetric heat capacity & $[\mathsf{J/(m^{\,3}.K)}]$ \\
  $c$ & specific heat & $[\mathsf{J/(kg.K)}]$ \\
  $g$ & liquid flux by rain &  $[\mathsf{kg/(s.m^{\,2})}]$ \\
  $h{}$ & specific enthalpy  & $[\mathsf{J/kg}]$ \\
  $I$ & volumetric capacity of source/sink & $[\mathsf{kg/(m^{\,3}.s)}]$\\
  $\ja{}$ & moisture flux by advection &  $[\mathsf{kg/(s.m^{\,2})}]$ \\
  $\jd{}$ & moisture flux by diffusion & $[\mathsf{kg/(s.m^{\,2})}]$ \\
  $\jq{}$ & heat flux & $[\mathsf{W/m^{\,2}}]$ \\
  $L$ & length & $[\unit{m}]$ \\
  $k$ & vapor or liquid permeability & $[\mathsf{s}]$ \\
  $k_{\,m}$ & moisture permeability & $[\mathsf{s}]$ \\
  $k_{\,q}$ & thermal conductivity & $[\mathsf{W/(m.K)}]$ \\
  $k_{\,qm}$ & heat transf. coeff. under vap. press. grad. & $[\mathsf{W.s^{\,2}/kg}]$ \\
  $\Pc$ & capillary pressure & $[\mathsf{Pa}]$ \\
  $\Ps$ & saturation pressure & $[\mathsf{Pa}]$ \\
  $\Pv$ & vapor pressure & $[\mathsf{Pa}]$ \\
  $r_{\,12}$ & latent heat of evaporation & $[\mathsf{J/kg}]$ \\
  $\Rv$ & water gas constant & $[\mathsf{J/(kg.K)}]$\\
  $T$ & temperature & $[\unit{K}]$ \\
  $t$ & time coordinate & $[\mathsf{s}]$ \\
  $x$ & space coordinate & $[\mathsf{m}]$ \\
  $\vi$ & mass average velocity & $[\mathsf{m/s}]$ \\
  $V$ & volume & $[\mathsf{m^{\,3}}]$ \\
  $w$ & volumetric concentration & $[\mathsf{kg/m^{\,3}}]$ \\
  \hline
\end{tabular*}

\begin{tabular*}{0.7\textwidth}{@{\extracolsep{\fill}} |@{} >{\scriptsize} c >{\scriptsize} l >{\scriptsize} l| }
  \hline
  \multicolumn{3}{|c|}{\emph{Greek letters}} \\
  $\alpha_{\,m}$ & convective vapour transfer coefficient & $[\mathsf{s/m}]$ \\
  $\alpha_{\,q}$ & convective heat transfer coefficient & $[\mathsf{W/(m^2.K)}]$ \\
  $\phi$ & relative humidity & $[-]$ \\
  $\rho$ & specific mass & $[\mathsf{kg/m^3}]$ \\
  \hline
\end{tabular*}


\appendix
\section{Dispersion Relation}
\label{sec:annex_disp_relation}

We consider the linear advection--diffusion Equation~\eqref{eq:conv_diff} written as: 
\begin{align}\label{annx_eq:diff_adv}
  \pd{u}{t} \plus a \;\pd{u}{x} \moins d \;\pd{^{\,2} u}{x^{\,2}} \egal 0 \,.
\end{align}
It admits plane wave solutions \cite{Trefethen1996}:
\begin{align}\label{annx_eq:wave_sol}
  u \,(\,x\,,\,t\,) \egal \expo{\, \ii \, \bigl(\, k \, x \moins \omega \, t \,\bigr)} \,,
\end{align}
where $\omega$ is the frequency and $k$ the wave number. Inserting solution $u$ from Eq.~\eqref{annx_eq:wave_sol} into Eq.~\eqref{annx_eq:wave_sol} yields to the dispersion relation: 
\begin{align*}
  \omega \egal a \, k \moins \ii \, d \, k^{\,2} \,.
\end{align*}

The semi-discrete formulation using central differences approach applied to Eq.~\eqref{annx_eq:diff_adv} gives:
\begin{align}\label{annx_eq:diff_adv_FD}
  \dfrac{\mathrm{d} u_{\,j}}{\mathrm{d} t} \plus \dfrac{a}{\dx}\;\biggl(\, u_{\,j} \moins u_{\,j\,-\,1} \,\biggr) \moins \dfrac{d}{\dx^{\,2}}\;\biggl(\, u_{\,j\,+\,1} \moins 2 \, u_{\,j} \plus u_{\,j\,-\,1} \,\biggr) \egal 0 \,.
\end{align}
The solution is assumed as:
\begin{align}\label{annx_eq:sol}
  u_{\,j} \,(\,t\,) \egal \expo{-\ii \, \omega \, t } \, \expo{\ii \, j \, k \, \dx } \,.
\end{align}
Inserting Eq.~\eqref{annx_eq:sol} in Eq.~\eqref{annx_eq:diff_adv_FD} gives: 
\begin{align*}
  - \, \ii \, \omega \plus \dfrac{a}{\dx}\;\biggl(\, 1 \moins \expo{-\,\ii\,k\,\dx} \,\biggr) \moins \dfrac{d}{\dx^{\,2}}\;\biggl(\, \expo{\ii\,k\,\dx} \moins 2 \plus \expo{-\,\ii\,k\,\dx} \,\biggr) \egal 0 \,,
\end{align*}
which can be rewritten as:
\begin{align}\label{annx_eq:omega_FD}
  \omega \egal - \, \ii \, \dfrac{a}{\dx}\;\biggl(\, 1 \moins \expo{-\,\ii\,k\,\dx} \,\biggr) \plus \ii \, \dfrac{d}{\dx^{\,2}}\;\biggl(\, \expo{\ii\,k\,\dx} \moins 2 \plus \expo{-\,\ii\,k\,\dx} \,\biggr) \,.
\end{align}
From Eq.~\eqref{annx_eq:omega_FD} we obtain:
\begin{subequations}\label{annx_eq:DL_DF}
\begin{align}
  \lim_{\,\dx \, \rightarrow \, 0} \mathrm{Re} \Biggl(\, \dfrac{\omega}{a \, k} \,\Biggr) & \egal 1 \moins \dfrac{1}{6} \; k^{\,2} \, \dx^{\,2} \plus \O(\dx^{\,4}) \,, \\ 
  \lim_{\,\dx \, \rightarrow \, 0} \mathrm{Im} \Biggl(\, \dfrac{- \,\omega}{d \, k^{\,2}} \,\Biggr) & \egal 1 \plus \dfrac{1}{2} \; \dfrac{a}{d} \, \dx \moins \dfrac{1}{12}\; k^{\,2} \, \dx^{\,2} \plus \O(\dx^{\,4}) \,.
\end{align}
\end{subequations}

Substituting solution~\eqref{annx_eq:sol} to the semi-discrete formulation of the \SG ~approach applied to Eq.~\eqref{annx_eq:diff_adv} gives:
\begin{align*}
  - \, \ii \, \omega \plus \dfrac{d}{\dx^{\,2}}\;\Biggl(\,- \, \mathrm{B} \biggl(\, \dfrac{a \, \dx}{d} \,\biggr) \, \expo{\ii\,k\,\dx} \plus \mathrm{B} \biggl(\, - \, \dfrac{a \, \dx}{d} \,\biggr) \plus \mathrm{B} \biggl(\, \dfrac{a \, \dx}{d} \,\biggr) \moins \mathrm{B} \biggl(\, - \, \dfrac{a \, \dx}{d} \,\biggr) \, \expo{-\,\ii\,k\,\dx}\,\Biggr) \egal 0 \,,
\end{align*}
which can be rewritten as: 
\begin{align}\label{annx_eq:omega_SG}
  \omega \egal - \, \ii\;\dfrac{d}{\dx^{\,2}} \Biggl(\,- \, \mathrm{B} \biggl(\, \dfrac{a \, \dx}{d} \,\biggr) \, \expo{\ii\,k\,\dx} \plus \mathrm{B} \biggl(\, - \, \dfrac{a \, \dx}{d} \,\biggr) \plus \mathrm{B} \biggl(\, \dfrac{a \, \dx}{d} \,\biggr) \moins \mathrm{B} \biggl(\, - \, \dfrac{a \, \dx}{d} \,\biggr) \, \expo{-\,\ii\,k\,\dx} \,\Biggr) \,.
\end{align}
Using \eqref{annx_eq:omega_SG}, it appears that: 
\begin{subequations}\label{annx_eq:DL_SG}
\begin{align}
  \lim_{\,\dx \, \rightarrow \, 0} \mathrm{Re} \Biggl(\, \dfrac{\omega}{a \, k} \,\Biggr) & \egal 1 \moins \dfrac{1}{6} \; k^{\,2} \, \dx^{\,2} \plus \O(\dx^{\,4}) \,, \\
  \lim_{\,\dx \, \rightarrow \, 0} \mathrm{Im} \Biggl(\, \dfrac{- \,\omega}{d \, k^{\,2}} \,\Biggr) & \egal 1 \plus \dfrac{1}{12} \; \bigl(\, \dfrac{a^{\,2}}{d^{\,2}} \moins k^{\,2} \,\bigr) \, \dx^{\,2} \plus \O(\dx^{\,4}) \,.
\end{align}
\end{subequations}

\begin{figure}
  \centering
  \subfigure[a][\label{fig_Annx:Re_disp}]{\includegraphics[width=.48\textwidth]{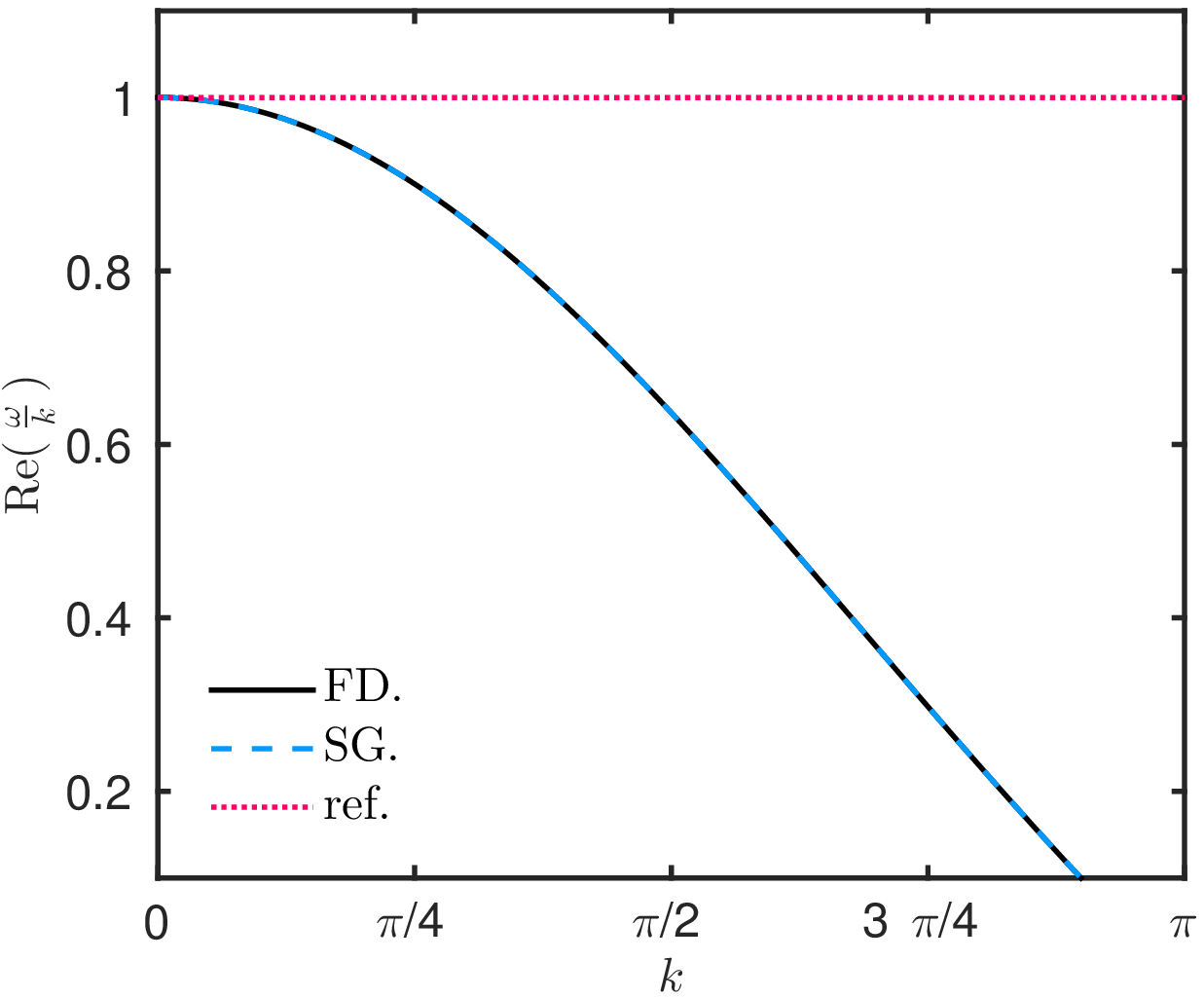}} 
  \subfigure[b][\label{fig_Annx:Im_disp}]{\includegraphics[width=.48\textwidth]{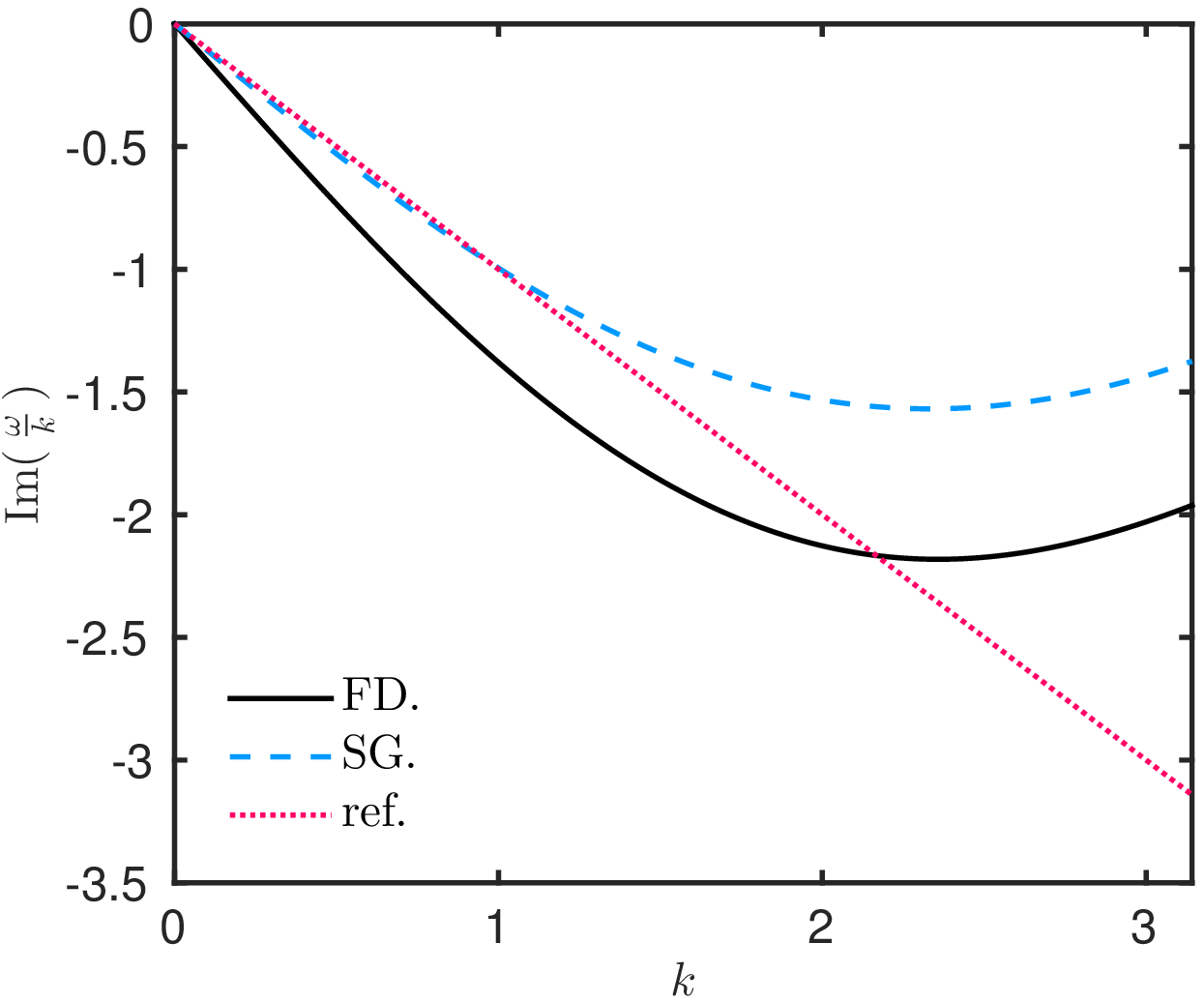}}
  \caption{\emph{\small{Dispersion relation for \Eu ~and \SG ~numerical schemes ($a \egal d \egal \dx \egal 1$).}}}
\end{figure}

By comparing Eqs.~\eqref{annx_eq:DL_DF} and \eqref{annx_eq:DL_SG}, it can be noted that the discrete approximation of both approaches, central differences and \SG, has a similar tendency for the real part of the phase velocity $c\ \eqdef\ \dfrac{w}{k}\,$. On the contrary, the imaginary part of the phase velocity is second-order for the \SG ~scheme and only first-order for the \Eu ~one. The dispersion relation for both approaches is illustrated in Figures~\ref{fig_Annx:Re_disp} and \ref{fig_Annx:Im_disp}. For each case, the dispersion relation has an accurate approximation for small values of the wave number $k\,$. The accuracy of \SG ~increases with $k\,$.

\bigskip
\addcontentsline{toc}{section}{References}
\bibliographystyle{abbrv}
\bibliography{biblio}
\bigskip\bigskip

\end{document}